\RequirePackage{fix-cm}
\documentclass[smallextended]{svjour3wide} 
\smartqed  

\usepackage{amsmath,amssymb,bm}
\usepackage{mathrsfs,amsbsy,amssymb,latexsym,amsfonts,amsmath,fancyhdr,lastpage}
\usepackage[dvipdfmx]{graphicx}
\usepackage{epsfig,amsfonts,amsbsy}
\usepackage{fancyhdr}
\usepackage{color}
\usepackage{url}
\usepackage[hidelinks]{hyperref}

\usepackage{ulem}

\newcommand{\der}[2]{\frac{d #1}{d  #2}}

\newcommand{\pder}[2]{\frac{\partial #1}{\partial  #2}}
\newcommand{\pdert}[2]{\frac{\partial^2 #1}{\partial  #2^2}}
\newcommand{\pderc}[3]{\frac{\partial^2 #1}{\partial  #2 \partial  #3}}
\newcommand{\pderf}[3]{\left(\frac{\partial #1}{\partial  #2}\right)_{#3}}
\newcommand{\pdertf}[3]{\left( \frac{\partial^2 #1}{\partial  #2^2}\right)_{#3}}

\newcommand{\bv}[1]{{\boldsymbol #1}}

\newcommand{\nm}{\nonumber\\}

\newcommand{\sbkt}[1]{\langle#1\rangle}
\newcommand{\bbkt}[1]{\bigl\langle#1\bigr\rangle}
\newcommand{\Bbkt}[1]{\Bigl\langle#1\Bigr\rangle}

\DeclareMathOperator*{\argmin}{arg\,min} 
 
\newcommand{\ep}{\varepsilon}

\newcommand{\kB}{k_\mathrm{B}}

\newcommand{\can}{\mathrm{eq}}

\newcommand{\sep}{\mathrm{sep}}

\newcommand{\homo}{\mathrm{homo}}
\newcommand{\hetero}{\mathrm{hetero}}

\newcommand{\originX}{x_\mathrm{o}}
\newcommand{\topX}{x_\mathrm{t}}
\newcommand{\botX}{x_\mathrm{b}}

\newcommand{\subO}{\mathrm{o}}
\newcommand{\subC}{\mathrm{c}}
\newcommand{\subM}{\mathrm{m}}
\newcommand{\subL}{\mathrm{L}}
\newcommand{\subG}{\mathrm{G}}

\newcommand{\subl}{\mathrm{\ell}}
\newcommand{\subu}{\mathrm{u}}
\newcommand{\sublu}{\mathrm{\ell/u}}
\newcommand{\subuR}{\mathrm{abv}}
\newcommand{\sublR}{\mathrm{blw}}
\newcommand{\submR}{\mathrm{mdl}}

\newcommand{\xm}{x_\mathrm{m}}
\newcommand{\xint}{x_\theta}

\newcommand{\Ps}{p_\mathrm{s}}
\newcommand{\Pm}{p_\mathrm{m}}

\newcommand{\bP}{\bar p}
\newcommand{\bmu}{\bar\mu}
\newcommand{\tmu}{\tilde\mu}

\makeatletter
\@addtoreset{equation}{section}
\makeatother
\renewcommand{\theequation}{\thesection.\arabic{equation}}

\journalname{Journal of Statistical Physics}

\begin{document}

\title{Global thermodynamics for isothermal fluids under weak gravity}

\author{
Naoko Nakagawa \and Shin-ichi Sasa \and Takamichi Hirao \and Tsuyoshi Shiina \and Kyosuke Tachi \and Akira Yoshida}
\authorrunning{Nakagawa, Sasa et al} 

\institute{
N. Nakagawa \at
  Department of Physics, Ibaraki University, Mito 310-8512, Japan             
             \email{naoko.nakagawa.phys@vc.ibaraki.ac.jp}
\and 
S.-i. Sasa \at
              Department of Physics, Kyoto University, Kyoto 606-8502, Japan 
              \email{sasa@scphys.kyoto-u.ac.jp}
\and
T. Hirao, T. Shiina, K. Tachi, A. Yoshida\at
  Department of Physics, Ibaraki University, Mito 310-8512, Japan             
}

\date{\today}

\maketitle

\begin{abstract}

 We develop a formulation of global thermodynamics for equilibrium systems under the influence of weak gravity. The free energy for simple fluids is extended to include a dependence on $(T, V, N, mgL)$, where $L$ represents the vertical system length in the direction of gravity. A central idea in this formulation is to uniquely fix the reference point of the gravitational potential, ensuring a consistent thermodynamic framework. Using this framework, we derive the probability density of thermodynamic quantities, which allows us to define a variational function for determining equilibrium liquid-gas coexistence under gravity when the interface is flat. The resulting free energy landscape, derived from the variational function, reveals the local stability of liquid-gas configurations. Specifically, the liquid phase resides at the lower portion of the system due to gravity, while the inverted configuration (with liquid on top) is also locally stable in this landscape. Furthermore, we characterize the transition between these liquid-gas configurations as a first-order phase transition using the thermodynamic free energy of $(T,V,N,mgL)$. Finally, we validate the predictions of global thermodynamics through molecular dynamics simulations, demonstrating the applicability and accuracy of the proposed framework.
\end{abstract}


\setcounter{tocdepth}{3}
\tableofcontents

\clearpage

\section{Introduction}
\label{intro}
Equilibrium thermodynamics characterizes fluids using a limited set of extensive and intensive variables, such as $(T, V, N)$ with temperature $T$, volume $V$, and particle number $N$, based on the assumption of spatial uniformity \cite{Callen,Landau-Lifshitz-Statphys}.
Local descriptions are generally favored to account for the non-uniformity and time evolution of fluids
with local thermodynamic quantities such as local particle density $\rho(\bm{r})$, temperature $T(\bm{r})$, and pressure $p(\bm{r})$ \cite{Landau-Lifshitz-Fluid,Prigogine-Kondepudi}.
We refer to the former description as ``global thermodynamics," contrasting the latter with ``local thermodynamics," typically represented by hydrodynamics. 

Recent research has proposed that global thermodynamics could extend equilibrium thermodynamics to non-equilibrium systems \cite{Global-PRL,Global-JSP,Global-PRR,KNS,Global-modelB}. For instance, heat conduction systems can be described using global thermodynamic variables, regardless of their non-uniformity. Although this global description neglects spatial inhomogeneity, it accurately predicts phenomena such as the deviation of the liquid-gas interface temperature from the equilibrium transition point and the stabilization of metastable local states. Thus, global thermodynamics offers a profound yet effective framework for describing thermodynamic phenomena in heat conduction systems.
However, the predictions of global thermodynamics should be validated based on microscopic dynamics. Nonequilibrium statistical mechanics, though, remains too formal to enable quantitative calculations of macroscopic quantities \cite{Zubarev,Mclennan,Maes-rep}.

Equilibrium systems under gravity show a notable resemblance to heat conduction systems: both exhibit one-dimensional structures, such as gradients in density, pressure, or temperature. Without detailed measurements of local temperature or pressure, the density distributions of equilibrium systems under gravity and those of heat conduction systems may appear indistinguishable. This resemblance suggests that equilibrium systems under gravity could potentially be described using global thermodynamics in a manner similar to heat conduction systems.

\begin{figure}[b] 
\begin{center} 
\includegraphics[width=10cm]{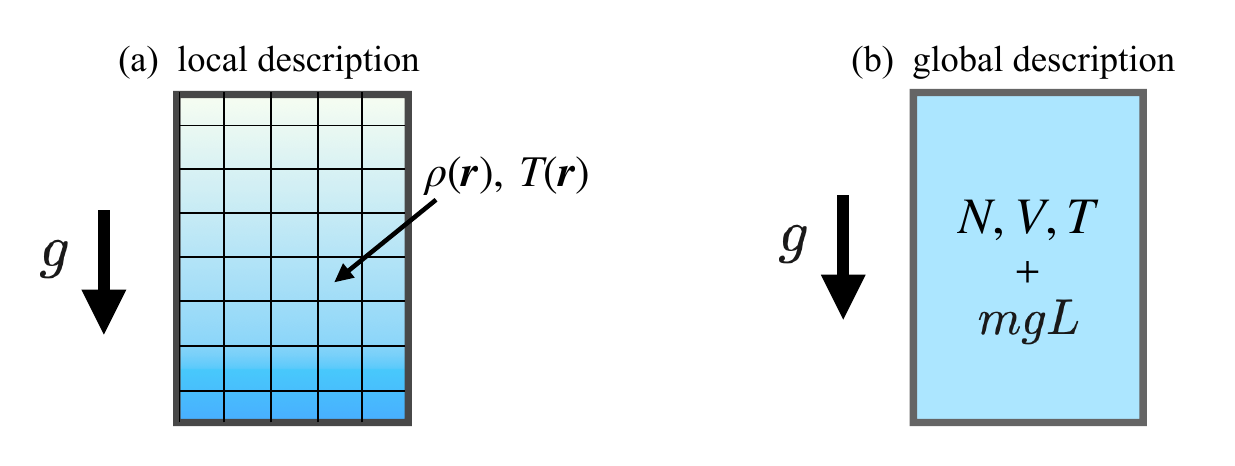}
\end{center} 
\caption{Local and global descriptions of a fluid under weak gravity in (a) and (b). 
Gravity related global thermodynamic variable $mgL$ is the gravitational potential energy from the bottom to the top. See Sec. \ref{s:GlobalThermo}.}
\label{fig:intro}
\end{figure}

Historically, fluids under gravity have been studied primarily within a hydrodynamic framework as shown in Fig.~\ref{fig:intro}(a), where the global perspective is often disregarded due to its lack of spatial detail. 
It is worth noting, however, that many fundamental experiments in equilibrium thermodynamics are conducted under the influence of gravity.
This might have led to the assumption that  ``weak" gravity has no effect on global thermodynamics.
While this may be true in simple gases or liquids, the situation for liquid-gas coexistence is significantly complex.
This raises an important question: Can systems under gravity be ``comprehensively" described from a global thermodynamic viewpoint?
In this context, the extension of thermodynamics includes the reformulation of the variational principles, such as entropy maximization and free energy minimization.
It is well known that gravity plays a crucial role in determining the configuration of liquid and gas,
while the variational principles without gravity determine the distribution of liquid and gas. 
This suggests a potential discrepancy between the phenomena observed in the presence of gravity and the results predicted by conventional variational principles.
By revising these variational principles, we can examine whether the gravitational force modifies the balance condition between liquid and gas.
This approach also enables a comparison of the effect of gravity and heat flow on equilibrium conditions of liquid-gas systems.

In this paper, we aim to incorporate weak gravitational effects into the equilibrium thermodynamics of liquid-gas coexistence. A useful example of such an extension is found in magnetic fluids, where the thermodynamic state is expressed by $(T, V, N, H)$, with the fundamental relation given by $dF = -SdT - pdV + \mu dN - MdH$. 
Here, $H$ denotes the applied magnetic field, and $M$ represents the magnetization \cite{Kubo}. 
Similarly, extending thermodynamics to include gravitational effects requires identifying the appropriate conjugate pairs of extensive and intensive variables. 
Unlike heat conduction systems, equilibrium statistical mechanics can be directly applied to systems under gravity in a straightforward manner. 
This provides a robust framework for addressing non-uniform states, whether in liquid, gas, or liquid-gas coexistence, and facilitates the identification of new conjugate pairs of extensive and intensive gravity-related variables.
Figure ~\ref{fig:intro}(b) illustrates the global description for the fluid under weak gravity, which will be concluded in Sec. \ref{s:GlobalThermo},  in comparison with the local description for the same fluid in Fig.~\ref{fig:intro}(a).

It is important to note that variational functions in equilibrium thermodynamics correspond to the logarithm of the probability densities of macroscopic fluctuations, which can be derived from canonical distributions \cite{Einstein,Landau-Lifshitz-Statphys}. 
The minimum (or maximum) value of a variational function determines the thermodynamic free energy (or thermodynamic entropy) for the entire system, for which the fundamental relation of thermodynamics remains valid. 
Consequently, any newly identified conjugate pair would be incorporated into this framework. 
The consistency among thermodynamic functions, variational functions, and the probability densities of macroscopic fluctuations serves as a guiding principle for extending the thermodynamic framework to non-uniform systems under gravity. 
Moreover, the probability densities of macroscopic fluctuations underpin the Onsager relations and non-equilibrium thermodynamics \cite{Onsager1931,Groot-Mazur}. 
An extension of thermodynamics that preserves consistency among these three elements could pave the way for the development of non-equilibrium thermodynamics under gravity.

The paper is organized as follows. Section 2 presents an overview of the main results derived from equilibrium statistical mechanics. The setup for equilibrium statistical mechanics is introduced in Section 3. Section 4 defines the free energy that characterizes the thermodynamic state of a system under gravity. In Section 5, a formulation of global thermodynamics under gravity is proposed based on this free energy. Sections 6 and 7 focus on constructing a variational principle for determining the spatial configuration of liquid and gas. The variational equations are then solved in Section 8. Section 9 is for the stability of the solutions. The free energy is compared among respective solutions in Section 10.
The free energy landscape and thermodynamic free energy are presented in Section 11 with a singularity at $g=0$ exhibiting the feature of the first-order transition. Numerical verification of these theoretical results is provided in Section 12, followed by concluding remarks in Section 13.

\section{Main results}

We focus on a single component fluid of $N$ particles enclosed in a cuboid container of volume $V$.  
At constant temperature $T$, the system without gravity is characterized by free energy $F(T,V,N)$,  
satisfying a fundamental relation of thermodynamics,  
\begin{align}
dF = -SdT - pdV + \mu dN.  
\label{e:F-relation-g0-main}
\end{align}
From the extensive property, we define the free energy density as  
\begin{align}
f(T,\rho) = F(T,V,N)/V,  
\end{align}
where $\rho = N/V$. Similarly, we have the entropy density $s(T,\rho) = S(T,V,N)/V$, etc.  
These densities form the basis of the local thermodynamics adopted in the hydrodynamic description.

\begin{figure}[bt] 
\begin{center} 
\includegraphics[width=7cm]{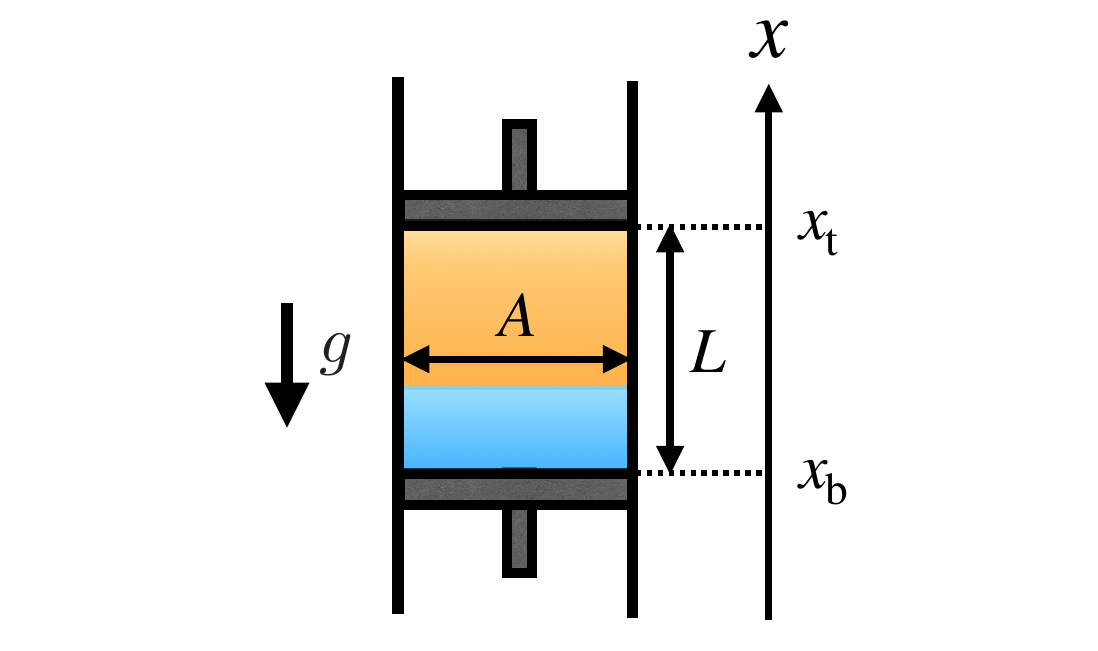}
\end{center} 
\caption{Configuration of the system under gravity. $\botX$ and $\topX$ are the positions of the bottom and the top of the container with $L=\topX-\botX$.}
\label{fig:config}
\end{figure}

Impose weak gravity in the opposite direction to the $x$-axis.  
With the mass $m$ of each particle and gravitational acceleration $g$, the gravitational potential is $mg(x-x_\subO)$ with an arbitrary reference point at $x_\subO$.  
Suppose the cuboid container has a vertical length $L$ in the $x$-direction and a horizontal area $A$.
The positions of the bottom and the top are fixed at $\botX$ and $\topX$, respectively,  
where $L = \topX - \botX$.  See Fig.~\ref{fig:config}.
All walls are made of the same materials and interact with particles in a short range.  
In the thermodynamic limit, the equilibrium state becomes horizontally uniform under gravity, and all local quantities are functions of $x$, such as $f(T, \rho(x))$, where $T(x) = T$ everywhere.

Set a dimensionless parameter for representing the effect of the gravitational acceleration as  
\begin{align}
\ep = \frac{mgL}{\kB T}.  
\label{e:ep-eq}
\end{align}
This paper concentrates on weak gravity cases and neglects the contribution of $O(\ep^2)$.  
Our first result is the extension of thermodynamics obtained from equilibrium statistical mechanics.  
It is revealed that an extra pair of extensive and intensive variables is required, such that  
\begin{align}
(T, V, N) \rightarrow (T, V, N, mgL)  
\end{align}
with a fundamental relation of global thermodynamics as  
\begin{align}
&dF_g = -SdT - \bar{p} dV + \bmu dN - N\frac{\xm - X}{L}d(mgL).  
\label{e:Fg-relation-main}
\end{align}
Here, $F_g$ is the free energy under gravity, which is given by  
\begin{align}
F_g &= A\int_{\botX}^{\botX+L}dx~[f(T,\rho(x)) + mg(x-\xm)\rho(x)]  
\label{e:Fg-eq}
\end{align}
with the local free energy density $f(T,\rho(x))$.  
$\xm$ is the midpoint of the system,  
\begin{align}
\xm = \frac{\botX + \topX}{2},  
\end{align}
and $X$ is the center of mass,  
\begin{align}
X = \frac{A}{N}\int_{\botX}^{\botX+L}dx~x \rho(x).  
\end{align}
Due to gravity, the pressure and chemical potential become dependent on $x$,  
while they are homogeneous without gravity.  
Then $p$ and $\mu$ in \eqref{e:F-relation-g0-main} are replaced by the spatial averages  
\begin{align}
&\bP \equiv \frac{1}{L}\int_{\botX}^{\botX+L} dx~p(x),\qquad  
\bmu \equiv \frac{1}{L}\int_{\botX}^{\botX+L} dx~\mu(x)  
\end{align}
in \eqref{e:Fg-relation-main}.  
We remark that fluids in a single-phase state, such as pure gas or pure liquid, follow thermodynamics without gravity, characterized by $(T,V,N)$, because $X = \xm + O(\ep^2)$.  
At the phase coexistence, $X \neq \xm$ as the liquid is placed below the gas.  
The new term appearing in \eqref{e:Fg-relation-main} becomes essential.  

The second result is a free energy landscape for determining the equilibrium configuration of the liquid-gas coexistence under gravity. 
In the light of global thermodynamics with $(T,V,N,mgL)$ in \eqref{e:Fg-relation-main},  
we restate the common phenomenon that liquid is placed below gas as the emergence of the heterogeneous structure via a first-order transition.  
From this point of view, the system without gravity is just at the transition point.  
Let $\subl$ and $\subu$ be the respective phases occupying the lower and upper regions in the container,  
and ${\mathcal N}^\subl$ and ${\mathcal V}^\subl$ be the number of particles and the volume of the phase $\subl$.  
We derive the variational function for determining the equilibrium state in liquid-gas coexistence,  
\begin{align}
&{\mathcal F}_g({\mathcal N}^\subl,{\mathcal V}^\subl;T,V,N,mgL) = F(T,{\mathcal V}^\subl,{\mathcal N}^\subl) + F(T,V-{\mathcal V}^\subl,N-{\mathcal N}^\subl) + \frac{NmgL}{2}\left(\frac{{\mathcal V}^\subl}{V} - \frac{{\mathcal N}^\subl}{N}\right).
\label{e:Fg-var-main}
\end{align}
The third term in \eqref{e:Fg-var-main} coincides with the spatial integration of the gravitational term in \eqref{e:Fg-eq}.  
The landscape of ${\mathcal F}_g$ in the space of $({\mathcal N}^\subl, {\mathcal V}^\subl)$ possesses two local minima:  
one is the configuration $(\subl, \subu) = (\subL, \subG)$ where the liquid is below the gas.  
The other is $(\subl, \subu) = (\subG, \subL)$ where the liquid is above the gas.  
Comparing the values of ${\mathcal F}_g$ for the two configurations, we find that $(\subl, \subu) = (\subL, \subG)$ is the equilibrium state corresponding to the global minimum for $g > 0$ due to the extra third term in \eqref{e:Fg-var-main}.  
The reverse configuration $(\subl, \subu) = (\subG, \subL)$ becomes the metastable state belonging to the local minimum.  
When changing the value of $mgL$ from positive to negative, we observe the interchange of the two minima  
at $mgL = 0$, providing a sudden change in the configuration as the upside-down of liquid and gas.  
This phenomenon, while well known, is formulated in global thermodynamics as a first-order transition.

Global thermodynamics provides a consistent result with hydrodynamics, showing that all local states, including the vicinity of the liquid-gas interface, are thermodynamically stable.  
The variational equations, which are satisfied by both the global and local minima, can be expressed as  
\begin{align}
\pderf{{\mathcal F}_g}{{\mathcal N}^\subl}{T,V,N,mgL} = 0, \qquad  
\pderf{{\mathcal F}_g}{{\mathcal V}^\subl}{T,V,N,mgL} = 0,  
\end{align}
which result in the continuity of the pressure profile with  
\begin{align}
p_\theta = \Ps(T).  
\end{align}
Here, $p_\theta$ is the pressure at the liquid-gas interface, and $\Ps(T)$ is the saturation pressure at temperature $T$.  
This demonstrates that the coexistence conditions in conventional equilibrium thermodynamics are satisfied locally at the interface.

\section{Equilibrium statistical mechanics}
\subsection{Setup}

Let $\Gamma=(\bv{r}_1, \cdots,\bv{r}_N,\bv{p}_1,\cdots,\bv{p}_N)$ denote the microstate of an $N$-particle system.
The Hamiltonian without gravity is given by
\begin{align}
H(\Gamma;\botX,\topX,A,N)=H_0(\Gamma;N)+\sum_{i=1}^N\phi_{\rm w}(\bv{r}_i;\topX,\botX,A),
\label{e:Hamiltonian0}
\end{align}
where $H_0(\Gamma;N)$ includes the kinetic energy and potentials describing interactions among particles. 
The term $\phi_{\rm w}(\bv{r}_i;\topX,\botX,A)$ represents the interaction potential between a particle and the walls. 
We assume that these interactions are short-ranged.
Given the symmetry in the $y$- and $z$-directions, the parameters specifying the positions of the four vertical walls are simplified and characterized by the horizontal area $A$.
For simplicity, we further assume symmetry between the bottom and top walls, such that $\phi_{\rm w}(\bv{r}_i;\topX,\botX,A)=\phi_{\rm w}(\bv{r}_i;\botX,\topX,A)$.
Any breakdown of this spatial symmetry is attributed solely to the presence of gravity.

The Hamiltonian under gravity is expressed as
\begin{align}
H_g^\subO(\Gamma;\botX,\topX,A,N,mg)=H(\Gamma;\botX,\topX,A,N)+\sum_{i=1}^N mg (x_i-\originX),
\label{e:Hamiltonian-o}
\end{align}
where $\originX$ is an arbitrary reference point for the gravitational potential, such as the position of the ground.
We consider an isothermal system at temperature $T$ under gravity.
The equilibrium statistical mechanics framework provides the canonical distribution, partition function, and the corresponding free energy as
\begin{align}
&\rho^\can(\Gamma)=\frac{1}{Z_g^\subO}e^{-\beta H_g^\subO(\Gamma)},\label{e:rho0-can}\\
&Z_g^\subO=\int d\Gamma~ e^{-\beta H_g^\subO(\Gamma)},\label{e:ZgO-def}\\
&F_g^\subO=-\kB T~(~\ln Z_g^\subO-\ln N!~),\label{e:FgO-def}
\end{align}
where $\beta=1/\kB T$ is the inverse temperature, and $\kB$ denotes the Boltzmann constant.

\subsection{Total differential of $F_g^\subO$}

The quantity $F_g^\subO$, defined from statistical mechanics, may be regarded as free energy. However, it depends on several macroscopic parameters $(T,\botX,\topX,A,N,mg)$ as well as the arbitrary reference point $x_\subO$.
To explore the physical meaning of $F_g^\subO$, we derive the total differential relation for $F_g^\subO$.
In the following, ensemble averages taken with the canonical distribution \eqref{e:rho-can} are denoted by $\bbkt{\cdot}$.

Using \eqref{e:ZgO-def} and \eqref{e:FgO-def}, 
the average energy is given by
\begin{align}
E^\subO\equiv \bbkt{H_g^\subO}=-\pderf{\ln Z_g^\subO}{\beta}{\botX,\topX,A,N,mg}, 
\label{e:E-def-mech}
\end{align}
from which the entropy of the system is written as
\begin{align}
S=\frac{E^\subO-F_g^\subO}{T}=-\pderf{F_g^\subO}{T}{\botX,\topX,A,N,mg}.
\label{e:S-def-mech}
\end{align}
For the dependence of $F_g^\subO$ on $mg$, we have
\begin{align}
\pderf{F_g^\subO}{(mg)}{T,\botX,\topX,A,N}=\Bbkt{\pder{H_g^\subO}{(mg)}}=N(X-\originX),
\end{align}
where $X$ is the center of mass,
\begin{align}
X=\Bbkt{\frac{1}{N}\sum_{i=1}^N x_i}.
\end{align}

The pressure in the system depends on the height $x$ due to the gravitational force.
Operationally, we consider the pressure $p_{\rm b}$ at the bottom, $p_{\rm t}$ at the top, and the spatially averaged pressure $\bP$ over the system.
While the pressure varies with height $x$, the mean total force exerted by the particles on the side walls corresponds to $\bP$.
Using \eqref{e:ZgO-def} and \eqref{e:FgO-def}, the pressures are expressed as
\begin{align}
&p_{\rm b} A=\Bbkt{\pder{H_g^\subO}{\botX}}=\pderf{F_g^\subO}{\botX}{T,\topX,A,N,mg},\\
&p_{\rm t} A=-\Bbkt{\pder{H_g^\subO}{\topX}}=-\pderf{F_g^\subO}{\topX}{T,\botX,A,N,mg},\\
&\bar{p} L=-\Bbkt{\pder{H_g^\subO}{A}}=-\pderf{F_g^\subO}{A}{T,\botX,\topX,N,mg}.
\end{align}

It should be noted that the system is not extensive in the direction of gravity but remains extensive and homogeneous in directions perpendicular to gravity.
This property is expressed as
\begin{align}
F_g^\subO(T,\botX,\topX,\lambda A,\lambda N,mg)=\lambda F_g^\subO(T,\botX,\topX,A,N,mg),
\end{align}
for any positive $\lambda$.
Consequently, we obtain
\begin{align}
\mu_g^\subO=\frac{F_g^\subO+\bar{p} LA}{N},
\label{e:mug-ES}
\end{align}
where $\mu_g^\subO$ is defined as
\begin{align}
\mu_g^\subO\equiv\pderf{F_g^\subO}{N}{T,\botX,\topX,A,mg}.
\label{e:mug-def}
\end{align}

Summarizing the above relations, the total differential of $F_g^\subO$ is given by
\begin{align}
dF_g^\subO=&-SdT+p_{\rm b} Ad\botX-p_{\rm t} Ad\topX-\bar{p} LdA +\mu_g^\subO dN+N(X-\originX)d(mg).
\label{e:Fg-relation-o}
\end{align}
While $F_g^\subO$, $S$, and $N$ are global extensive quantities, $\botX$ and $\topX$ and their conjugate quantities $p_{\rm b}$ and $p_{\rm t}$ are local quantities.

\section{Free energy characterizing thermodynamic states under gravity}

The total differential \eqref{e:Fg-relation-o} arises from the Hamiltonian \eqref{e:Hamiltonian-o} in accordance with equilibrium statistical mechanics.  
We now consider whether $F_g^\subO$ functions as a thermodynamic free energy for fluids under gravity.  
A challenge lies in the arbitrary reference point $x_\subO$ for the gravitational potential energy.  
Since thermodynamic properties are independent of the choice of $x_\subO$, thermodynamic functions must also be free from $x_\subO$.  
Thus, $F_g^\subO$ is not considered a thermodynamic function as it stands, because the relation \eqref{e:Fg-relation-o} explicitly involves $x_\subO$.  

To examine this further, we analyze two operations illustrated in Fig.~\ref{fig:shift}, which leave the thermodynamic states of the systems unchanged.  
These operations expose non-thermodynamic contributions in $F_g^\subO$.  
We also demonstrate the unique determination of $x_\subO$ to define a free energy $F_g$ that accurately characterizes the thermodynamic states of systems under gravity.  

\subsection{Translational symmetry}
\label{s:Elevation}

\begin{figure}[bt] 
\begin{center} 
\includegraphics[width=13cm]{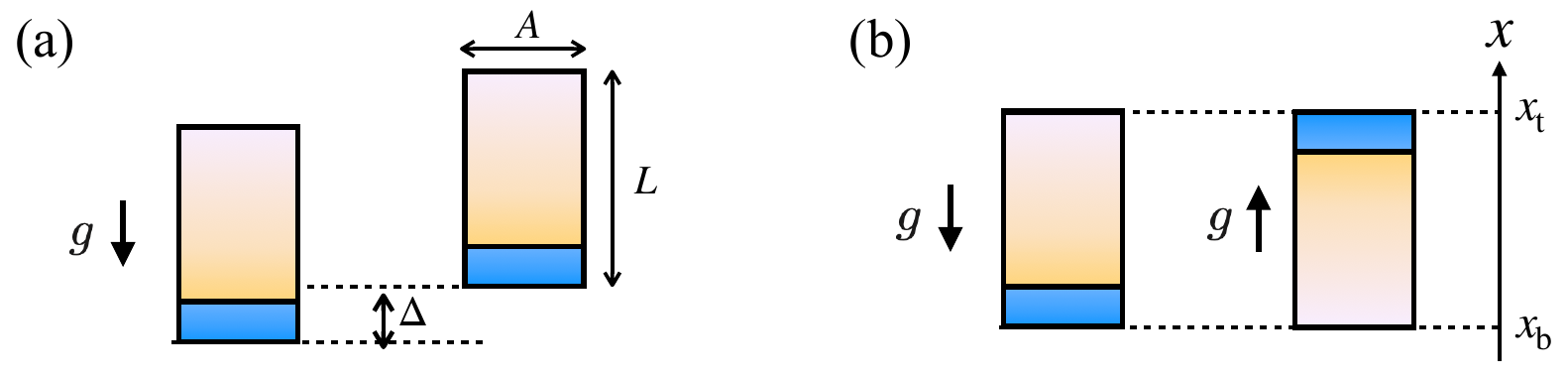}
\end{center} 
\caption{Thermodynamic invariance for two operations. (a) Invariance due to the translational symmetry for the shift along the $x$-axis. The operation produces $\Delta F_g^\subO=(p_{\rm b}-p_{\rm t})A\Delta$.
(b) Invariance due to the reflection symmetry for reversing the direction of gravity. The operation produces $\Delta F_g^\subO=2Nmg(\xm-x_\subO)$.}
\label{fig:shift}
\end{figure}

Consider elevating the container vertically, such that $\botX \to \botX + \Delta$ and $\topX \to \topX + \Delta$, as illustrated in Fig.~\ref{fig:shift}(a).  
This operation does not alter the thermodynamic state of the system but increases the system's potential energy by $Nmg\Delta$.  

Explicitly, under this elevation, the relation \eqref{e:Fg-relation-o} gives the change in $F_g^\subO$ as
\begin{align}
\Delta^{\rm ev} F_g^\subO \equiv (p_{\rm b} - p_{\rm t}) A \Delta.
\label{e:buoyancy}
\end{align}
Using the force balance 
\begin{align}
(p_{\rm b} - p_{\rm t}) A = Nmg,
\label{e:buoyancy-gravity}
\end{align}
\eqref{e:buoyancy} simplifies to 
\begin{align}
\Delta^{\rm ev} F_g^\subO = Nmg \Delta.
\label{e:dFg^o-shift}
\end{align}
Thus, $\Delta^{\rm ev} F_g^\subO$ solely represents the increase in potential energy caused by the container's elevation, without any change in the thermodynamic state.

\subsection{Reflectional symmetry}
\label{s:Reverse}

Now, consider the thermodynamic state when the direction of gravity is reversed, i.e., $g \to -g$.  
The thermodynamic state remains identical for $g$ and $-g$, as shown in Fig.~\ref{fig:shift}(b).  
Let $X(g)$ denote the center of mass for positive $g$ (left panel).  
The center of mass for negative gravity, $X(-g)$ (right panel), satisfies
\begin{align}
X(g) - \botX = \topX - X(-g).
\end{align}
The gravitational potential for the system at $g$ is given by $Nmg(X(g) - x_\subO)$.  
Thus, with $(T, \botX, \topX, A, N)$ fixed, the transformation from $-g$ to $g$ changes $F_g^\subO$ by
\begin{align}
\Delta^{\rm rv} F_g^\subO &\equiv Nmg(X(g) - x_\subO) - Nm(-g)(X(-g) - x_\subO) \nonumber \\
&= 2Nmg(\xm - x_\subO).
\label{e:dFg^o-reverse}
\end{align}
Here, $\xm$ is the midpoint of the container,
\begin{align}
\xm = \frac{\botX + \topX}{2}.
\label{e:xm-def}
\end{align}
This midpoint is independent of the system's thermodynamic state.

\subsection{Free energy under gravity}

Based on the considerations above, we define the free energy $F_g$ by subtracting the non-thermodynamic contributions \eqref{e:dFg^o-shift} and \eqref{e:dFg^o-reverse} from $F_g^\subO$.  
Given the arbitrariness of the reference point $x_\subO$, we define
\begin{align}
F_g \equiv F_g^\subO - Nmg (c - x_\subO),
\label{e:Fg-m-0}
\end{align}
where $c$ depends on $\botX$ and $\topX$ but is independent of $(T, A, L, N, mg)$.  
$F_g$ should not change in the two operations above.
The change in the elevation $ (\botX , \topX)\to (\botX +\Delta, \topX+\Delta)$ leads to
\begin{align}
0 = \Delta^{\rm ev} F_g^\subO - Nmg \left[ c(\botX+\Delta,\topX+\Delta) - c(\botX,\topX) \right],
\label{con1}
\end{align}
and in the reversal of $g$,
\begin{align}
0 = \Delta^{\rm rv} F_g^\subO - \left[ Nmg \left( c(\botX,\topX) - \originX \right) - Nm \left( -g \right) \left( c(\botX,\topX) - \originX \right) \right].
\label{con2}
\end{align}
From \eqref{e:dFg^o-shift} and \eqref{con1}, the equation for $c$ becomes
\begin{align}
c(\botX + \Delta, \topX + \Delta) = c(\botX, \topX) + \Delta,
\label{e:c-Delta}
\end{align}
for any $\Delta$.  
Comparing \eqref{con2} with \eqref{e:dFg^o-reverse} leads directly to
\begin{align}
c(\botX, \topX) = \xm,
\end{align}
which satisfies \eqref{e:c-Delta}.  
Thus, we uniquely determine 
\begin{align}
F_g \equiv F_g^\subO - Nmg (\xm - x_\subO).
\label{e:Fg-m}
\end{align}
Accordingly, the chemical potential $\mu_g$ is defined as
\begin{align}
\mu_g \equiv \pderf{F_g}{N}{T, \botX, \topX, A, mg} = \mu_g^\subO - mg (\xm - x_\subO).
\label{e:mug-m}
\end{align}
Equation \eqref{e:Fg-m} is interpreted as reassigning the reference point for the gravitational potential from $x_\subO$ to $\xm$.

In light of \eqref{e:Fg-m}, the Hamiltonian $H_g^\subO$ in \eqref{e:Hamiltonian-o} is replaced by
\begin{align}
H_g(\Gamma; \botX, \topX, A, N, mg) = H(\Gamma; \botX, \topX, A, N) + \sum_{i=1}^N mg(x_i - \xm).
\label{e:Hamiltonian}
\end{align}
This replacement does not alter the canonical distribution or ensemble averages.  
Using the new Hamiltonian \eqref{e:Hamiltonian}, the partition function and corresponding free energy are redefined as
\begin{align}
Z_g &= \int d\Gamma~ e^{-\beta H_g(\Gamma)},
\label{e:Zg-def} \\
F_g &= -k_B T \left( \ln Z_g - \ln N! \right).
\label{e:Fg-def}
\end{align}
The free energy $F_g$ in \eqref{e:Fg-def} is equivalent to that in \eqref{e:Fg-m}.  
The canonical distribution \eqref{e:rho0-can} is rewritten as
\begin{align}
\rho^\can(\Gamma) = \frac{1}{Z_g} e^{-\beta H_g(\Gamma)}.
\label{e:rho-can}
\end{align}

\subsection{Total differential of $F_g$}

We now show that the total differential of $F_g$, as defined in \eqref{e:Fg-m}, is independent of the reference point $x_\subO$.  
The definition \eqref{e:Fg-m} gives
\begin{align}
dF_g = dF_g^\subO - mg (\xm - x_\subO) dN - N (\xm - x_\subO) d(mg) - Nmg d\xm.
\label{e:Fg-relation-00}
\end{align}
Substituting \eqref{e:Fg-relation-o} into \eqref{e:Fg-relation-00}, we obtain
\begin{align}
dF_g = -S dT + p_{\rm b} A d\botX - p_{\rm t} A d\topX - \bar{p} L dA + \mu_g dN + N (X - \xm) d(mg) - Nmg d\xm.
\label{e:Fg-relation-0}
\end{align}

Using $L = \topX - \botX$ and \eqref{e:xm-def}, we have $\botX = \xm - L/2$ and $\topX = \xm + L/2$.  
Thus,
\begin{align}
p_{\rm b} A d\botX - p_{\rm t} A d\topX = -\Pm A dL + Nmg d\xm,
\end{align}
where 
\begin{align}
\Pm = \frac{p_{\rm b} + p_{\rm t}}{2}.
\label{e:pm-def}
\end{align}
The total differential \eqref{e:Fg-relation-0} is now expressed as
\begin{align}
dF_g = -S dT - \Pm A dL - \bar{p} L dA + \mu_g dN - N (\xm - X) d(mg).
\label{e:Fg-relation-m}
\end{align}
Remarkably, \eqref{e:Fg-relation-m} is unaffected by the container's shift, represented by changes in $\xm$.  
Only the difference $X - \xm$ is relevant from a thermodynamic perspective.

The midpoint $\xm$ shifts when $\topX$ or $\botX$ changes.  
Equation \eqref{e:Fg-m} implies that the updated value of $\xm$ must be adopted for the free energy to accurately characterize the thermodynamic state.  
General operations altering $T$, $L$, $A$, or $mg$ do not change the position of $\xm$, but they may shift $X$.  
Such shifts result in $\Delta F_g \neq 0$, reflecting a change in the thermodynamic state.

\section{Global thermodynamics under weak gravity}
\label{s:GlobalThermo}

We further transform \eqref{e:Fg-relation-m} into
\begin{align}
dF_g = -S dT - \bP dV + \mu_g dN - \Psi_g d(mgL) 
+ (\bar{p} - \Pm + mg \bar{\rho} (\xm - X)) A dL,
\label{e:Fg-relation-pre}
\end{align}
where $V = AL$, $\bar{\rho} = N/V$, and
\begin{align}
\Psi_g = N \frac{\xm - X}{L}.
\label{e:Psig-def}
\end{align}
Because no particle currents flow in equilibrium, forces balance at any local point $x$, and, therefore, the pressure profile $p(x)$ becomes continuous throughout the system.  
With this constraint, we have
\begin{align}
\bar{p} - \Pm = mg \bar{\rho} (X - \xm), 
\label{e:bP-Pm}
\end{align}
as shown in Sec.~\ref{s:bP-Pm}.  
Then, the relation \eqref{e:Fg-relation-pre} forms a fundamental thermodynamic relation as
\begin{align}
dF_g = -S dT - \bar{p} dV + \mu_g dN - \Psi_g d(mgL),
\label{e:Fg-relation}
\end{align}
with an error of $O(\ep^2)$.  
The fourth term is newly added and is essential for describing the nonuniformity of the local states brought about by gravity.  
We thus extend the equilibrium thermodynamic variables as
\begin{align}
(T, V, N) \rightarrow (T, V, N, mgL)
\label{e:thermo-extension}
\end{align}
for equilibrium systems under gravity.  

In a pure liquid or pure gas, $X - \xm = O(\ep)$.  
Accordingly, in the single-phase state, we have $\Psi_g = O(\ep)$, and therefore, the fourth term of \eqref{e:Fg-relation} is negligible.  
The fundamental relation \eqref{e:Fg-relation} then becomes equivalent to the fundamental relation of thermodynamics without gravity.  
In liquid-gas coexistence, $X - \xm$ becomes $O(\ep^0)$, where $X < \xm$ for $g > 0$ and $X > \xm$ for $g < 0$.  
The fourth term in \eqref{e:Fg-relation} remains $O(\ep)$ because $\Psi_g = O(\ep^0)$.  
Thus, it is essential in the liquid-gas coexistence to extend thermodynamics as in \eqref{e:thermo-extension}.  

\subsection{Scaling relation}

\begin{figure}[tb] 
\begin{center} 
\includegraphics[width=8cm]{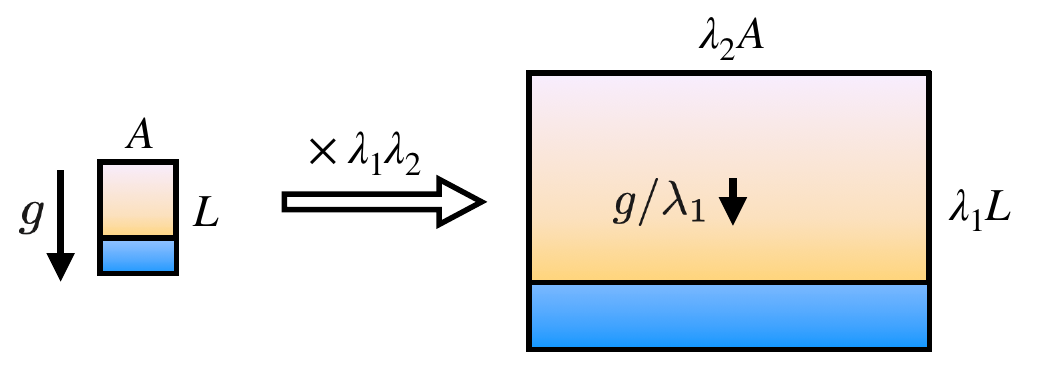}
\end{center} 
\caption{Scaling property formulated in \eqref{e:extensivity}.
}
\label{fig:extensive}
\end{figure}

The scaling properties connecting thermodynamic functions of different system sizes are anisotropic due to gravity.  
Consider two systems with $(L, A, N)$ and $(\lambda_1 L, \lambda_2 A, \lambda_1 \lambda_2 N)$, with arbitrary positive constants $(\lambda_1, \lambda_2)$.  
The fundamental relation \eqref{e:Fg-relation} indicates 
\begin{align}
F_g(T, \lambda_1 L, \lambda_2 A, \lambda_1 \lambda_2 N, mg / \lambda_1) 
= \lambda_1 \lambda_2 F_g(T, L, A, N, mg),
\label{e:extensivity}
\end{align}
where the variables are denoted as $(T, L, A, N, mg)$ instead of $(T, V, N, mgL)$ to clarify the point.  
The standard scaling property $(N, V, F_g) \rightarrow (\lambda_1 \lambda_2 N, \lambda_1 \lambda_2 V, \lambda_1 \lambda_2 F_g)$ holds only when the magnitude of gravity is changed as $g \rightarrow g / \lambda_1$ in accordance with the change $L \rightarrow \lambda_1 L$.  
See Fig.~\ref{fig:extensive}.  

\begin{figure}[tb] 
\begin{center} 
\includegraphics[width=10cm]{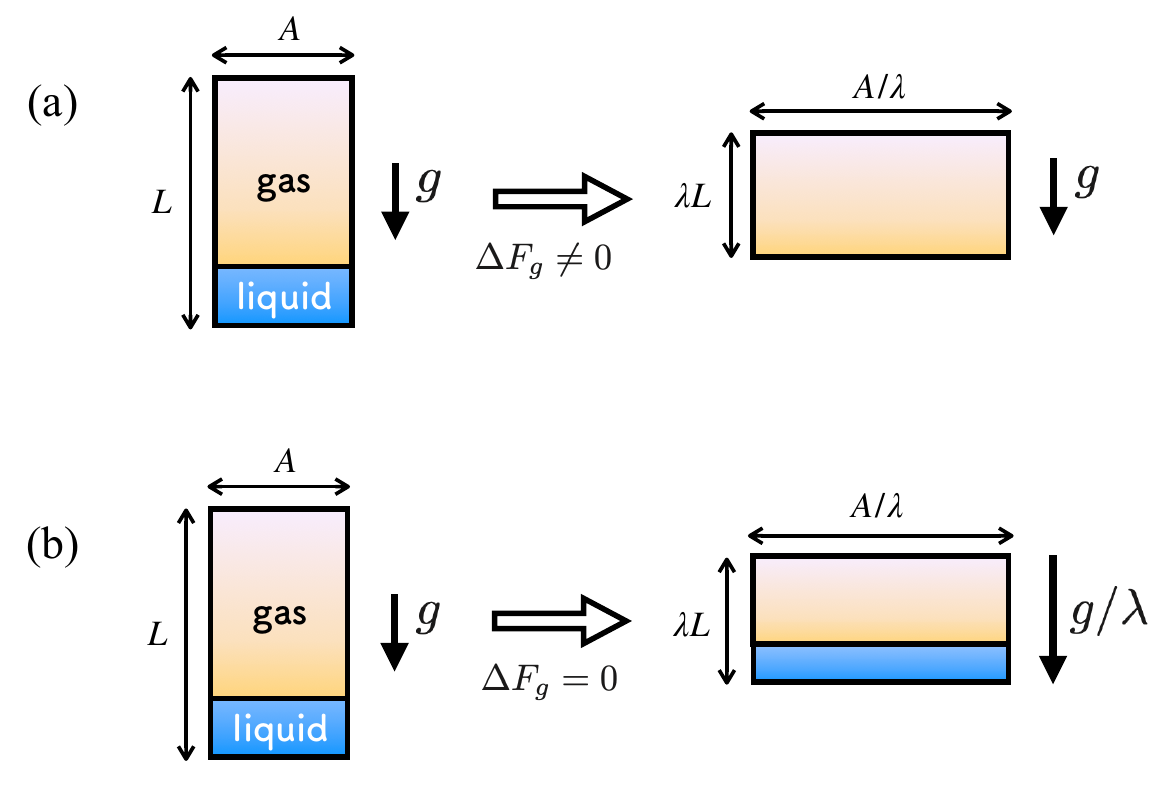}
\end{center} 
\caption{
Two operations that alter the aspect ratio of a cuboid container while keeping the volume constant. 
(a) The liquid-gas coexistence observed in the initial equilibrium state transitions to a gas phase when the container is flattened under constant gravity. 
(b) The amount of liquid remains unchanged when the gravitational acceleration is increased to compensate for the flattening operation.
}
\label{fig:mgL-const}
\end{figure}

To aid understanding, let us compare two operations to change the shape of the container with $V$ fixed, as shown in Fig.~\ref{fig:mgL-const}.  
We set $(T, V, N, mgL)$ such that liquid coexists with gas under gravity.  
We change the aspect ratio of the container by a factor $\lambda$ such that the volume $V$ of the container remains unchanged, i.e., $V = LA = (\lambda L)(A / \lambda)$.  
In this change, the scaling relation \eqref{e:extensivity} is written as 
$F_g(T, \lambda L, A / \lambda, N, mg / \lambda) = F_g(T, L, A, N, mg)$.  

The value of $F_g$ may change during the operation shown in Fig.~\ref{fig:mgL-const}(a) but remains unchanged in the operation shown in Fig.~\ref{fig:mgL-const}(b).  
Suppose $\lambda < 1$.  
When the vertical length becomes shorter with $g$ fixed, as shown in Fig.~\ref{fig:mgL-const}(a), the pressure difference $p_{\rm b} - p_{\rm t}$ decreases.  
This change causes the system's state to deviate from the initial state; for instance, the liquid may evaporate, or the gas may condense.  
Correspondingly, \eqref{e:Fg-relation} leads to $\Delta F_g \neq 0$ in the operation in Fig.~\ref{fig:mgL-const}(a).  

The fundamental relation \eqref{e:Fg-relation} shows that $\Delta F_g$ can vanish only if the operation is accompanied by a change in gravitational acceleration.  
In the operation shown in Fig.~\ref{fig:mgL-const}(b), with $g \rightarrow g / \lambda$, the fourth term in \eqref{e:Fg-relation} does not contribute to $\Delta F_g$.  
In this case, $p_{\rm b}$, $p_{\rm t}$, and the liquid-gas distribution remain unchanged from the initial state.

\subsection{Derivation of \eqref{e:bP-Pm}}
\label{s:bP-Pm}

When the system is in the single-phase state, \eqref{e:bP-Pm} holds trivially because we have $X = \xm + O(\ep)$ and $\bP = \Pm + O(\ep^2)$.  
These results can be derived by applying the trapezoidal rule to the nearly linear profiles of $\rho(x)$ and $p(x)$.  

We now focus on the liquid-gas coexistence.  
Assuming that the liquid-gas interface is single and flat, we label the phases below and above the interface as $\subl$ and $\subu$, respectively, and use superscripts $\subl$ and $\subu$ for the quantities in each phase, such as $N^\subl$, $N^\subu$, $X^\subl$, $X^\subu$, etc.  
Under weak gravity, the density and pressure profiles in $\subl$ and $\subu$ are approximately linear with respect to $x$.  
Thus,  
\begin{align}
X^\sublu = \xm^\sublu + O(\ep), \qquad
\bar{p}^\sublu = \Pm^\sublu + O(\ep^2).  
\label{e:x-p-mean}
\end{align}  

Because the force balance holds at every point $x$, the pressure $p(x)$ is continuous across the liquid-gas interface.  
Let $p_\theta$ denote the pressure at the interface.  
The definitions of $\bar{p}$ and $\Pm$ are as follows:  
\begin{align}
&\bar{p}= \frac{1}{L} \int_{\botX}^{\topX} dx \, p(x) = \frac{V^\subl}{V} \bar{p}^\subl + \frac{V^\subu}{V} \bar{p}^\subu,  \\  
&\Pm = \frac{p_{\rm b} + p_{\rm t}}{2},  
\end{align}
where $p_{\rm b} = p(\botX)$ and $p_{\rm t} = p(\topX)$.  
Subtracting the above expressions and using \eqref{e:x-p-mean}, we find  
\begin{align}
\bar{p} - \Pm = \frac{V^\subl}{2V}(p_\theta - p_{\rm t}) + \frac{V^\subu}{2V}(p_\theta - p_{\rm b}),  
\end{align}
where we used $\Pm^\subl = (p_{\rm b} + p_\theta)/2$ and $\Pm^\subu = (p_{\rm t} + p_\theta)/2$.  
The force balance, as expressed in \eqref{e:buoyancy-gravity}, holds in each region:  
\begin{align}
p_{\rm b} - p_\theta = \frac{N^\subl mg}{A}, \qquad 
p_\theta - p_{\rm t} = \frac{N^\subu mg}{A}.  
\label{e:buoyancy-gravity-lu}  
\end{align}
Substituting these relations into the expression for $\bar{p} - \Pm$, we obtain  
\begin{align}
\bar{p} - \Pm = -\frac{mg L}{2} \frac{N}{V} \left( \frac{N^\subl}{N} - \frac{V^\subl}{V} \right),  
\label{e:bP-Pm-NL}  
\end{align}
where we used $V^\subl + V^\subu = V$, $N^\subl + N^\subu = N$, and $V = AL$.  

Next, consider the center of mass $X$ and the midpoint $\xm$,
\begin{align}
&X = \frac{N^\subl}{N} \xm^\subl + \frac{N^\subu}{N} \xm^\subu, \notag \\  
&\xm = \frac{V^\subl}{V} \xm^\subl + \frac{V^\subu}{V} \xm^\subu.  
\end{align}
Subtracting these equations gives  
\begin{align}
\frac{\xm - X}{L} = \frac{1}{2} \left( \frac{N^\subl}{N} - \frac{V^\subl}{V} \right),  
\label{e:X-xm}  
\end{align}
where we used $\xm^\subu - \xm^\subl = L/2$.  

Combining \eqref{e:X-xm} and \eqref{e:bP-Pm-NL}, we recover \eqref{e:bP-Pm}.

\section{Effective Hamiltonian for the distribution of particles}
\label{s:eff-Hamiltonian}

By coarse-graining the canonical distribution \eqref{e:rho-can}, we derive the probability density for the particle number in a specified region.  
The most probable value corresponds to the equilibrium state, determining the system's equilibrium configuration.

\begin{figure}[bt] 
\begin{center} 
\includegraphics[width=12cm]{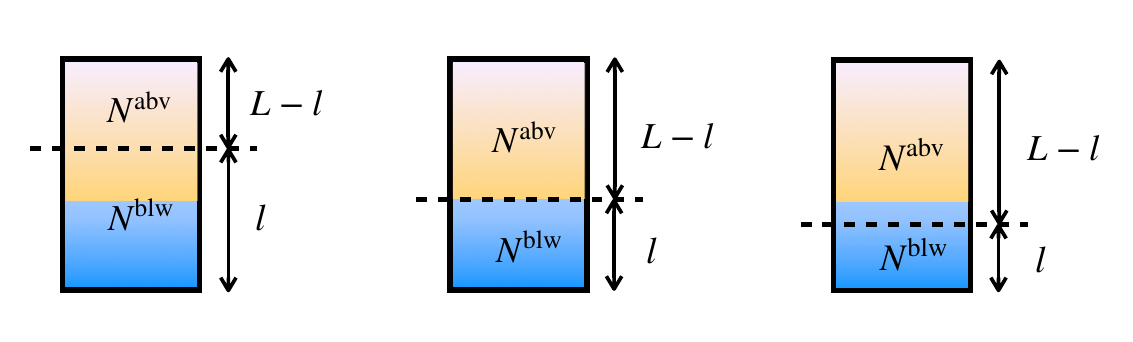}
\end{center} 
\caption{Three variants of setting up a virtual wall. 
The thermodynamic states of the lower and upper regions do not depend on the position of the virtual wall.}
\label{fig:divide}
\end{figure}

We virtually divide the fluid in the container with a section at an arbitrary position $x = \botX + l$, defining the regions below ($x < \botX + l$) and above ($x > \botX + l$) in this section as ``$\sublR$" and ``$\subuR$," respectively. These regions may contain liquid, gas, or a coexistence of both (see Fig.~\ref{fig:divide}).
Because the wall at $x=\botX+l$ is virtual, the number of particles in the region $x < \botX + l$ fluctuates.  
The probability density for observing the particle number ${\mathcal N}^\sublR$ is formally given by
\begin{align}
\rho({\mathcal N}^\sublR;l) = \int d\Gamma~\rho^\can(\Gamma)\delta({\mathcal N}^\sublR - \hat N^\sublR(\Gamma;l)),
\label{e:rhoNl-def}
\end{align}
where $\rho^\can(\Gamma)$ represents the canonical distribution in \eqref{e:rho-can}, and $\hat N^\sublR(\Gamma;l)$ is the particle number in the region $x < \botX + l$ for a given microstate $\Gamma$, expressed as
\begin{align}
\hat N^\sublR(\Gamma;l) = \sum_{i=1}^N \int_{\botX}^{\botX+l} dx~\delta(x_i - x).
\end{align}
We define an effective Hamiltonian ${\mathcal F}_g({\mathcal N}^\sublR; l)$ to represent the probability density $\rho({\mathcal N}^\sublR;l)$ as
\begin{align}
\rho({\mathcal N}^\sublR;l) = e^{\beta(F_g - {\mathcal F}_g({\mathcal N}^\sublR;l))}
\label{e:rhoNl}
\end{align}
with 
\begin{align}
{\mathcal F}_g({\mathcal N}^\sublR; l) 
= F_g^\sublR(T, l, A, {\mathcal N}^\sublR, mg) 
+ F_g^\subuR(T, L - l, A, N - {\mathcal N}^\sublR, mg) 
+ \frac{mg}{2}(lN - L{\mathcal N}^\sublR),
\label{e:eff-Hamiltonian}
\end{align}
as derived in Appendix A.  
Here, $F_g$, $F_g^\sublR$, and $F_g^\subuR$ denote the free energies for the total system and respective regions, while contributions of $o(N)$ terms are neglected.  
The functional dependencies of $F_g^\sublR$ and $F_g^\subuR$ are consistent with the total differential in \eqref{e:Fg-relation-m}.

Let ${N}^\sublR$ represent the most probable value of \eqref{e:rhoNl}, i.e.,
\begin{align}
N^\sublR(l) = \argmin_{{\mathcal N}^\sublR}{\mathcal F}_g({\mathcal N}^\sublR; l),
\label{e:var-Nl-max}
\end{align}
which implies
\begin{align}
\pderf{{\mathcal F}_g}{{\mathcal N}^\sublR}{*} = 0,
\label{e:var-Nl}
\end{align}
where $(\cdot )_*$ is an abbreviation for $( \cdot )_{T,V,N,mgL}$.
The equilibrium state is specified by ${N}^\sublR(l)$, and the thermodynamic free energy corresponds to the minimum value
\begin{align}
F_g = {\mathcal F}_g(N^\sublR(l); l).
\label{e:Fg-Fgvar}
\end{align}

As illustrated in Fig. \ref{fig:divide}, the equilibrium states are independent of $l$.  
Consequently, the free energy $F_g$ in \eqref{e:Fg-Fgvar} satisfies
\begin{align}
\frac{dF_g}{dl} = 0
~~\Leftrightarrow~~
\pderf{{\mathcal F}_g}{{\mathcal N}^\sublR}{*}\frac{dN^\sublR}{dl} + \pderf{{\mathcal F}_g}{l}{*} = 0.
\label{e:var-l-pre}
\end{align}
Substituting \eqref{e:var-Nl} into \eqref{e:var-l-pre}, we obtain
\begin{align}
\pderf{{\mathcal F}_g}{l}{*} = 0.
\label{e:var-l}
\end{align}

The relations \eqref{e:var-Nl} and \eqref{e:var-l}, combined with \eqref{e:eff-Hamiltonian}, yield
\begin{align}
&\pderf{F_g^\sublR}{{\mathcal N}^\sublR}{*} + \pderf{F_g^\subuR}{{\mathcal N}^\sublR}{*} - \frac{mgL}{2} = 0, \\
&\pderf{F_g^\sublR}{l}{*} + \pderf{F_g^\subuR}{l}{*} + \frac{mgN}{2} = 0.
\end{align}
The total differential relations \eqref{e:Fg-relation-m} for the respective regions are expressed as
\begin{align}
dF_g^\sublR &= -\Pm^\sublR A dl + \mu_g^\sublR d{\mathcal N}^\sublR, \\
dF_g^\subuR &= -\Pm^\subuR A d(L - l) + \mu_g^\subuR d{\mathcal N}^\subuR
\end{align}
with $T$, $A$, and $mg$ fixed.  
These lead to specific expressions for \eqref{e:var-Nl} and \eqref{e:var-l} as
\begin{align}
&\mu_g^\sublR - \mu_g^\subuR = \frac{mgL}{2}, \label{e:var2} \\
&(\Pm^\sublR - \Pm^\subuR)A = \frac{mgN}{2}. \label{e:var1}
\end{align}
These two equations provide the necessary conditions for the equilibrium state.

\subsection{Continuity of the pressure profile provided by \eqref{e:var1}} \label{s:varV-eq}

Let $p_-$ and $p_+$ denote the pressures immediately below and above the divided position $x = \botX + l$, respectively. 
We begin by considering the mid values of the pressure in the lower and upper regions:
\begin{align}
 \Pm^\sublR &= \frac{p_{\rm b} + p_-}{2}, \qquad \Pm^\subuR = \frac{p_{\rm t} + p_+}{2}. 
 \end{align} 
Substituting these expressions into \eqref{e:var1}, we obtain 
\begin{align} 
p_{\rm b} - p_{\rm t} + p_- - p_+ = \frac{mg N}{A}. 
\end{align} 
By applying the force balance condition \eqref{e:buoyancy-gravity}, it follows that 
\begin{align} 
p_- = p_+. \label{e:p-continuity} 
\end{align} 
Since \eqref{e:p-continuity} holds for any arbitrarily chosen value of $l$, the pressure profile $p(x)$ must be continuous throughout the system. This continuity is independent of whether the dividing position $l$ coincides with the liquid-gas interface. The result is consistent with the mechanical equilibrium condition assumed in Sec.~\ref{s:GlobalThermo}  in the absence of fluid flow.

\subsection{Linearity of the chemical potential profile concluded from \eqref{e:var2}} 
\label{s:varN-eq}

\begin{figure}[b] 
\begin{center} 
\includegraphics[width=7cm]{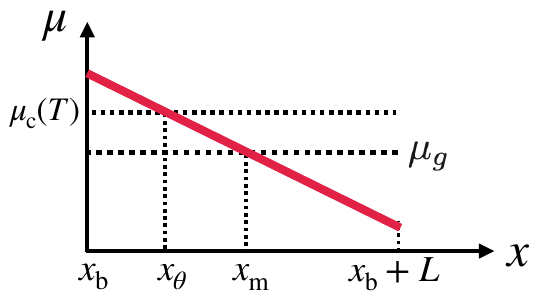}
\end{center} 
\caption{
Profile of the chemical potential $\mu(x)$ under gravity, shown as the red solid line. 
The slope of $\mu(x)$ is consistently $-mg$ regardless of whether the system is in a pure liquid, pure gas, or liquid-gas coexistence state. 
The difference between these equilibrium states is reflected only in the value of $\mu_g = \bmu = \mu(\xm)$. 
}
\label{fig:mu-eq}
\end{figure}

Combining the fundamental thermodynamic relation \eqref{e:Fg-relation} with \eqref{e:Fg-Fgvar},
we find 
\begin{align} 
\mu_g = \pderf{{\mathcal F}_g}{N}{*}. 
\end{align} 
Using \eqref{e:eff-Hamiltonian}, this can be rewritten as 
\begin{align} 
\mu_g = \mu_g^\subuR + \frac{mgl}{2}. \label{e:mug-mugu} 
\end{align} 
Substituting \eqref{e:mug-mugu} into \eqref{e:var2}, we obtain 
\begin{align} 
\mu_g + mg \xm = \mu_g^\sublR + mg \xm^\sublR = \mu_g^\subuR + mg \xm^\subuR, \label{e:mug-mugl} 
\end{align} 
where $\xm^\subuR - \xm = l/2$ and $\xm^\sublR - \xm = (l - L)/2$ have been used. The relation \eqref{e:mug-mugl} holds irrespective of whether the liquid-gas interface exists within the defined regions.

Except at points corresponding to the liquid-gas interface, 
the pressure profile $p(x)$ is smooth, and the force balance can be expressed as 
\begin{align} 
\der{p}{x} = -mg \rho, 
\end{align} 
which can be rewritten using the Gibbs-Duhem relation $d\mu = dp / \rho$ at constant temperature as 
\begin{align} 
\der{\mu}{x} = -mg. \label{e:mu-slope} 
\end{align} 
Here, the local chemical potential is defined by the free energy density as $\mu(x) \equiv \partial_\rho f(T, \rho(x))$. Equation \eqref{e:mu-slope} indicates that $\mu(x)$ is linear with a constant slope in $x$.

To examine the continuity of $\mu(x)$ at the discontinuities of $\rho(x)$, we divide the system into regions separated by these discontinuities and denote each region by a superscript ``${\rm i}$''. Within each region ${\rm i}$, which is in a single-phase state, the local thermodynamic quantities at the midpoint $\xm^{\rm i}$ determine the chemical potential, satisfying $\mu_g^{\rm i} = \bar\mu^{\rm i} = \mu(\xm^{\rm i})$. Combining this result with the slope $-mg$ in \eqref{e:mu-slope}, the profile in each region ${\rm i}$ is expressed as \begin{align} \mu^{\rm i}(x) = -mg x + \mu_g^{\rm i} + mg \xm^{\rm i}. \end{align} The intercept $\mu_g^{\rm i} + mg \xm^{\rm i}$ corresponds to the term appearing in \eqref{e:mug-mugl}. Therefore, the equality in \eqref{e:mug-mugl} ensures that the functional forms of $\mu^\subuR(x)$ and $\mu^\sublR(x)$ are equivalent, including both the slope and the intercept.

Thus, the chemical potential profile, irrespective of the liquid-gas interface, is universally described by 
\begin{align} 
\mu(x) = -mg(x - \xm) + \bar\mu
 \label{e:mug-local} 
\end{align} 
for $\botX < x < \topX$, where 
\begin{align}
 \mu_g = \bar\mu \equiv \frac{1}{L} \int_{\botX}^{\botX+L} dx~\mu(x). \label{e:def-bar-mu} 
 \end{align} 
 The formula \eqref{e:mug-local} applies to both single-phase states and liquid-gas coexistence. As \eqref{e:mug-local} is linear, it follows that $\bar\mu = \mu(\xm)$. Figure~\ref{fig:mu-eq} illustrates the relationships among $\mu(x)$, $\mu_g$ and $\mu_\subC(T)\equiv \mu^\subL(T, \Ps(T)) = \mu^\subG(T, \Ps(T))$, where $\mu^\subL$ and $\mu^\subG$ are the equilibrium chemical potentials for the liquid and gas phases, respectively.

Finally, we emphasize that the linear profile \eqref{e:mug-local} provides no direct information about the liquid-gas interface. For example, $\mu_g$ does not coincide with the saturation chemical potential $\mu_\subC(T)$.

\section{Variational principle for the phase separation under gravity}
\label{s:variational}

The effective Hamiltonian ${\mathcal F}_g({\mathcal N}^\sublR;l)$ in \eqref{e:eff-Hamiltonian} 
has established the continuity of $p(x)$ and the linearity of $\mu(x)$.
Since both $T(x)$ and $\mu(x)$ are linear, they lack any special point in $x$, rendering them incapable of identifying the interface position.
While the pressure profile $p(x)$ may exhibit piecewise linearity due to the liquid-gas configuration, 
Eqs.~\eqref{e:var2} and \eqref{e:var1} do not specify the local pressures at the interfaces.
This section examines the case where the number of liquid-gas interfaces is at most one, constructing a variational principle that leads to consistent equations with \eqref{e:var2} and \eqref{e:var1}.
The equilibrium conditions derived are then transformed into the chemical potential balance at the interface.
The case of multiple interfaces is discussed in Appendix~B.

\subsection{Constraints to refine each region in a single-phase state}
\label{s:X-xm-constrains}

\begin{figure}[bt] 
\begin{center} 
\includegraphics[width=14cm]{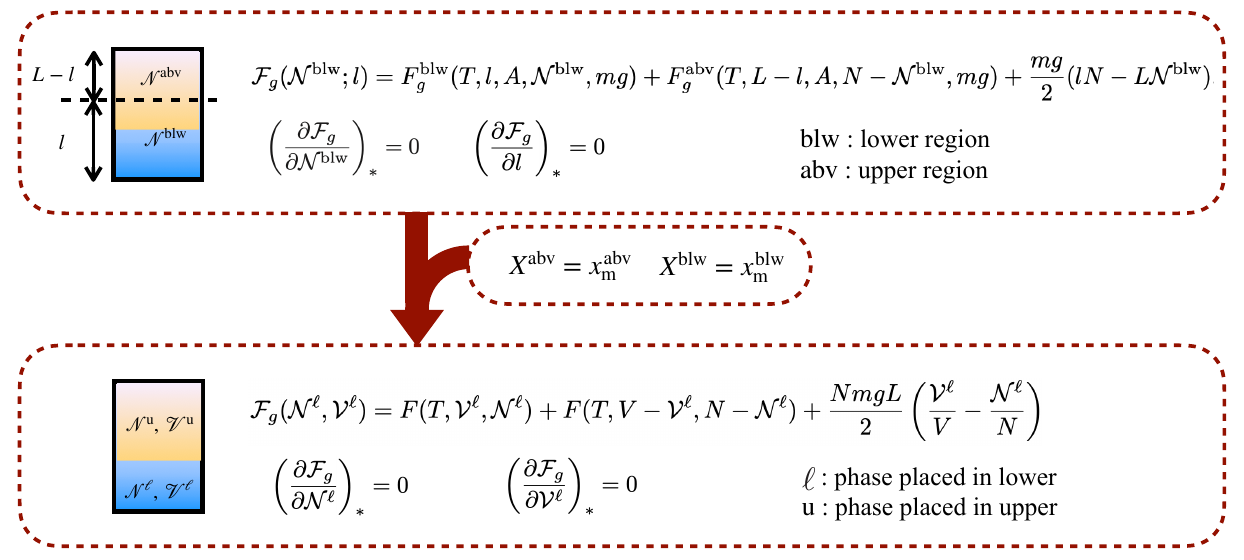}
\end{center} 
\caption{Construction of the variational principle to determine liquid and gas distribution.
}
\label{fig:scheme-var}
\end{figure}

To this point, the dividing position $\botX+l$ has been set arbitrarily.
We now impose the constraint 
\begin{align}
X^\sublR = \xm^\sublR + O(\ep), \qquad
X^\subuR = \xm^\subuR + O(\ep),
\label{e:Xl-Xu}
\end{align}
which, referring to \eqref{e:x-p-mean}, implies that both regions are in a single-phase state.
This section examines this implication in detail.

Starting with the fundamental relation \eqref{e:Fg-relation}, 
the following equations hold for each region above or below the virtual section:
\begin{align}
dF_g^\sublR &= -S^\sublR dT - \bar{p}^\sublR d{\mathcal V}^\sublR + \bar{\mu}^\sublR d{\mathcal N}^\sublR +{\mathcal N}^\sublR \frac{X^\sublR - \xm^\sublR}{l} d(mgl), \label{e:dFg-l} \\
dF_g^\subuR &= -S^\subuR dT - \bar{p}^\subuR d{\mathcal V}^\subuR + \bar{\mu}^\subuR d{\mathcal N}^\subuR + {\mathcal N}^\subuR\frac{X^\subuR - \xm^\subuR}{L-l} d(mg(L-l)). \label{e:dFg-u}
\end{align}
Here, ${\mathcal V}^\sublR = Al$ and ${\mathcal V}^\subuR = A(L-l)$.
Equations \eqref{e:dFg-l} and \eqref{e:dFg-u} yield Maxwell relations, such as
\begin{align}
&\pderf{\bar{\mu}^\sublR}{(mgl)}{T, {\mathcal V}^\sublR, {\mathcal N}^\sublR} 
= \left(\pder{}{{\mathcal N}^\sublR}\left({\mathcal N}^\sublR\frac{X^\sublR - \xm^\sublR}{l}\right)\right)_{T, {\mathcal V}^\sublR, mgl}, \label{e:Maxwell-mu1} \\
&\pderf{\bar{\mu}^\subuR}{(mg(L-l))}{T, {\mathcal V}^\subuR, {\mathcal N}^\subuR} 
= \left(\pder{}{{\mathcal N}^\subuR}\left({\mathcal N}^\subuR\frac{X^\subuR - \xm^\subuR}{L-l}\right)\right)_{T, {\mathcal V}^\subuR, mg(L-l)}.
\label{e:Maxwell-mu2}
\end{align}
These relations elucidate how $\bar{\mu}^\sublR$ and $\bar{\mu}^\subuR$ respond to gravity.

Under the constraint \eqref{e:Xl-Xu}, the right-hand sides of \eqref{e:Maxwell-mu1} and \eqref{e:Maxwell-mu2} are $O(\ep)$.
Thus, the Maxwell relations indicate $\bmu^\sublR(g)=\bmu^\sublR(g=0)+mgL\cdot O(\ep)$,
and therefore, $\bar{\mu}^\sublR$ and $\bar{\mu}^\subuR$ do not contain $O(\ep)$ terms.
Together with \eqref{e:dFg-l} and \eqref{e:dFg-u}, the constraint \eqref{e:Xl-Xu} refines $\bar{\mu}^\sublR$ and $\bar{\mu}^\subuR$ as functions of $(T, v^\sublR)$ and $(T, v^\subuR)$ by neglecting $O(\ep^2)$ terms, where $v^\sublR = {\mathcal V}^\sublR/{\mathcal N}^\sublR$ and $v^\subuR = {\mathcal V}^\subuR/{\mathcal N}^\subuR$.
Moreover, the constraint \eqref{e:Xl-Xu} simplifies the free energies of the respective regions
\begin{align}
F_g^\sublR = F(T, {\mathcal V}^\sublR, {\mathcal N}^\sublR), \qquad
F_g^\subuR = F(T, {\mathcal V}^\subuR, {\mathcal N}^\subuR),
\label{e:Fg-X=xm}
\end{align}
where $F(T, V, N)$ denotes the free energy at $g = 0$.
When $g \neq 0$, \eqref{e:Fg-X=xm} holds if the regions are in single-phase states, or the system is entirely in a single-phase state such as liquid or gas.

\subsection{Reformulation of the variational principle}
\label{s:varN-eq-LG}

The effective Hamiltonian 
$\mathcal{F}_g(\mathcal{N}^\sublR; l)$ in \eqref{e:eff-Hamiltonian} was derived as a single-variable function of $\mathcal{N}^\sublR$, 
and its minimization led to \eqref{e:var-Nl}.
Subsequently, the arbitrariness of $l$ was addressed in \eqref{e:var-l}.
We reformulate these statements under the constraint \eqref{e:Xl-Xu}, while preserving the form of the effective Hamiltonian.

Hereafter, the indices $\subl$ and $\subu$ denote the ``phases" in the lower and upper regions, respectively. 
Specifically, there are four possible configurations,
\begin{align}
(\subl,\subu)=(\subL,\subG), ~(\subG, \subL), ~(\subL,\subL), ~(\subG,\subG),
\label{e:config}
\end{align}
where $\subL$ and $\subG$ represent the liquid and gas phases. 
Consistent with the effective Hamiltonian \eqref{e:eff-Hamiltonian}, we introduce the following variational function
\begin{align}
&\mathcal{F}_g(\mathcal{N}^\subl,\mathcal{V}^\subl;T,V,N,mgL) \equiv F(T,\mathcal{V}^\subl,\mathcal{N}^\subl) + F(T,V-\mathcal{V}^\subl,N-\mathcal{N}^\subl)  + \frac{NmgL}{2}\left(\frac{\mathcal{V}^\subl}{V}-\frac{\mathcal{N}^\subl}{N}\right),
\label{e:Fg-var-lu}
\end{align}
where the variables $(\mathcal{N}^\subl,\mathcal{V}^\subl)$ represent the number of particles and the volume in the lower phase.

The equilibrium state in the liquid-gas coexistence can be determined by the following variational principle
\begin{align}
&(N^\subl,V^\subl) = \argmin_{\mathcal{N}^\subl,\mathcal{V}^\subl} \mathcal{F}_g(\mathcal{N}^\subl,\mathcal{V}^\subl),
\label{e:var-lu-max}
\end{align}
which leads to the variational equations as conditions for equilibrium
\begin{align}
\pderf{\mathcal{F}_g}{\mathcal{N}^\subl}{*} = 0, \qquad
\pderf{\mathcal{F}_g}{\mathcal{V}^\subl}{*} = 0.
\label{e:var-eq-lu}
\end{align}

Using the global variational variables $(\mathcal{N}^\subl, \mathcal{V}^\subl)$, the variational free energy \eqref{e:Fg-var-lu} results in the two variational equations \eqref{e:var-eq-lu}, replacing \eqref{e:var2} and \eqref{e:var1} with the constraint \eqref{e:Xl-Xu}. 
The relationship between the two variational principles, \eqref{e:var-Nl-max} and \eqref{e:var-lu-max}, is summarized in Fig. \ref{fig:scheme-var}.

\subsection{Balance of pressure and chemical potential at the interface} 

Substituting \eqref{e:Fg-var-lu} into \eqref{e:var-eq-lu}, we obtain
\begin{align}
&\pder{F(T,\mathcal{V}^\subl,\mathcal{N}^\subl)}{\mathcal{V}^\subl}+\pder{F(T,V-\mathcal{V}^\subl,N-\mathcal{N}^\subl)}{\mathcal{V}^\subl}+\frac{NmgL}{2V}=0, \label{e:7-1}\\
&\pder{F(T,\mathcal{V}^\subl,\mathcal{N}^\subl)}{\mathcal{N}^\subl}+\pder{F(T,V-\mathcal{V}^\subl,N-\mathcal{N}^\subl)}{\mathcal{N}^\subl}-\frac{mgL}{2}=0. \label{e:7-2}
\end{align}

Because $F(T,\mathcal{V}^\subl,\mathcal{N}^\subl)$ and $F(T,V-\mathcal{V}^\subl,N-\mathcal{N}^\subl)$ represent the free energies in single-phase states, \eqref{e:7-1} and \eqref{e:7-2} can be rewritten as
\begin{align}
&(\bP^\subl - \bP^\subu)A = \frac{mgN}{2}, \qquad
\bmu^\subl - \bmu^\subu = \frac{mgL}{2}.
\label{e:var-re}
\end{align}
Since $\bP^\subl = \Pm^\subl$ and $\bP^\subu = \Pm^\subu$, the first equality in \eqref{e:var-re} leads to the continuity of pressure
\begin{align}
p_\theta = p_+ = p_-.
\label{e:p-continuous-lu}
\end{align}

From the force balance in \eqref{e:buoyancy-gravity-lu}, we find
\begin{align}
&\bP^\subl = p_\theta + \frac{p_{\mathrm{b}} - p_\theta}{2} = p_\theta + \frac{mg N^\subl}{2A}, \label{e:pl}\\
&\bP^\subu = p_\theta + \frac{p_{\mathrm{t}} - p_\theta}{2} = p_\theta - \frac{mg N^\subu}{2A}. \label{e:pu}
\end{align}
Because the chemical potential at $g=0$ depends on $(T,p)$, we write
\begin{align}
\bmu^\subl = \mu^\subl(T,\bP^\subl), \qquad
\bmu^\subu = \mu^\subu(T,\bP^\subu),
\label{e:mug-mu-lu}
\end{align}
where $\mu^\subl(T,p)$ and $\mu^\subu(T,p)$ represent the chemical potentials for liquid or gas phases.

Substituting \eqref{e:pl} and \eqref{e:pu} into \eqref{e:mug-mu-lu} and expanding $\mu^\subl$ and $\mu^\subu$ in terms of $mg$, we obtain
\begin{align}
\bmu^\subl = \mu^\subl(T,p_\theta) + \frac{mgl}{2}, \qquad
\bmu^\subu = \mu^\subu(T,p_\theta) - \frac{mg(L-l)}{2},
\end{align}
where we have used $\pderf{\mu}{p}{T} = 1/\rho$.
Thus, the second equality in \eqref{e:var-re} becomes
\begin{align}
\mu^\subl(T,p_\theta) = \mu^\subu(T,p_\theta),
\label{e:sol-var-eq}
\end{align}
indicating the continuity of the chemical potential at the interface.

In Appendix B, we extend the above arguments to systems with two or multiple interfaces, demonstrating that the thermodynamically stable state contains at most one interface.

\section{Heterogeneous and homogeneous solutions of the variational equations}

We have shown in the previous section that the variational equations \eqref{e:var-eq-lu} result in \eqref{e:p-continuous-lu}
and \eqref{e:sol-var-eq}. 
Below, we demonstrate that \eqref{e:sol-var-eq} provides two types of solutions: heterogeneous and homogeneous. 

\subsection{Thermodynamics at $g=0$}

\begin{figure}[bt] 
\begin{center} 
\includegraphics[width=12cm]{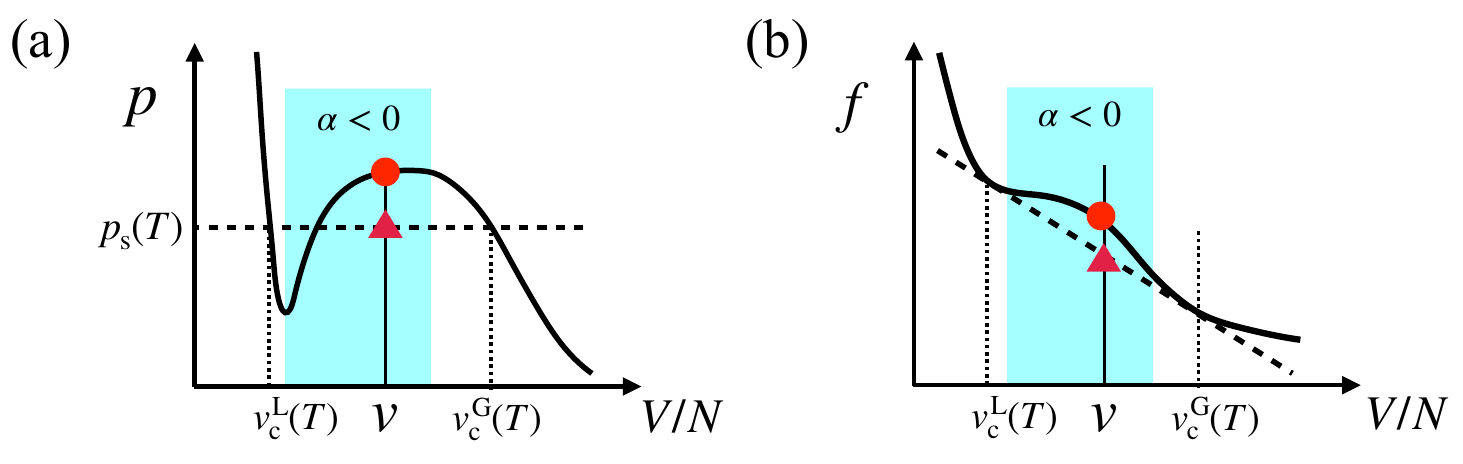}
\end{center} 
\caption{
The solid lines in (a) and (b) represent the equation of state $p=p(T, v)$ and the free energy $f(T, v)$, respectively, for a temperature $T$ significantly below the critical temperature. 
The dotted line indicates the values corresponding to the liquid-gas coexistence, where the pressure $p$ is equal to the saturation pressure $\Ps(T)$. 
In the blue-shaded region, which lies inside the spinodal line, the compressibility for homogeneous states becomes negative. 
}
\label{fig:homo-g0}
\end{figure}

We begin by considering the thermodynamic states when $g=0$.
Let the equation of state be $p=p(T,v)$ with $v=V/N$, as illustrated in Fig.~\ref{fig:homo-g0}(a).
The thermodynamic state on the solid line is assumed to represent a single phase, either liquid or gas,
even though it is rarely observed within the spinodal region, shaded in light blue.
By integrating $p=p(T,v)$ with respect to $v$, the free energy $f(T,v)$ is obtained, represented by the solid line in Fig.~\ref{fig:homo-g0}(b).
The single-phase state becomes unstable in the region $v_\subC^\subL(T) < v < v_\subC^\subG(T)$, 
where a portion of the gas transitions into liquid, or vice versa. 
The free energy of the liquid-gas coexistence is depicted as the dotted line in Fig.~\ref{fig:homo-g0}(b),
with its slope corresponding to the saturation pressure $\Ps(T)$.
This satisfies $\Ps(T)=p(T,v_\subC^\subL(T))$ and $\Ps(T)=p(T,v_\subC^\subG(T))$, 
where $v_\subC^\subL(T)$ and $v_\subC^\subG(T)$ denote the specific volumes of liquid and gas at saturation.

We emphasize that the state satisfying the equation of state corresponds to a single phase, referred to as the \textit{homogeneous} state. 
Conversely, the state represented by the dotted line is termed \textit{heterogeneous}, as it comprises a mixture of liquid and gas.
The free energy in the heterogeneous state, distinct from $f(T,v)$, is defined as
\begin{align}
&f^\hetero(T,v) \equiv \frac{N_0^\subL}{N}f(T,v_\subC^\subL(T)) + \frac{N_0^\subG}{N}f(T,v_\subC^\subG(T)),
\label{e:F-hetero}
\end{align}
where $N_0^\subL/N$ and $N_0^\subG/N$ represent the fractions of liquid and gas, respectively.

The number of particles and volumes,
$(N_0^\subL, V_0^\subL)$ and $(N_0^\subG, V_0^\subG)$,
for the liquid and gas are determined using the lever rule. Specifically, the condition
\begin{align}
&V_0^\subL = v_\subC^\subL(T)N_0^\subL, \quad
V_0^\subG = v_\subC^\subG(T)N_0^\subG, \quad
N_0^\subL + N_0^\subG = N, \quad 
V_0^\subL + V_0^\subG = V,
\end{align}
yields
\begin{align}
&\frac{N_0^\subL}{N} = \frac{v_\subC^\subG(T) - v}{v_\subC^\subG(T) - v_\subC^\subL(T)}, \quad
\frac{V_0^\subL}{V} = \frac{v_\subC^\subL(T)}{v} \frac{v_\subC^\subG(T) - v}{v_\subC^\subG(T) - v_\subC^\subL(T)}, \label{e:NV0-L}\\
&\frac{N_0^\subG}{N} = \frac{v - v_\subC^\subL(T)}{v_\subC^\subG(T) - v_\subC^\subL(T)}, \quad
\frac{V_0^\subG}{V} = \frac{v_\subC^\subG(T)}{v} \frac{v - v_\subC^\subL(T)}{v_\subC^\subG(T) - v_\subC^\subL(T)}. \label{e:NV0-G}
\end{align}
Thus, $(N_0^\subL/N, V_0^\subL/V)$ and $(N_0^\subG/N, V_0^\subG/V)$ are uniquely determined for a given $(T,v)$.

In Fig.~\ref{fig:homo-g0}(b),
$f(T,v)$ and $f^\hetero(T, v)$ are depicted as red circular and triangular points, respectively, for a given $v$.
Clearly, $f^\hetero(T, v) < f(T,v)$ for $v$ in $v_\subC^\subL(T) < v < v_\subC^\subG(T)$.

\subsection{Heterogeneous solution where phases \( \subl \) and \( \subu \) are distinct}
\label{s:solution-var-hetero}

\begin{figure}[bt] 
\begin{center} 
\includegraphics[width=14cm]{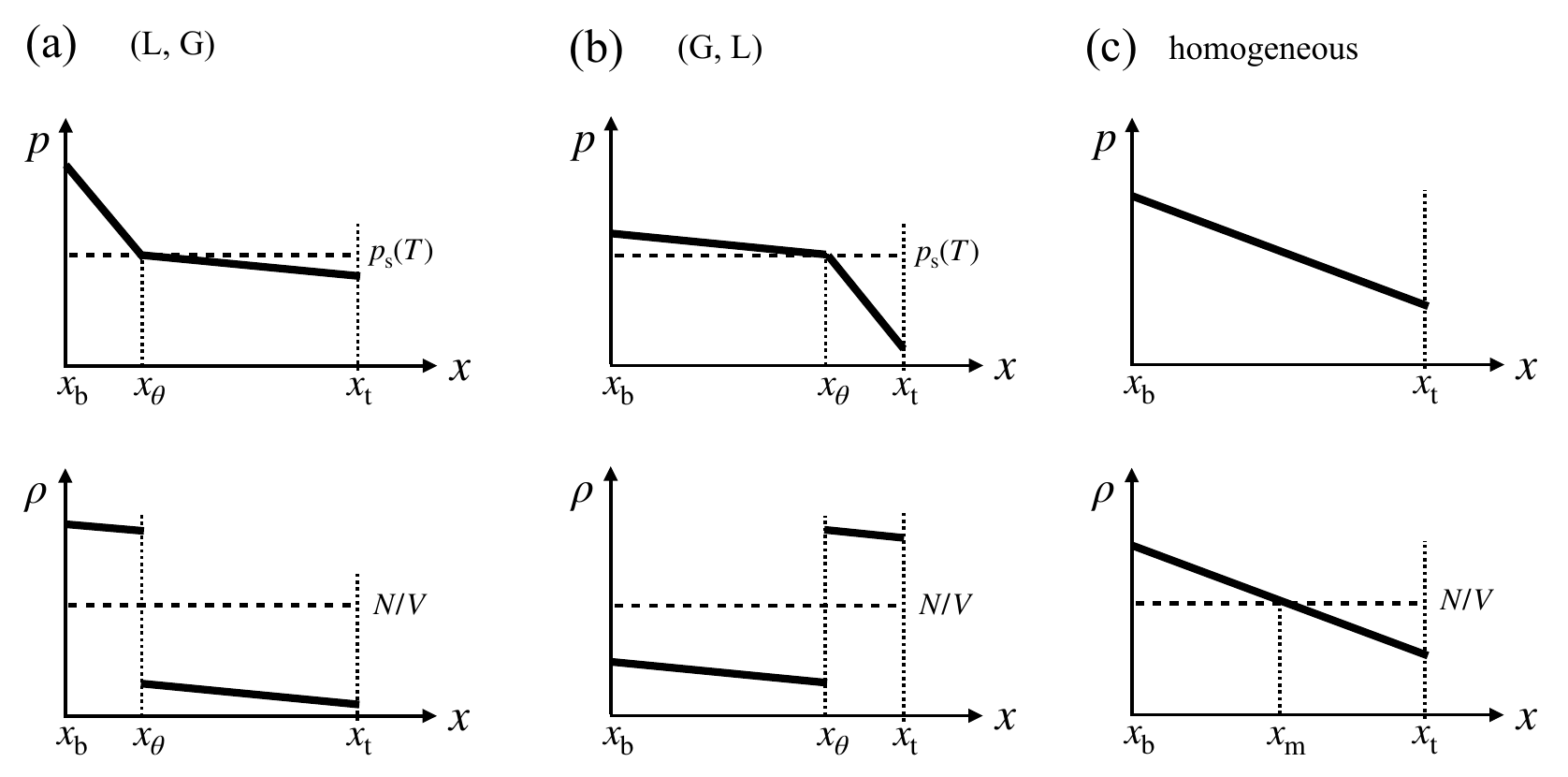}
\end{center} 
\caption{
Pressure and density profiles for two heterogeneous solutions with a liquid-gas interface located at $x=\xint$, as shown in (a) and (b). 
In (a), the liquid lies below the gas, corresponding to $(\subl, \subu) = (\subL, \subG)$, whereas in (b), the gas lies below the liquid, corresponding to $(\subl, \subu) = (\subG, \subL)$. 
(c) Profiles for a homogeneous solution. 
$g > 0$ is assumed for all cases.
}
\label{fig:profiles-eq}
\end{figure}

When the phases \( \subl \) and \( \subu \) are distinct, such that
\begin{align}
(\subl, \subu) = (\subL, \subG), \quad (\subG, \subL),
\label{e:config-hetero}
\end{align}
\eqref{e:sol-var-eq} yields
\begin{align}
p_\theta = \Ps(T),
\label{e:ptheta-Ps}
\end{align}
where \( \Ps(T) \) is the saturation pressure at temperature \( T \).
The relation \eqref{e:sol-var-eq} represents the balance of chemical potentials at the liquid-gas interface.
Henceforth, we refer to the solution comprising two distinct phases, as given in \eqref{e:config-hetero}, as a \textit{heterogeneous solution}.
Local states in the heterogeneous solution are schematically depicted as profiles of \( p(x) \) and \( \rho(x) \) in Figs.~\ref{fig:profiles-eq}(a) and (b) for a given \( (T,V,N,mgL) \),
where (a) and (b) correspond to the configurations \( (\subl, \subu) = (\subL, \subG) \) and \( (\subG, \subL) \), respectively.

We distinguish the two heterogeneous solutions as \( (N^\subl_{(\subL,\subG)}, V^\subl_{(\subL,\subG)}) \) and \( (N^\subl_{(\subG,\subL)}, V^\subl_{(\subG,\subL)}) \),
which are expressed as
\begin{align}
\frac{{V}^\subl_{(\subL,\subG)}}{{N}^\subl_{(\subL,\subG)}} = v_\subC^\subL(T) + O(\ep), \qquad
\frac{{V}^\subl_{(\subG,\subL)}}{{N}^\subl_{(\subG,\subL)}} = v_\subC^\subG(T) + O(\ep).
\label{e:hetero}
\end{align}
The pressure and free energy at \( |g| \to 0 \) in the two solutions are represented by the red triangular point in Fig.~\ref{fig:homo-g0}.

Using \( N_0^\subL \), \( V_0^\subL \), \( N_0^\subG \), \( V_0^\subG \) in \eqref{e:NV0-L} and \eqref{e:NV0-G},
one of the heterogeneous solutions in \( (\subl,\subu) = (\subL,\subG) \) is given as
\begin{align}
V^\subl_{(\subL,\subG)} = V_0^\subL + O(\ep), \quad
N^\subl_{(\subL,\subG)} = N_0^\subL + O(\ep),
\label{e:hetero-LG}
\end{align}
while the other in \( (\subl,\subu) = (\subG,\subL) \) is
\begin{align}
V^\subl_{(\subG,\subL)} = V_0^\subG + O(\ep), \quad
N^\subl_{(\subG,\subL)} = N_0^\subG + O(\ep),
\label{e:hetero-GL}
\end{align}
where the terms \( O(\ep) \) account for the effects of gravity, which will be analyzed in Sec.~\ref{s:NlVl-g}.

\subsection{Homogeneous solution where phases \( \subl \) and \( \subu \) are not distinct}
\label{s:solution-var-homo}

The homogeneous solution at \( |g| \to 0 \) is represented by the red circular point in Fig.~\ref{fig:homo-g0}(a) and (b) 
on the curve satisfying the equation of state.
For clarity, we conceptually divide the uniform state into \( \subl \) and \( \subu \),
and denote the number of particles and volumes in \( \subl \) as \( (N^\subl_\mathrm{homo}, V^\subl_\mathrm{homo}) \).
Since the system is uniform, we have
\begin{align}
\frac{{V}^\subl_\mathrm{homo}}{{N}^\subl_\mathrm{homo}} = \bar{v},
\label{e:homo0}
\end{align}
where \( \bar{v} = V/N \) is the average specific volume.
This type of solution is referred to as \textit{homogeneous}.
In this case, the position of the interface \( \xint \) cannot be specified.

To extend the homogeneous solution to \( g \neq 0 \), we consider
\begin{align}
\frac{{V}^\subl_\mathrm{homo}}{{N}^\subl_\mathrm{homo}} = \bar{v} + O(\ep).
\label{e:homo}
\end{align}
Comparing \eqref{e:hetero} with \eqref{e:homo}, the difference between heterogeneous and homogeneous solutions becomes apparent in terms of the specific volume in each region.
In the homogeneous solution, the system is composed entirely of either liquid or gas, even if the liquid or gas is thermodynamically unstable.

The profiles \( p(x) \) and \( \rho(x) \) in the homogeneous solution are sloped due to gravity, as shown in Fig.~\ref{fig:profiles-eq}(c) for a given \( (T, V, N, mgL) \).
Notably, any local region of the system cannot be distinguished from others in terms of phase.
Moreover, the pressure \( p_\theta \) at the dividing position is not necessarily equal to \( \Ps(T) \),
and \( (N^\subl_\mathrm{homo}, V^\subl_\mathrm{homo}) \) are unrelated to \( (N_0^\subL, V_0^\subL) \) or \( (N_0^\subG, V_0^\subG) \).

\subsection{Precise expression of $(N^\subl, V^\subl)$ in the heterogeneous solutions}
\label{s:NlVl-g}

Let $\Delta N$ and $\Delta V$ denote the correction terms of $O(\ep)$ in the heterogeneous solutions \eqref{e:hetero-LG}. Specifically, in the configuration $(\subl,\subu)=(\subL,\subG)$, we set
\begin{align}
  N^\subl_{(\subL,\subG)} = N_0^\subL - \Delta N, \quad
  V^\subl_{(\subL,\subG)} = V_0^\subL - \Delta V.
  \label{e:LG-NlVl}
\end{align}
We can then derive 
\begin{align}
&\frac{\Delta N}{N} \equiv
\frac{mg L}{2} \left[
\alpha^\subL \frac{v^\subL_\subC(T)}{\bar{v}}
\left(\frac{N^\subL_0}{N}\right)^2
+
\alpha^\subG \frac{v^\subG_\subC(T)}{\bar{v}}
\left(\frac{N^\subG_0}{N}\right)^2
\right] \frac{1}{v^\subG_\subC(T) - v^\subL_\subC(T)},
\label{e:DeltaN-def}\\
&\frac{\Delta V}{V} \equiv
\frac{mgL}{2} \frac{v^\subL_\subC(T) v^\subG_\subC(T)}{\bar{v}^2} \left[
\alpha^\subL \left(\frac{N^\subL_0}{N}\right)^2 + \alpha^\subG \left(\frac{N^\subG_0}{N}\right)^2
\right] \frac{1}{v^\subG_\subC(T) - v^\subL_\subC(T)},
\label{e:DeltaV-def}
\end{align}
where $\alpha^\subL$ and $\alpha^\subG$ are the isothermal compressibilities for the respective phases:
\begin{align}
\alpha^\subL = -\frac{1}{V^\subL} \pderf{V^\subL}{p}{*}, \qquad
\alpha^\subG = -\frac{1}{V^\subG} \pderf{V^\subG}{p}{*}.
\label{e:compressibility}
\end{align}
In $(\subl,\subu)=(\subG,\subL)$, we have
\begin{align}
  N^\subl_{(\subG,\subL)} = N_0^\subG + \Delta N, \quad
  V^\subl_{(\subG,\subL)} = V_0^\subG + \Delta V,
  \label{e:GL-NlVl}
\end{align}
using the same $\Delta N$ and $\Delta V$ as those defined in \eqref{e:LG-NlVl}.

Note that $\Delta N$ and $\Delta V$ are proportional to $mgL$, as shown in \eqref{e:DeltaN-def} and \eqref{e:DeltaV-def}. Because $v^\subG_\subC(T) > v^\subL_\subC(T)$, the signs of $\Delta N$ and $\Delta V$ depend on $\alpha^\subL$ and $\alpha^\subG$, as well as $g$.

\subsubsection{Derivation of \eqref{e:DeltaN-def} and \eqref{e:DeltaV-def}}
\label{s:state-hetero}

Let $v^\subl(T, \bP^\subl)$ and $v^\subu(T, \bP^\subu)$ be the specific volumes for the phases $\subl$ and $\subu$, respectively. Then, $V^\subl = v^\subl(T, \bP^\subl)N^\subl$ and $V^\subu = v^\subu(T, \bP^\subu)N^\subu$.  
The conservation of particle number and volume,  
\begin{align}
N^\subl + N^\subu = N, \quad v^\subl(T, \bP^\subl)N^\subl + v^\subu(T, \bP^\subu)N^\subu = \bar{v}N,
\end{align}
with $\bar{v} = V / N$, yields
\begin{align}
\frac{N^\subl}{N} = \frac{v^\subu(T, \bP^\subu) - \bar{v}}{v^\subu(T, \bP^\subu) - v^\subl(T, \bP^\subl)}, \quad
\frac{N^\subu}{N} = \frac{\bar{v} - v^\subl(T, \bP^\subl)}{v^\subu(T, \bP^\subu) - v^\subl(T, \bP^\subl)}.
\label{e:N-lu}
\end{align}
Note that $\bP^\subl$ and $\bP^\subu$ are connected to $p_\theta$ via \eqref{e:pl} and \eqref{e:pu}.  
Since $p_\theta = \Ps(T)$, the specific volumes can be expanded around $(T, \Ps(T))$ as
\begin{align}
&v^\subl(T, \bP^\subl) = v^\subl_\subC(T) -
\frac{mgL}{2} \frac{v^\subl_\subC(T)}{\bar{v}} \frac{N^\subl}{N} \alpha^\subl, \label{e:vl-eq}\\
&v^\subu(T, \bP^\subu) = v^\subu_\subC(T) +
\frac{mgL}{2} \frac{v^\subu_\subC(T)}{\bar{v}} \frac{N^\subu}{N} \alpha^\subu, \label{e:vu-eq}
\end{align}
neglecting $O(\ep^2)$ terms, where $v^\subl_\subC(T) = v^\subl(T, \Ps(T))$ and $v^\subu_\subC(T) = v^\subu(T, \Ps(T))$. The isothermal compressibilities, $\alpha^\subl$ and $\alpha^\subu$, are defined as in \eqref{e:compressibility}.

Below, we formulate a perturbative expression of $(N^\subl, V^\subl)$. We recall that $N_0^\subl$ and $V_0^\subl$ are the number of particles and the volume in the phase $\subl$ at $g=0$. We denote the increases induced by gravity as $\Delta N_{(\subl,\subu)}$ and $\Delta V_{(\subl,\subu)}$. Then, we have
\begin{align}
&\frac{N^\subl}{N}=\frac{N^\subl_0}{N}+\frac{\Delta N_{(\subl,\subu)}}{N}, \qquad
\frac{N^\subu}{N}=\frac{N^\subu_0}{N}-\frac{\Delta N_{(\subl,\subu)}}{N},\label{e:Nl-expand}\\
&\frac{V^\subl}{V}=\frac{V^\subl_0}{V}+\frac{\Delta V_{(\subl,\subu)}}{V}, \qquad
\frac{V^\subu}{V}=\frac{V^\subu_0}{V}-\frac{\Delta V_{(\subl,\subu)}}{V}.\label{e:Vl-expand}
\end{align}
Substitute \eqref{e:vl-eq} and \eqref{e:vu-eq} into \eqref{e:N-lu} and expand in $mgL$. The leading order corresponds to $N_0^\subl$, and the next order provides $\Delta N_{(\subl,\subu)}$. We thus obtain
\begin{align}
&\frac{N^\subl_0}{N}=\frac{v^\subu_\subC(T)-\bar v}{v^\subu_\subC(T)-v^\subl_\subC(T)}, \qquad
\frac{N^\subu_0}{N}=\frac{\bar v-v^\subl_\subC(T)}{v^\subu_\subC(T)-v^\subl_\subC(T)},\label{e:N0-lu}\\
&\frac{\Delta N_{(\subl,\subu)}}{N}=
-\frac{mg L}{2}\left[
\frac{v^\subl_\subC(T)}{\bar v}
\left(\frac{N^\subl_0}{N}\right)^2\alpha^\subl
+
\frac{v^\subu_\subC(T)}{\bar v}
\left(\frac{N^\subu_0}{N}\right)^2\alpha^\subu
\right]\frac{1}{v^\subu_\subC(T)-v^\subl_\subC(T)}.\label{e:DeltaN-lu}
\end{align}
To derive $V^\subl$, we start with
\begin{align}
\frac{V^\subl}{V}=\frac{v^\subl(T,\bP^\subl)}{\bar v}\frac{N^\subl}{N}, \qquad
\frac{V^\subu}{V}=\frac{v^\subu(T,\bP^\subu)}{\bar v}\frac{N^\subu}{N}.
\end{align}
Substituting \eqref{e:vl-eq} and $N^\subl=N_0^\subl+\Delta N_{(\subl,\subu)}$ yields
\begin{align}
\frac{V^\subl}{V}
=\frac{v_\subC^\subl(T)}{\bar v}\left(1-\frac{mgL}{2}\frac{N^\subl_0}{N}\frac{\alpha^\subl}{\bar v}\right)
\left(\frac{N^\subl_0}{N}+\frac{\Delta N_{(\subl,\subu)}}{N}\right).
\end{align}
We further substitute \eqref{e:DeltaN-lu} and obtain the expanded form of $V^\subl$. The leading and the next order are calculated as
\begin{align}
&\frac{V_0^\subl}{V}=\frac{v^\subl_\subC(T)}{\bar v}\frac{N_0^\subl }{N}, \qquad
\frac{V_0^\subu}{V}=\frac{v^\subu_\subC(T)}{\bar v}\frac{N_0^\subu }{N},\label{e:V0-lu}\\
&\frac{\Delta V_{(\subl,\subu)}}{V}
=
-\frac{mgL}{2}\frac{v^\subl_\subC(T) v^\subu_\subC(T)}{\bar v^2}\left[
\alpha^\subl\left(\frac{N^\subl_0}{N}\right)^2+\alpha^\subu\left(\frac{N^\subu_0}{N}\right)^2
\right]\frac{1}{v^\subu_\subC(T)-v^\subl_\subC(T)}.\label{e:DeltaV-lu}
\end{align}
\eqref{e:DeltaN-lu} and \eqref{e:DeltaV-lu} indicate 
$-\Delta N_{(\subL,\subG)}=\Delta N_{(\subG,\subL)}$ and $-\Delta V_{(\subL,\subG)}=\Delta V_{(\subG,\subL)}$. Thus, letting
\begin{align}
&\Delta N\equiv -\Delta N_{(\subL,\subG)}= \Delta N_{(\subG,\subL)},\\
&\Delta V\equiv -\Delta V_{(\subL,\subG)}= \Delta V_{(\subG,\subL)},
\end{align}
we obtain \eqref{e:DeltaN-def} and \eqref{e:DeltaV-def}.

\section{Stability of each solution}
\label{s:stability-eq}

The variational equations provide necessary conditions for the equilibrium states. Here, we examine the thermodynamic stability of each solution. Let us denote the solution of \eqref{e:var-re} as $(N^\subl, V^\subl)$ and consider a small fluctuation $(\delta{\mathcal N}^\subl, \delta{\mathcal V}^\subl)$ around $(N^\subl, V^\subl)$. The probability of a fluctuation $(\delta{\mathcal N}^\subl, \delta{\mathcal V}^\subl)$ is formally represented by the Einstein formula \cite{Einstein,Onsager1931}
\begin{align}
\rho(\delta{\mathcal N}^\subl,\delta{\mathcal V}^\subl) \propto e^{-\beta{\delta{\mathcal F}_g}},
\label{e:rhoNl-dev}
\end{align}
which is equivalent to \eqref{e:rhoNl}, derived from equilibrium statistical mechanics. 

The corresponding fluctuation of the free energy is expressed as
\begin{align}
\delta {\mathcal F}_g(\delta{\mathcal N}^\subl, \delta{\mathcal V}^\subl) 
= {\mathcal F}_g(N^\subl+\delta {\mathcal N}^\subl, V^\subl+\delta{\mathcal V}^\subl; T, V, N, mgL) 
- {\mathcal F}_g(N^\subl, V^\subl; T, V, N, mgL).
\label{e:delta-Fg-eq}
\end{align}
Since the linear terms of $O(\delta{\mathcal N}^\subl, \delta{\mathcal V}^\subl)$ cancel out, we obtain
\begin{align}
\delta{\mathcal F}_g(\delta{\mathcal N}^\subl, \delta{\mathcal V}^\subl) = \delta^2(F^\subl+F^\subu),
\label{e:delta2-F-eq}
\end{align}
neglecting higher-order terms. Here, $F^\subl=F(T,{\mathcal V}^\subl, {\mathcal N}^\subl)$ and $F^\subu=F(T, V-{\mathcal V}^\subl, N-{\mathcal N}^\subl)$, while
\begin{align}
\delta^2(F^\subl+F^\subu) = 
\begin{pmatrix}
\delta({\mathcal N}^\subl/N) & \delta({\mathcal V}^\subl/V)
\end{pmatrix}
\begin{pmatrix}
\left(\pdert{(F^\subl+F^\subu)}{{({\mathcal N}^\subl/N)}}\right)_* & \left(\pderc{(F^\subl+F^\subu)}{({\mathcal N}^\subl/N)}{({\mathcal V}^\subl/V)}\right)_* \\[0.5em]
\left(\pderc{(F^\subl+F^\subu)}{({\mathcal N}^\subl/N)}{({\mathcal V}^\subl/V)}\right)_* & \left(\pdert{(F^\subl+F^\subu)}{{({\mathcal V}^\subl/V)}}\right)_*
\end{pmatrix}
\begin{pmatrix}
\delta({\mathcal N}^\subl/N) \\[0.5em]
\delta({\mathcal V}^\subl/V)
\end{pmatrix}.
\label{e:matrix-stability}
\end{align}
The formula \eqref{e:rhoNl-dev}, combined with \eqref{e:delta2-F-eq}, indicates that the solution $(N^\subl, V^\subl)$ is locally stable in the space of $({\mathcal N}^\subl, {\mathcal V}^\subl)$ if 
\begin{align}
\delta^2(F^\subl+F^\subu) > 0.
\label{e:stability-matrix}
\end{align}

The matrix elements in \eqref{e:matrix-stability} can be expressed in terms of the isothermal compressibilities for each phase, $\alpha^\subl$ and $\alpha^\subu$. The first diagonal element transforms as
\begin{align}
\pdert{(F^\subl+F^\subu)}{{({\mathcal N}^\subl/N)}}
= N^2\pder{\bmu^\subl}{{\mathcal N}^\subl}
+ N^2\pder{\bmu^\subu}{{\mathcal N}^\subu}
= -N^2\left(\frac{{\mathcal V}^\subl}{{\mathcal N}^\subl}\right)^2\pder{\bP^\subl}{{\mathcal V}^\subl} 
- N^2\left(\frac{{\mathcal V}^\subu}{{\mathcal N}^\subu}\right)^2\pder{\bP^\subu}{{\mathcal V}^\subu},
\end{align}
where the second equality follows from the Gibbs-Duhem relation
\begin{align}
{\mathcal N}^\subl \pder{\bmu^\subl}{{\mathcal N}^\subl} = {\mathcal V}^\subl \pder{\bP^\subl}{{\mathcal N}^\subl}.
\end{align}
Thus, we have
\begin{align}
\pdertf{(F^\subl+F^\subu)}{{({\mathcal N}^\subl/N)}}{*}
= \frac{N^2}{\alpha^\subl V^\subl} \left(\frac{{V}^\subl}{{N}^\subl}\right)^2
+ \frac{N^2}{\alpha^\subu V^\subu} \left(\frac{{V}^\subu}{{N}^\subu}\right)^2.
\end{align}
The second diagonal element is given by
\begin{align}
\pdert{(F^\subl+F^\subu)}{{({\mathcal V}^\subl/V)}}
= -V^2 \pder{\bP^\subl}{{\mathcal V}^\subl} - V^2 \pder{\bP^\subu}{{\mathcal V}^\subu},
\end{align}
and can be expressed as
\begin{align}
\pdertf{(F^\subl+F^\subu)}{{({\mathcal V}^\subl/V)}}{*}
= \frac{V^2}{\alpha^\subl V^\subl} + \frac{V^2}{\alpha^\subu V^\subu}.
\end{align}
The off-diagonal element is transformed as
\begin{align}
\pderc{(F^\subl+F^\subu)}{({\mathcal N}^\subl/N)}{({\mathcal V}^\subl/V)}
= -NV \pder{\bP^\subl}{{\mathcal N}^\subl} - NV \pder{\bP^\subu}{{\mathcal N}^\subu}.
\end{align}
To rewrite this expression, we utilize the fact that the pressure is an intensive quantity, which implies the following relation for each phase:
\begin{align}
{\mathcal V}^\subl \pder{\bP^\subl}{{\mathcal V}^\subl} + {\mathcal N}^\subl \pder{\bP^\subl}{{\mathcal N}^\subl} = 0.
\end{align}
Applying this result to each phase, we find
\begin{align}
\pderc{(F^\subl+F^\subu)}{({\mathcal N}^\subl/N)}{({\mathcal V}^\subl/V)}
= NV \frac{{\mathcal V}^\subl}{{\mathcal N}^\subl} \pder{\bP^\subl}{{\mathcal V}^\subl} + 
NV \frac{{\mathcal V}^\subu}{{\mathcal N}^\subu} \pder{\bP^\subu}{{\mathcal V}^\subu}.
\end{align}
Thus, the off-diagonal element becomes
\begin{align}
\left(\pderc{(F^\subl+F^\subu)}{({\mathcal N}^\subl/N)}{({\mathcal V}^\subl/V)}\right)_*
= -\frac{NV}{\alpha^\subl N^\subl} - \frac{NV}{\alpha^\subu N^\subu}.
\end{align}

The eigenvalues of the matrix in \eqref{e:matrix-stability} are given by 
\begin{align}
\lambda_+ = A + \sqrt{A^2 - D}, \qquad \lambda_- = A - \sqrt{A^2 - D},
\end{align}
using the half-trace $A$ and determinant $D$ of the matrix, where $A^2 > D \geq 0$. 
Explicitly, $2A$ and $D$ are calculated as
\begin{align}
2A &\equiv \left(\pdert{(F^\subl+F^\subu)}{{({\mathcal N}^\subl/N)}} + \pdert{(F^\subl+F^\subu)}{{({\mathcal V}^\subl/V)}}\right)_*
= \frac{N^2}{\alpha^\subl V^\subl} 
\left(\left(\frac{V}{N}\right)^2 + \left(\frac{V^\subl}{N^\subl}\right)^2\right)
+ \frac{N^2}{\alpha^\subu V^\subu} 
\left(\left(\frac{V}{N}\right)^2 + \left(\frac{V^\subu}{N^\subu}\right)^2\right),
\label{e:stability-1} \\
D &\equiv \left(\pdert{(F^\subl+F^\subu)}{{({\mathcal N}^\subl/N)}}
\pdert{(F^\subl+F^\subu)}{{({\mathcal V}^\subl/V)}} - 
\left(\pderc{(F^\subl+F^\subu)}{({\mathcal N}^\subl/N)}{({\mathcal V}^\subl/V)}\right)^2\right)_*
= \frac{N^2 V^2}{\alpha^\subl \alpha^\subu V^\subl V^\subu} 
\left(\frac{V^\subl}{N^\subl} - \frac{V^\subu}{N^\subu}\right)^2.
\label{e:stability-2}
\end{align}

\subsection{Heterogeneous solutions}
\label{s:hetero-stability}

The values of $\alpha^\subl$ and $\alpha^\subu$ in the heterogeneous solutions \eqref{e:hetero}  under weak gravity are approximated by their respective values at the saturation pressure $\Ps(T)$ without gravity, where $\alpha^\subl > 0$ and $\alpha^\subu > 0$ as shown in Fig.~\ref{fig:homo-g0}.
We then conclude $A>0$ and $D>0$, and therefore, 
\begin{align}
\lambda_+ > 0, \qquad \lambda_- > 0
\end{align}
in both configurations: $(\subl,\subu) = (\subL,\subG)$ and $(\subG,\subL)$. These solutions are locally stable, regardless of whether $g > 0$ or $g < 0$.
It is important to note that hydrodynamic instabilities, such as the Rayleigh-Taylor instability, lie outside the scope of this stability analysis. This is because the definition of local stability in this context is confined to the phase space $({\mathcal N}^\subl, {\mathcal V}^\subl)$.

\subsection{Homogeneous solutions}
\label{s:homo-stability}

The homogeneous solutions \eqref{e:homo} satisfy
\begin{align}
\frac{V^\subl}{N^\subl} = \frac{V^\subu}{N^\subu} + O(\ep), \qquad
\alpha^\subl = \alpha^\subu + O(\ep),
\end{align}
which implies $D = O(\ep^2)$ and
\begin{align}
A = \frac{1}{\alpha^\homo} \frac{V^3}{V^\subl V^\subu}.
\end{align}
The eigenvalues are then estimated as
\begin{align}
\lambda_+ = A + |A|, \qquad \lambda_- = A - |A|
\label{e:eigen-homo}
\end{align}
with errors of $O(\ep^2)$. Thus, either $\lambda_+ = 0$ or $\lambda_- = 0$ depending on the sign of $\alpha^\homo$.

\section{Free energies in heterogeneous and homogeneous solutions}
\label{s:unique-min}

We have shown in Sec.~\ref{s:stability-eq} that the two heterogeneous solutions are locally stable. 
Correspondingly, the variational function ${\mathcal F}_g({\mathcal N}^\subl, {\mathcal V}^\subl)$ should possess two minima. 
Below, we determine the values of ${\mathcal F}_g$ for the two heterogeneous and homogeneous solutions and provide the landscape of ${\mathcal F}_g({\mathcal N}^\subl, {\mathcal V}^\subl)$. 

\subsection{Comparison of states in local and global minima}

Note that Eqs.~\eqref{e:LG-NlVl} and \eqref{e:GL-NlVl} provide 
\begin{align}
&N^\subl_{(\subL,\subG)}=N^\subu_{(\subG,\subL)}=N_0^\subL-\Delta N, \\
&V^\subl_{(\subL,\subG)}=V^\subu_{(\subG,\subL)}=V_0^\subL-\Delta V,
\end{align}
where $\Delta N$ and $\Delta V$ are given in \eqref{e:DeltaN-def} and \eqref{e:DeltaV-def}. 
These relations indicate that the amount and volume of liquid, $N^\subL$ and $V^\subL$, are the same between the two heterogeneous solutions. Specifically, 
\begin{align}
&N^\subL=N_0^\subL-\Delta N, \quad~
V^\subL=V_0^\subL-\Delta V, \\
&N^\subG=N_0^\subG+\Delta N, \quad
V^\subG=V_0^\subG+\Delta V.
\label{e:NL-VL-def}
\end{align}

\subsection{Free energy of the heterogeneous solutions}
\label{s:state-hetero}

For the heterogeneous solutions described in Sec.~\ref{s:solution-var-hetero}, we define their free energies as
\begin{align}
&F_g^{(\subL,\subG)}\equiv {\mathcal F}_g(N^\subl_{(\subL,\subG)}, V^\subl_{(\subL,\subG)})
={\mathcal F}_g(N_0^\subL-\Delta N, V_0^\subL-\Delta V),
\label{e:Fg-LG-def} \\
&F_g^{(\subG,\subL)}\equiv {\mathcal F}_g(N^\subl_{(\subG,\subL)}, V^\subl_{(\subG,\subL)})={\mathcal F}_g(N_0^\subG+\Delta N, V_0^\subG+\Delta V).
\label{e:Fg-GL-def}
\end{align}
Below, we use 
\begin{align}
&\psi_g^{(\subL, \subG)}\equiv \frac{1}{2}\left(\frac{{N}_0^\subL}{N}-\frac{{V}_0^\subL}{V}\right),
\label{e:Psig-LG0}
\end{align}
where $N_0^\subL$ and $V_0^\subL$ are the liquid values at $g=0$, determined by the lever rule in \eqref{e:NV0-L}. Substituting \eqref{e:NV0-L} into \eqref{e:Psig-LG0}, we obtain
\begin{align}
\psi_g^{(\subL,\subG)}=\frac{1}{2}\frac{(v^\subG_\subC(T)-\bar{v})(\bar{v}-v^\subL_\subC(T))}{\bar{v}(v^\subG_\subC(T)-v^\subL_\subC(T))}>0
\label{e:Psig-LG}
\end{align}
for $v_\subC^\subL(T)<\bar{v}<v_\subC^\subG(T)$.

Using \eqref{e:Fg-var-lu}, Eq.~\eqref{e:Fg-LG-def} can be written as
\begin{align}
{F}_g^{(\subL,\subG)}
&=F(T, V_0^\subL-\Delta V, N^\subL-\Delta N)+F(T, V^\subG+\Delta V, N^\subG+\Delta N) -NmgL\psi_g^{(\subL, \subG)} \notag \\
&=F(T,V_0^\subL,N_0^\subL)+F(T,V_0^\subG,N_0^\subG)+(\bP^\subL-\bP^\subG)\Delta V-(\bmu^\subL-\bmu^\subG)\Delta N-NmgL\psi_g^{(\subL, \subG)} \notag \\
&=F(T,V_0^\subL,N_0^\subL)+F(T,V_0^\subG,N_0^\subG)-NmgL\psi_g^{(\subL, \subG)},
\label{e:FtoF0-pre}
\end{align}
neglecting terms of $O(\ep^2)$. The third equality follows from \eqref{e:var-re}.
Similarly, we obtain the expression for \eqref{e:Fg-GL-def} as
\begin{align}
{F}_g^{(\subG,\subL)}
=F(T, V_0^\subL, N_0^\subL)+F(T, V_0^\subG, N_0^\subG)-NmgL\psi_g^{(\subL, \subG)}.
\end{align}

Letting $f_g^{(\subL,\subG)}\equiv {F}_g^{(\subL,\subG)}/N$ and $f_g^{(\subG,\subL)}\equiv {F}_g^{(\subG,\subL)}/N$, 
the state dependence of ${F}_g^{(\subL,\subG)}$ is expressed as
\begin{align}
&f_g^{(\subL,\subG)}(T,\bar{v},mgL)=f^\hetero(T,\bar{v})-mgL\psi_g^{(\subL, \subG)},
\label{e:Fg-LG} \\
&f_g^{(\subG,\subL)}(T,\bar{v},mgL)=
f^\hetero(T,\bar{v})+mgL\psi_g^{(\subL, \subG)}.
\label{e:Fg-GL}
\end{align}
Here, $f^\hetero(T,\bar{v})$ is the free energy per particle defined in \eqref{e:F-hetero}, depicted as the red triangular point in Fig.~\ref{fig:homo-g0}(b).

\subsection{Free energy of the homogeneous solutions}
\label{s:state-homo}

Because the profiles are linear, as shown in Fig.~\ref{fig:profiles-eq}(c), we have
\begin{align}
f_g^\homo(T,\bar{v},mgL)=f(T,\bar{v}),
\label{e:F-homo}
\end{align}
where $f_g^\homo\equiv{F_g^\homo}/{N}$, and $f(T, v)$ is the free energy per particle determined from the equation of states, represented by the red circular point in Fig.~\ref{fig:homo-g0}(b). 
Equation \eqref{e:F-homo} is independent of $V_0^\subL$ and $N_0^\subL$ and is unaffected by gravity.

\subsection{Comparison of the values of free energy}
\label{s:global-minima-eq}

From \eqref{e:Fg-LG} and \eqref{e:Fg-GL}, the difference in free energy between the two heterogeneous solutions is given by
\begin{align}
f_g^{(\subG, \subL)}-f_g^{(\subL, \subG)}
&=2mgL\psi_g^{(\subL, \subG)}.
\label{e:Fg-diff-twomin}
\end{align}
When $v_\subC^\subL(T)<\bar{v}<v_\subC^\subG(T)$, the second law of thermodynamics leads to
\begin{align}
f^\hetero(T,\bar{v})<f(T,\bar{v}),
\end{align}
as shown in Fig.~\ref{fig:homo-g0}. Consequently, we find
\begin{align}
f_g^{(\subL, \subG)}-f_g^\homo
&=f^\hetero(T,\bar{v})-f(T,\bar{v})-mgL\psi_g^{(\subL, \subG)}<-mgL\psi_g^{(\subL, \subG)},
\label{e:Fg-diff-homo1} \\
f_g^{(\subG, \subL)}-f_g^\homo
&=f^\hetero(T,\bar{v})-f(T,\bar{v})+mgL\psi_g^{(\subL, \subG)}<mgL\psi_g^{(\subL, \subG)}.
\label{e:Fg-diff-homo2}
\end{align}
Because $\psi_g^{(\subL, \subG)}>0$, we have
\begin{align}
&f_g^{(\subL, \subG)}<f_g^{(\subG, \subL)}, \quad f_g^{(\subL, \subG)}<f_g^\homo \qquad (g>0), \\
&f_g^{(\subL, \subG)}=f_g^{(\subG, \subL)}<f_g^\homo \qquad \qquad \qquad (g=0), \\
&f_g^{(\subG, \subL)}<f_g^{(\subL, \subG)}, \quad f_g^{(\subG, \subL)}<f_g^\homo \qquad (g<0).
\end{align}
When $g>0$, the heterogeneous solution with $(\subl,\subu)=(\subL,\subG)$ corresponds to the global minimum, whereas the solution with $(\subl,\subu)=(\subG,\subL)$ corresponds to the local minimum.

\begin{figure}[bt] 
\begin{center} 
\includegraphics[width=15cm]{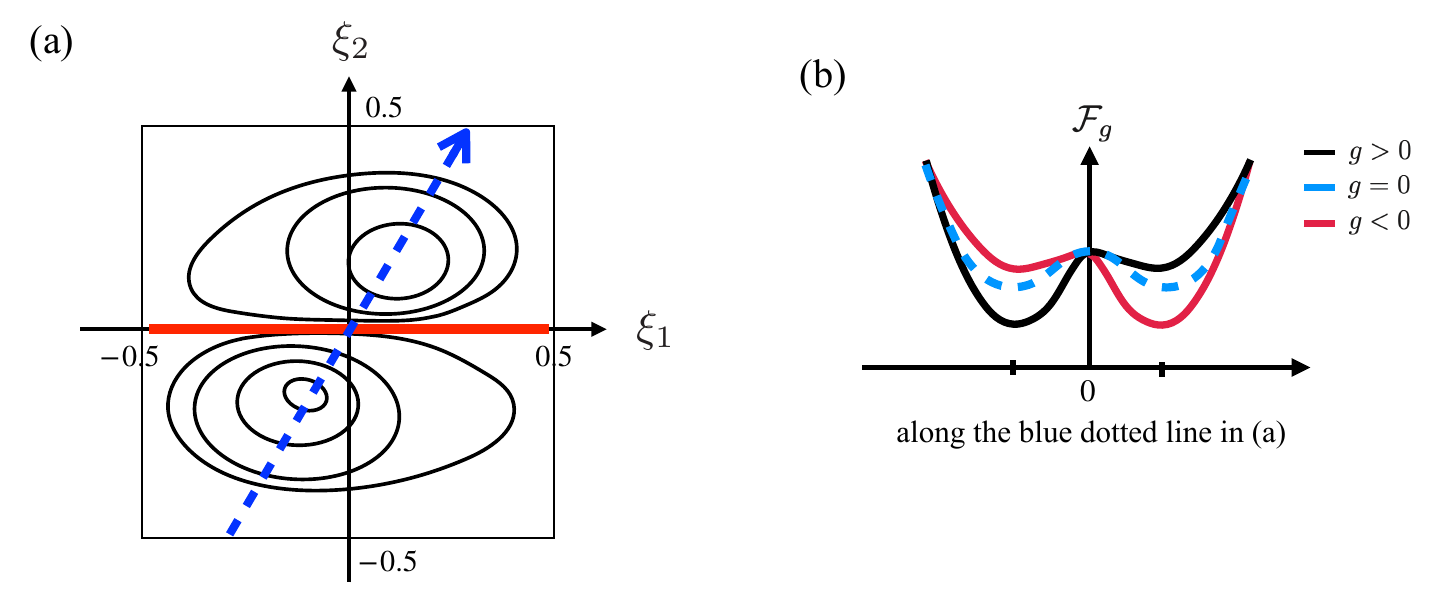}
\end{center} 
\caption{
Free energy landscape when $\bar v$ is chosen such that $\alpha^\homo < 0$, making the homogeneous solution unstable. 
(a) A contour plot of ${\mathcal F}_g$ for $g > 0$ in the space of $(\xi_1,\xi_2)$, which is linearly mapped from the space of $({\mathcal N}^\subl, {\mathcal V}^\subl)$ as described in \eqref{e:X-xm-var} and \eqref{e:xtheta-xm-var}. 
The two minima are symmetrically located about the origin, but differ in depth. 
The red solid line represents the homogeneous solution. 
(b) The dependence of ${\mathcal F}_g$ on $g$ along the blue dotted line in (a), which connects the two minima. 
}
\label{fig:reverse}
\end{figure}

\begin{figure}[bt] 
\begin{center} 
\includegraphics[width=15cm]{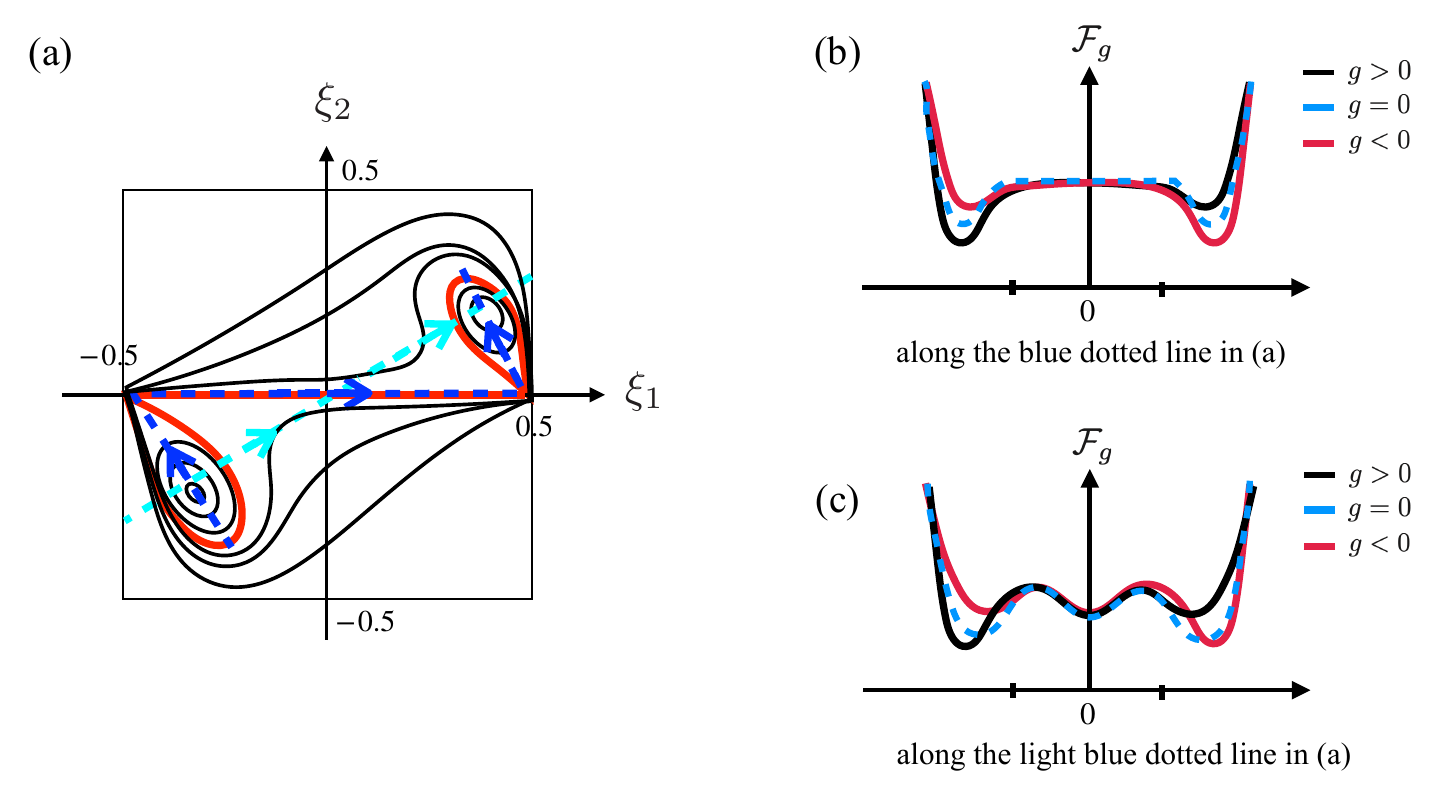}
\end{center} 
\caption{
Free energy landscape when $\bar v$ is chosen such that $\alpha^\homo > 0$, making the homogeneous solution marginally stable. 
(a) A contour plot of ${\mathcal F}_g$ for $g > 0$ in the space of $(\xi_1,\xi_2)$.
The two minima are symmetrically located about the origin, but differ in depth. 
The red solid line represents the contour for ${\mathcal F}_g = N f_g^\homo(T, \bar{v})$.
(b) The dependence of ${\mathcal F}_g$ on $g$ along the blue dotted line in (a), connecting the two minima. 
(c) The dependence of ${\mathcal F}_g$ on $g$ along the light-blue dotted line in (a). 
}
\label{fig:reverse-homo}
\end{figure}

\subsection{Free energy landscape}

As the local and global minima of the free energy have been determined, 
we can illustrate the free energy landscape ${\mathcal F}_g({\mathcal N}^\subl,{\mathcal V}^\subl)$. 
The space of $({\mathcal N}^\subl/N, {\mathcal V}^\subl/V)$ is linearly mapped to
$(\xi_1,\xi_2)$ as
\begin{align}
&\xi_1 = \frac{{\mathcal V}^\subl}{V} - \frac{1}{2},
\label{e:xtheta-xm-var}\\
&\xi_2 = \frac{1}{2} \left( \frac{{\mathcal V}^\subl}{V} - \frac{{\mathcal N}^\subl}{N} \right).
\label{e:X-xm-var}
\end{align}
$\xi_1$ and $\xi_2$ are connected to the position of the interface and the center of mass as
\begin{align}
\argmin_{\xi_1,\xi_2} {\mathcal F}_g(\xi_1,\xi_2;T,V,N,mgL)=\left(\frac{\xint-\xm}{L},\frac{X-\xm}{L}\right).
\end{align}
Figure \ref{fig:reverse}(a) and Figure \ref{fig:reverse-homo}(a) schematically show the contour plots of ${\mathcal F}_g$ for $g>0$ 
in the space of $(\xi_1,\xi_2)$.
A set of homogeneous solutions constitutes the horizontal axis $\xi_2=0$, 
while the two heterogeneous solutions correspond to the two minima.

Recall that the homogeneous solution possesses a zero eigenvalue as shown in \eqref{e:eigen-homo}.
The eigenvector for the zero eigenvalue is $(1,0)$, while the eigenvector for the other eigenvalue is $(1,-2/3)$ 
in the space of $(\xi_1,\xi_2)$.
The eigenvector $(1,0)$ indicates that the homogeneous solution is marginal against fluctuations preserving $\xi_2=0$,
irrespective of the sign of $\alpha^\homo$. 
On the contrary, the sign of the non-vanishing eigenvalue, and therefore the stability of the homogeneous solutions, depends on the sign of $\alpha^\homo$.
The sign of $\alpha^\homo$ depends on the value of $\bar{v}$, as shown in Fig.~ \ref{fig:homo-g0}.
When $\bar{v}$ lies inside the spinodal line, as shown by the blue shaded area in Fig.~\ref{fig:homo-g0},
we have $\alpha^\homo<0$, rendering the homogeneous solution unstable.
Conversely, when the liquid volume is relatively small such that $\bar{v}$ approaches $v_\subC^\subG(T)$, 
corresponding to the region between the binodal and spinodal lines in Fig.~\ref{fig:homo-g0},
we have $\alpha^\homo > 0$, where the homogeneous solution becomes locally stable.

Figure \ref{fig:reverse}(a) illustrates the landscape when $\alpha^\homo<0$, where the homogeneous solution is unstable. 
The red line represents the contour at ${\mathcal F}_g = N f_g^\homo(T,\bar{v})$, i.e., the set of homogeneous solutions. 
Figure~\ref{fig:reverse}(a) shows that the homogeneous solutions (red line) act as saddle points separating the basins of the two local minima. 
The free energy ${\mathcal F}_g$ exhibits a typical quadratic-like functional shape along a line connecting the two minima, 
as demonstrated in Figure \ref{fig:reverse}(b).

Figure \ref{fig:reverse-homo}(a) provides a schematic depiction of the landscape for the case $\alpha^\homo > 0$. 
Since the homogeneous state no longer acts as a saddle point, the landscape ${\mathcal F}_g$ is qualitatively distinct from Fig.~\ref{fig:reverse}. 
The red line represents the contour for ${\mathcal F}_g = N f_g^\homo(T,\bar{v})$. 
Remarkably, the red line connects to cycle-like contours surrounding each local minimum. 
This observation suggests that the heterogeneous solutions are less accessible than the homogeneous solution. 
As depicted in Fig.~\ref{fig:reverse-homo}(b), following a path along the blue dotted line, the single-phase state may be recognized as marginally unstable. 
However, the single-phase state may appear stable when considering paths like the light blue dotted line, as illustrated in Fig.~\ref{fig:reverse-homo}(c). 
Suppose the system is prepared at a high temperature and then cooled down. 
While a small density fluctuation may occur, it would be smeared out, allowing the homogeneous state to recover 
because the homogeneous solutions are locally stable. 
Thus, the system's state may remain in the homogeneous solution, corresponding to the gas phase, without transitioning into the heterogeneous solution representing liquid-gas coexistence. 
To reach the global minimum exhibiting phase separation, the system must deviate significantly to find the entrance to the global minimum. 
Such a scenario appears consistent with the hysteresis phenomena typically associated with first-order transitions.

\section{First-order transition occurring at $g=0$ and thermodynamic free energy}

\begin{figure}[bt] 
\begin{center} 
\includegraphics[width=8cm]{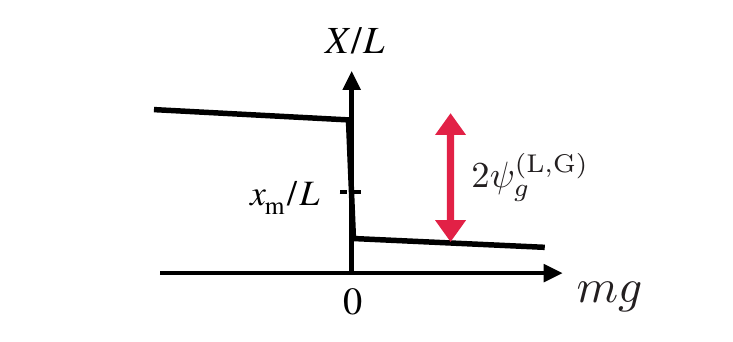}
\end{center} 
\caption{Center of mass $X$ as a step function of $g$.  The jump of $X$ at $g=0$ results from the interchange of the global and local minima.}
\label{fig:Xjump}
\end{figure}

The existence of two local minima gives rise to a first-order transition. 
For $g>0$, the equilibrium configuration is specified as $(\subl,\subu)=(\subL,\subG)$, 
and for $g<0$, it becomes $(\subl,\subu)=(\subG,\subL)$. 
Thus, we obtain
\begin{align}
\frac{\xm-X}{L} = \frac{g}{|g|} \psi_g^{(\subL,\subG)}
= \frac{g}{2|g|} \frac{(v^\subG_\subC(T)-\bar{v})(\bar{v}-v^\subL_\subC(T))}{\bar{v}(v^\subG_\subC(T)-v^\subL_\subC(T))},
\label{e:xm-X-g}
\end{align}
applying \eqref{e:Psig-LG}.
Since $\psi_g^{(\subL,\subG)} > 0$, the sign of $g$ determines the sign of $X-\xm$. 
As $g$ changes value, the center of mass jumps at $g=0$, as shown in Fig.~\ref{fig:Xjump}. 
This transition is observed as the inversion of liquid and gas phases. 
The well-known observation that heavier phases are located below lighter phases 
can now be interpreted thermodynamically as a phase transition between these two configurations.

This first-order transition occurs irrespective of the sign of $\alpha^\homo$. 
However, the relaxation dynamics may differ quantitatively depending on whether $\alpha^\homo < 0$ or $\alpha^\homo > 0$, 
as suggested by the differences of Fig.~\ref{fig:reverse}(b) from Figs.~\ref{fig:reverse-homo}(b), (c).

\subsection{Thermodynamic free energy}

The thermodynamic free energy is defined as the global minimum value of ${\mathcal F}_g$, given by
\begin{align}
F_g(T,V,N,mgL) = \min_{{\mathcal N}^\subl,{\mathcal V}^\subl} {\mathcal F}_g({\mathcal N}^\subl,{\mathcal V}^\subl; T, V, N, mgL).
\end{align}
From \eqref{e:Fg-var-lu}, we write
\begin{align}
F_g(T,V,N,mgL) = F(T,V^\subL,N^\subL) + F(T,V-V^\subL,N-N^\subL) - mgL \Psi_g(T,V,N),
\label{e:Fg-thermo-eq}
\end{align}
where
\begin{align}
\Psi_g(T,V,N) = N \frac{g}{|g|} \psi_g^{(\subL,\subG)}(T,V/N)
\label{e:Psig-absolute}
\end{align}
with $\psi_g^{(\subL,\subG)}(T,V/N)$ determined in \eqref{e:Psig-LG}.
It should be noted that \eqref{e:Fg-thermo-eq} is expressed in terms of $V^\subL$ and $N^\subL$, defined in \eqref{e:NL-VL-def}, rather than $V^\subl$ and $N^\subl$.

\begin{figure}[bt] 
\begin{center} 
\includegraphics[width=12cm]{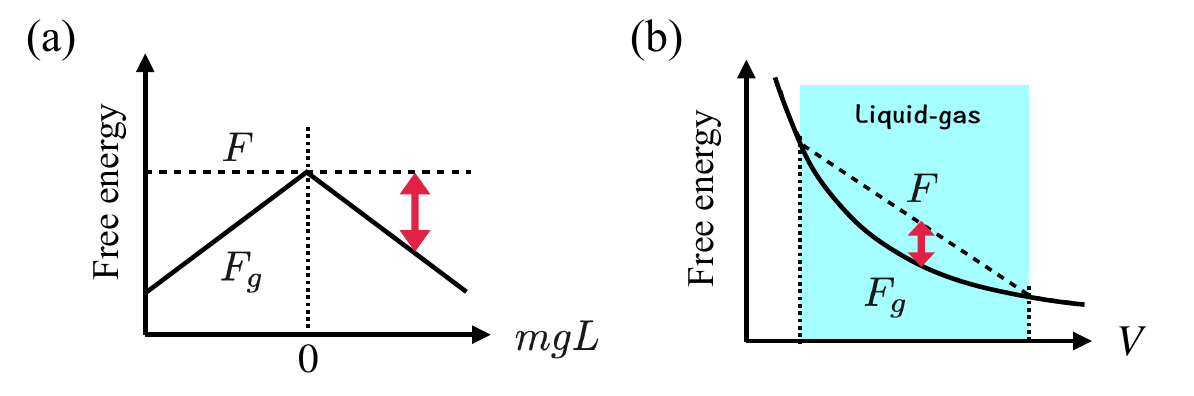}
\end{center} 
\caption{
(a) $F_g$ and $F$ as functions of $mgL$ for fixed $(T, V, N)$. 
(b) $F_g$ and $F$ as functions of $V$ for fixed $(T, N, mgL)$.
}
\label{fig:Fg-thermo-func}
\end{figure}

Referring to the transformation in \eqref{e:FtoF0-pre}, the first two terms in \eqref{e:Fg-thermo-eq} can be arranged as
\begin{align}
F(T,V^\subL,N^\subL) + F(T,V-V^\subL,N-N^\subL)
&= F(T,V_0^\subL,N_0^\subL) + F(T,V-V_0^\subL,N-N_0^\subL) \nonumber\\
&= F_g(T,V,N,mgL=0).
\end{align}
Since $F_g(T,V,N,mgL=0) = F(T,V,N)$, we conclude
\begin{align}
F_g(T,V,N,mgL) = F(T,V,N) - m|g|L N \psi_g^{(\subL,\subG)}(T,V/N).
\label{e:Fg-F-0}
\end{align}
Thus, $F_g(T,V,N,mgL)$ becomes a continuous but non-smooth function of $mgL$ at $g=0$, as illustrated in Fig.~\ref{fig:Fg-thermo-func}(a). 
These results confirm that $g=0$ is a first-order transition point. 
Consequently, thermodynamic properties in liquid-gas coexistence exhibit singular behavior in the small-gravity limit $|g| \rightarrow 0$.
From $\bar v$ dependence of $\psi_g^{(\subL,\subG)}$ in \eqref{e:xm-X-g}, the deviation of $F_g(T,V,N,mgL)$ from $F(T,V,N)$ can be calculated and schematically drawn as Fig.~\ref{fig:Fg-thermo-func}(b).

\subsection{Relation between $F_g$ and $F$}

Using \eqref{e:Psig-absolute}, Eq.~\eqref{e:Fg-F-0} is expressed as
\begin{align}
F_g(T,V,N,mgL) = F(T,V,N) - mgL \Psi_g(T,V,N),
\label{e:F-Fg}
\end{align}
despite the singularity at $g=0$. 
The total differential of $F_g$ forms the fundamental relation of thermodynamics, derived in \eqref{e:Fg-relation}, as
\begin{align}
dF_g = -S dT - \bar{p} dV + \bar{\mu} dN - \Psi_g d(mgL), 
\label{e:Fg-relation-eq}
\end{align}
which implies
\begin{align}
\pderf{F_g}{mgL}{T,V,N} = -\Psi_g,
\end{align}
even though $\Psi_g \propto g/|g|$ is singular at $g=0$.

Note that \eqref{e:F-Fg} provides a Legendre transformation. 
Applying \eqref{e:F-Fg} to \eqref{e:Fg-relation-eq}, we obtain
\begin{align}
dF = -S dT - \bar{p} dV + \tilde{\mu} dN + NmgL \, d\left(\frac{\xm-X}{L}\right).
\label{e:F-relation-eq0}
\end{align}
Here, we replaced $\bar{\mu}$ with a global chemical potential defined as
\begin{align}
\tilde{\mu} \equiv \frac{A}{N} \int_{\botX}^{\botX+L} dx \, \mu(x) \rho(x), 
\label{e:def-tilde-mu}
\end{align}
with which $\bar{\mu}$ is expressed as
\begin{align}
\bar{\mu} = \tilde{\mu} + mg(X - \xm),
\label{e:mg=bmu}
\end{align}
derived from the local relation \eqref{e:mug-local}.

Since $F$ is the free energy of the system without gravity, 
we rewrite \eqref{e:F-relation-eq0} as
\begin{align}
dF = -S_0 dT - p_0 dV + \mu_0 dN,
\label{e:F-relation-eq0-2}
\end{align}
where 
\begin{align}
  &S_0=S(T,V,N,mgL)-NmgL\pderf{(\xm-X)/L}{T}{V,N},\label{e:S0}\\
  &p_0=\bP(T,V,N,mgL)-NmgL\pderf{(\xm-X)/L}{V}{T,N},\label{e:p0}\\
  &\mu_0=\tmu(T,V,N,mgL)+NmgL\pderf{(\xm-X)/L}{N}{T,V}.\label{e:mu0}
 \end{align}
Note that $S_0$, $p_0$, and $\mu_0$ are independent of $g$ and should correspond to $S(T,V,N)$, $p(T,V,N)$, and $\mu(T,V,N)$ derived from $F(T,V,N)$ for the case $g=0$.
In the phase coexistence state, $S_0$, $p_0$, and $\mu_0$ can be expressed as
\begin{align}
S_0 &= S(T,V_0^\subL,N_0^\subL) + S(T,V_0^\subG,N_0^\subG),
\label{e:eq-values}
\end{align}
and
\begin{align}
p_0 = \Ps(T), \quad \mu_0 = \mu_\subC(T).
\label{e:eq-values-2}
\end{align}
This argument provides expressions for $S$, $\bP$, and $\bmu$ in terms of $(T,V,N,mgL)$ using \eqref{e:S0}, \eqref{e:p0}, and \eqref{e:mu0}.

\section{Numerical demonstration of global thermodynamics}
\label{s:numerical}

In this section, we aim to verify the validity of global thermodynamics for liquid-gas coexistence under gravity.
We present a formula predicted by global thermodynamics, namely Maxwell's relation.
We then perform molecular dynamics simulations on the liquid-gas coexistence and compare the numerical results with the theoretical predictions.

\subsection{Maxwell's relation}
\label{s:Maxwell}

The fundamental relation \eqref{e:Fg-relation-eq} contains an extra term $\Psi_g d(mgL)$, which introduces new Maxwell's relations such as
\begin{align}
\pderf{\bar{p}}{(mgL)}{T,V,N} &= \pderf{\Psi_g}{V}{T,mgL,N}
= \frac{g}{2|g|} \frac{\bar{\rho}^2 - \rho_\subC^\subL(T)\rho_\subC^\subG(T)}{\rho_\subC^\subL(T) - \rho_\subC^\subG(T)},
\label{e:Maxwell-g-eq}
\end{align}
by substituting \eqref{e:Psig-absolute}.
Here, we used $\bar{\rho}=1/\bar{v}$.
From \eqref{e:Maxwell-g-eq}, we obtain
\begin{align}
\bar{p}(T,\bar{\rho},mgL) = \Ps(T) + \frac{|mgL|}{2} \frac{\bar{\rho}^2 - \rho_\subC^\subL(T)\rho_\subC^\subG(T)}{\rho_\subC^\subL(T) - \rho_\subC^\subG(T)},
\label{e:barP-predict}
\end{align}
using $\bar{p} = \Ps(T)$ at $g=0$.
Note that \eqref{e:barP-predict} clearly shows the singularity at $g=0$, as expected.

\begin{figure}[bt] 
\begin{center} 
\includegraphics[width=13cm]{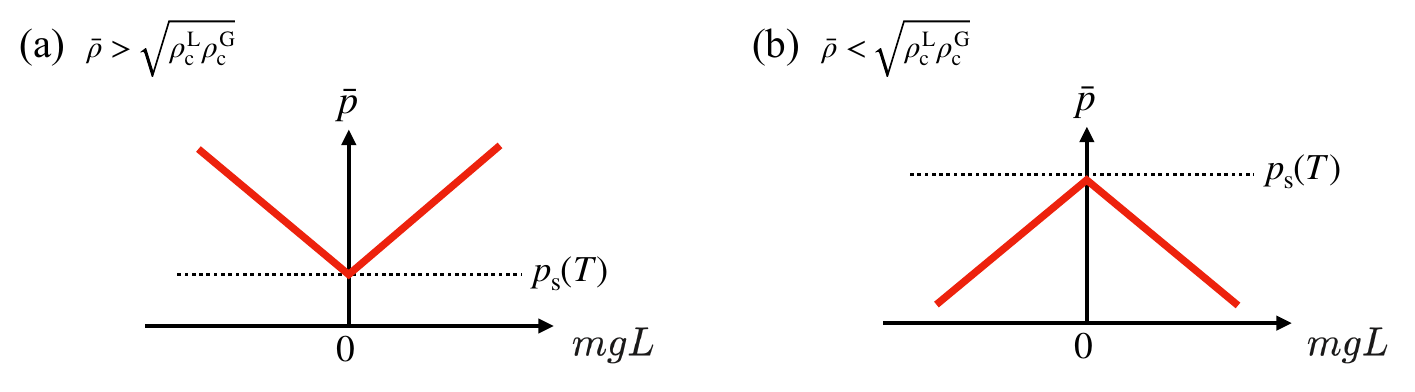}
\end{center} 
\caption{
Dependence of the spatially averaged pressure $\bP$ on $mgL$ as given by \eqref{e:barP-predict}. 
The temperature $T$ is fixed well below the critical temperature. 
(a) $\bP$ increases linearly with $|g|$ when $\bar\rho = N/V$ is greater than $\bar\rho_\subC$. 
(b) $\bP$ decreases with $|g|$ when $\bar\rho < \bar\rho_\subC$. 
}
\label{fig:Maxwell}
\end{figure}

\eqref{e:barP-predict} predicts that there is a special value for $\bar{\rho}=V/N$,
\begin{align}
\bar{\rho}_\subC \equiv \sqrt{\rho_\subC^\subL(T)\rho_\subC^\subG(T)},
\label{e:rho-cr}
\end{align}
at which the spatial average of the pressure, $\bar{p}$, remains unchanged with the change of the gravitational acceleration and is kept as 
\begin{align}
\bar{p} = \Ps(T).
\end{align}
$\bar{p}$ grows linearly from the saturation pressure $\Ps(T)$ with $|mgL|$ for $\bar{\rho} > \bar{\rho}_\subC$, while it decreases when $\bar{\rho} < \bar{\rho}_\subC$, as shown in Fig.~\ref{fig:Maxwell}.
Because the interface pressure is $p_\theta = \Ps(T)$, as derived in \eqref{e:ptheta-Ps}, we have $\bar{p} = p_\theta$.
Thus, $\bar{\rho}_\subC$ is the special density at which the local interface thermodynamics is observed as global thermodynamics.

Substituting \eqref{e:xm-X-g} into the relation $\Pm = \bar{p} + mg\bar{\rho}(\xm - X)$ in \eqref{e:bP-Pm}, 
and combining with $p_{\rm b} - p_{\rm t} = mgL\bar{\rho}$ from the force balance yields
\begin{align}
p_{\rm b} &= \bar{p} + \frac{mgL}{2} \left[\bar{\rho} + \frac{g}{|g|} \frac{(\bar{\rho}-\rho_\subC^\subG)(\rho_\subC^\subL-\bar{\rho})}{\rho_\subC^\subL-\rho_\subC^\subG}\right], \\
p_{\rm t} &= \bar{p} + \frac{mgL}{2} \left[-\bar{\rho} + \frac{g}{|g|} \frac{(\bar{\rho}-\rho_\subC^\subG)(\rho_\subC^\subL-\bar{\rho})}{\rho_\subC^\subL-\rho_\subC^\subG}\right].
\end{align}
Substituting \eqref{e:barP-predict}, we obtain
\begin{align}
p_{\rm b} &= \Ps(T) + mgL\bar{\rho} \frac{\rho_\subC^\subL}{\rho_\subC^\subL-\rho_\subC^\subG}, \\
p_{\rm t} &= \Ps(T) - mgL\bar{\rho} \frac{\rho_\subC^\subG}{\rho_\subC^\subL-\rho_\subC^\subG},
\end{align}
for $g>0$.

\subsection{Fluctuation-response relation}

Let the center of mass and its deviation from $\xm$ be
\begin{align}
\hat{X} = \frac{1}{N} \sum_{i=1}^{N} x_i, \qquad \hat{Y} \equiv \frac{\xm - \hat{X}}{L}.
\end{align}
Their means are denoted by $X = \langle \hat{X} \rangle$ and $Y = \langle \hat{Y} \rangle$.
Define a response
\begin{align}
\chi_g(T,V,N,mgL) \equiv \pder{Y}{(mgL)},
\label{e:chig-def}
\end{align}
which can be expressed as $\chi_g(T,\bar{\rho},mgL)$ because of its intensive nature. Using
\begin{align}
\chi_g = \frac{\partial}{\partial (mgL)} \int d\Gamma \frac{\hat{Y}}{Z_g} e^{-\beta H(\Gamma)} e^{\beta N mgL \hat{Y}},
\end{align}
and the relations
\begin{align}
&\left. \pderf{Z_g}{(mgL)}{T,L,A,N} \right|_{mgL} = Z_g \beta N \langle \hat{Y} \rangle, \\
&\left. \pdertf{Z_g}{(mgL)}{T,L,A,N} \right|_{mgL} = Z_g \beta^2 N^2 \langle \hat{Y}^2 \rangle,
\end{align}
we derive a fluctuation-response relation,
\begin{align}
\chi_g(T,\bar{\rho},mgL) = \beta N \sigma_Y^2(T,L,A,N,mgL),
\label{e:FRR}
\end{align}
where $\sigma_Y$ is the standard deviation of $\hat{Y}$,
\begin{align}
\sigma_Y(T,L,A,N,mgL) \equiv \sqrt{\langle (\hat{Y} - \langle \hat{Y} \rangle)^2 \rangle}.
\end{align}
The relation \eqref{e:FRR} is expressed as
\begin{align}
\sigma_Y(T,L,A,\bar{\rho},mgL) = \sqrt{\frac{1}{LA}}
\sqrt{\frac{k_B T \chi_g(T,\bar{\rho},mgL)}{\bar{\rho}}}.
\label{e:FRR-scaling}
\end{align}
The formula \eqref{e:FRR-scaling} indicates that fluctuations can be ignored in the thermodynamic limit $N \to \infty$, $L \to \infty$, $A \to \infty$ with $mgL$ and $\bar{\rho}$ fixed.
Note that $L \to \infty$ should be taken with $mg \to 0$ to keep $mgL$ at a small finite value.
Moreover, \eqref{e:FRR-scaling} leads to a scaling relation
\begin{align}
LA\sigma^2_Y(T,L,A,\bar\rho,\phi) =
L'A'\sigma^2_Y(T,L',A',\bar\rho,\phi),
\label{e:FRR-scaling2}
\end{align}
where $\phi=mgL=mg'L'$ holds on both sides.

\subsection{Model and parameters}
\label{s:Nmodel}

We adopt two-dimensional systems under gravity for the numerical demonstration.
We confine $N$ particles in a rectangular container with side lengths $L$ and $A$.
The direction of gravity is opposite to the $x$-axis.
We set $\botX=0$ and $\topX=L$ without loss of generality.
Because the system is two-dimensional, the horizontal area $A$ in the previous sections is identified with the size $A$.
For a collection of particle positions and momenta,
$\Gamma=(\bv{r}_1, \bv{r}_2, \cdots,\bv{r}_N, \bv{p}_1, \bv{p}_2, \cdots,\bv{p}_N)$, we set the Hamiltonian as
\begin{equation}
H_g(\Gamma) = \sum_{i=1}^N \left[ \frac{|\bv{p}_i|^2}{2m} + \sum_{j<i} \phi(|\bv{r}_i - \bv{r}_j|) + m g x_i + V_{\rm wall}(\bv{r}_i;L,A) \right],
\end{equation}  
where $m$ is the mass of the particles, and $g$ is the gravitational acceleration.
The two-body interaction potential $\phi$ is the 12-6 Lennard--Jones interaction 
\begin{align}
\phi(r) = 4\epsilon \left[ \left(\frac{\sigma}{r}\right)^{12} - \left(\frac{\sigma}{r}\right)^6 \right]\theta(r_c - r). 
\label{e:LJ}
\end{align}
Here, $r$ is the distance between two particles, $\epsilon$ is the well depth, $\sigma$ is the particle diameter, $r_c$ is the cutoff length of
the interaction, and $\theta(r_{\rm c} - r)$ is the Heaviside step function. 
We set $r_{\rm c} = 3\sigma$. 
We assume a fixed boundary condition for the four walls using a soft-core repulsive wall represented by
$V_{\rm wall}(\bv{r}_i)$ with the Weeks--Chandler--Andersen potential 
truncating the attractive interaction in \eqref{e:LJ}, while replacing $\sigma$ with $0.5\sigma$ to weaken the repulsive force from the walls.
Below, we take the Lennard--Jones unit such that $\sigma=1$, $\epsilon=1$, and $m=1$ with 
the Boltzmann constant $k_B=1$.

We perform the molecular dynamics simulations with the Nose--Hoover thermostat.
We fix $T=0.43$. 
To be specific, at $T=0.43$, we estimate numerically as $\rho_\subC^\subG \simeq 0.025$ and $\rho_\subC^\subL \simeq 0.71$.
We then have
\begin{align}
\bar{\rho}_\subC \simeq 0.133
\end{align}
according to \eqref{e:rho-cr}.
Here, the volume ratio of the gas and liquid is estimated as $82:18$.

\subsection{Spatial average of the pressure}

\begin{figure}[bt] 
\begin{center} 
\includegraphics[width=13cm]{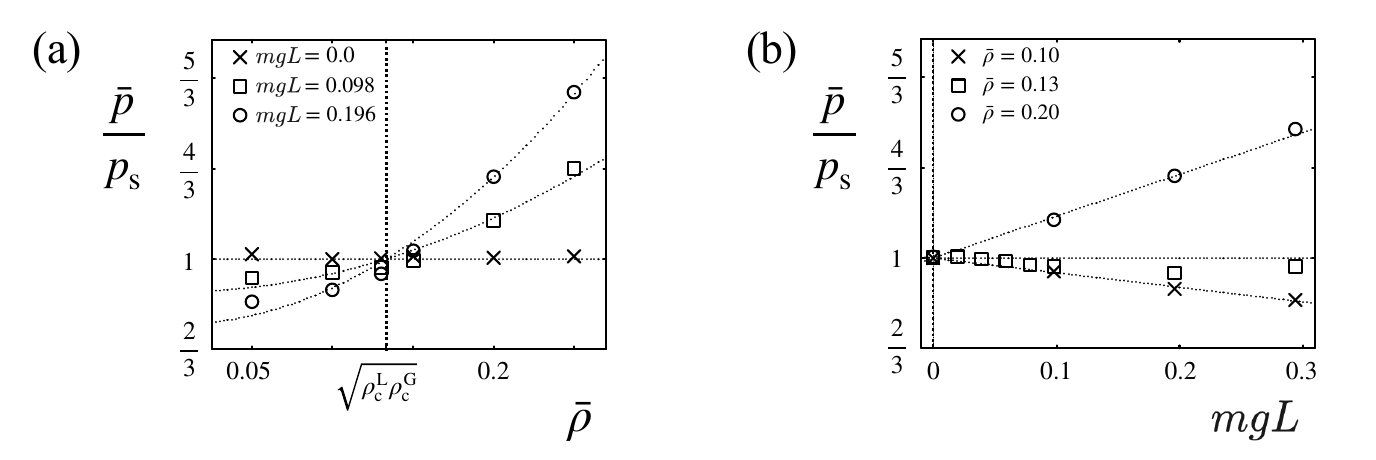}
\end{center} 
\caption{Normalized virial pressure $\bP/\Ps(T)$ for a system with vertical size $L = 196$, temperature $T=0.43$, and $N = 5000$ particles. The saturation pressure used for normalization, $\Ps(T)$, was estimated to be $0.0105$ from simulations at $g = 0$. (a) Dependence on mean density $\bar\rho$ for different gravitational strengths: $mgL = 0$ (crosses), $0.098$ (squares), and $0.196$ (circles). Higher density $\bar\rho$ corresponds to smaller horizontal size $A = N / (\bar\rho L)$. (b) Dependence on the gravitational parameter $mgL$ for fixed mean densities: $\bar\rho = 0.1$ (crosses), $0.13$ (squares), and $0.2$ (circles). In both panels, dotted lines show the theoretical prediction from Eq. \eqref{e:barP-predict} for the thermodynamic limit, using parameter values $\rho_\subC^\subG = 0.025$ and $\rho_\subC^\subL = 0.71$.
}
\label{fig:mgL-P}
\end{figure}

The instantaneous virial pressure is defined as
\begin{align}
\hat p = \frac{1}{2L A} \left[ \sum^N_{i=1} \frac{|\bv{p}_i |^2}{m} - \sum^N_{i=1}\sum^N_{j>i} \frac{\partial \phi(r_{ij})}{\partial \bv{r}_{ij}} \cdot \bv{r}_{ij}  \right].
\end{align}
This quantity corresponds to the total force exerted by all particles on the walls, divided by the total wall length $2(L+A)$. Thus, $\hat p$ represents the spatially averaged pressure $\bP$ over the system. Specifically, 
\begin{align}
\bP = \bbkt{\hat p},
\end{align}
where the ensemble and long-term averages are taken after sufficient relaxation.

Figure \ref{fig:mgL-P}(a) shows numerical results for $\bP$ as a function of the horizontal size $A$ for three different values of $mg$, with a vertical size $L=196$ and a fixed number of particles $N=5000$. The mean density is defined as $\bar\rho=N/(LA)$. The numerical results agree well with the theoretical predictions from \eqref{e:barP-predict}, which are plotted as dotted lines for the corresponding $mgL$ values. However, deviations from the theoretical curves occur at small $\bar\rho$ values (i.e., large $A$) because the liquid tends to form a droplet at the bottom of the container when $\bar\rho$ is small. This droplet formation scenario is outside the scope of the current theory, which assumes a flat liquid-gas interface.

Figure \ref{fig:mgL-P}(b) illustrates the dependence of $\bP$ on the gravitational parameter $mgL$ for three different $\bar\rho$ values, again with $L=196$ and $N=5000$. It is observed that $\bP$ deviates linearly from the saturation pressure $\Ps(T)$ as $mgL$ increases. At the specific density $\bar\rho=\bar\rho_\subC$ defined in \eqref{e:rho-cr}, $\bP$ becomes invariant with respect to changes in $mgL$.

\subsection{Center of mass}
\label{s:X-mgL}

\begin{figure}[bt] 
\begin{center} 
\includegraphics[width=10cm]{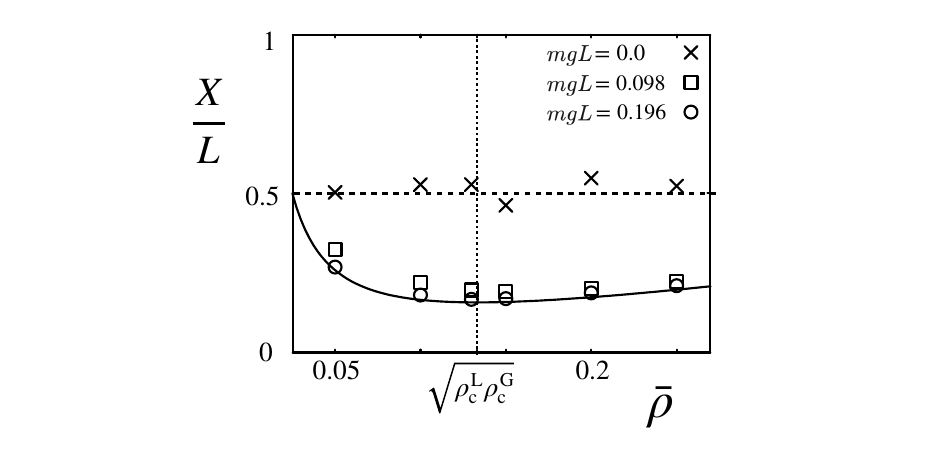}
\end{center} 
\caption{ 
Scaled center of mass $X/L$ versus mean density $\bar\rho$ for different gravitational strengths: $mgL=0$ (crosses), $mgL=0.098$ (squares), and $mgL=0.196$ (circles). Data are obtained from the simulations presented in Fig. \ref{fig:mgL-P}(a). The solid line shows the theoretical prediction for the thermodynamic limit \eqref{e:xm-X-g}. The dotted line represents $X=\xm$, the expected value for $g=0$.
}
\label{fig:rho-X}
\end{figure}

\begin{figure}[bt] 
\begin{center} 
\includegraphics[width=13cm]{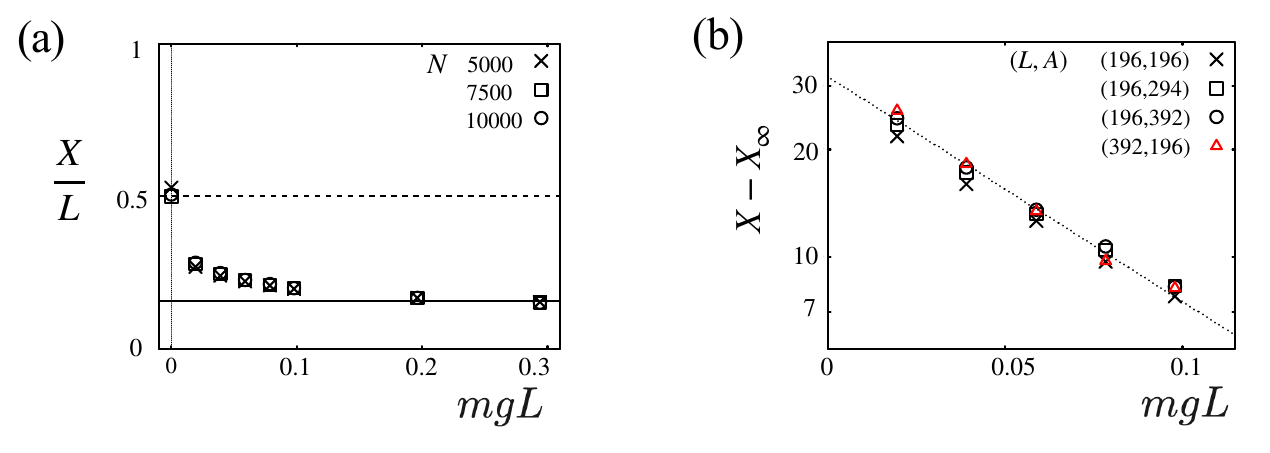}
\end{center} 
\caption{
(a) Scaled center of mass $X/L$ versus the gravitational parameter $mgL$ (varied by changing $mg$) at fixed density $\bar\rho=0.13\simeq\bar\rho_\subC$ and fixed vertical height $L=196$. Symbols represent different horizontal sizes: $(A,N)=(196,5000)$ (crosses), $(294,7500)$ (rectangles), and $(392,10000)$ (circles). The solid line is $X(L)/L=X_\infty(L)$, where $X_\infty$ is the theoretical prediction in \eqref{e:xm-X-g}, and the dotted line is $X=\xm$.
(b) Semi-logarithmic plot of the deviation $X-X_\infty$ versus $mgL$ (lateral axis is logarithmic). Symbols correspond to the systems shown in (a); red triangles represent additional data for increased height $(L,A,N)=(392,196,10000)$. The dotted line shows the exponential fit $X-X_\infty=L_0 e^{-0.16\beta mgL}$ with $L_0=32$.
}
\label{fig:mgL-X}
\end{figure}

Figure \ref{fig:rho-X} presents the center of mass, $X=\sbkt{\hat X}$, as a function of the mean density $\bar\rho$, obtained from the numerical simulations described in Fig. \ref{fig:mgL-P}(a). The solid line represents the theoretical prediction from  \eqref{e:xm-X-g} in the thermodynamic limit (where $\bar v=1/\bar\rho$), while the dotted line indicates the trivial result $X=\xm$ expected in the absence of gravity. At $g=0$ (corresponding to $mgL=0.0$), the simulation results fluctuate around $X\simeq\xm$. For non-zero gravity ($mgL=0.098$ and $0.196$), the results approach the theoretical curve, although some deviations persist.

Compared to the good agreement observed for the pressure (Fig. \ref{fig:mgL-P}), the center of mass $X$ presented in Fig. \ref{fig:rho-X} exhibits larger deviations from its theoretical prediction \eqref{e:xm-X-g} for $mgL>0$. While these deviations might be partly influenced by the effects occurring at small $\bar\rho$ (large $A$), as discussed previously, their magnitude suggests that this is not the complete explanation. Notably, Fig. \ref{fig:rho-X} itself shows that for a given $mgL>0$, the results for $X$ appear relatively converged with respect to changes in $A$ (implicit in the variation of $\bar\rho$). This hints that the system might be reasonably close to the thermodynamic limit concerning the horizontal dimension $A$. Therefore, we hypothesize that the primary source of the remaining, significant deviation observed in Fig. \ref{fig:rho-X} is the finite size of the system in the vertical direction, $L$. To test this hypothesis and further analyze the interplay between gravity and finite system height, we examine the behavior of the scaled center of mass $X/L$ as a function of the gravitational parameter $mgL$, while maintaining a fixed mean density near the critical value, 
$\bar\rho=N/(LA)=0.13\simeq\bar\rho_\subC$.

Figure \ref{fig:mgL-X}(a) shows the dependence of $X/L$ on $mgL$ (varied by changing $mg$) for a fixed vertical size $L=196$ and fixed density $\bar\rho=0.13$. Different symbols correspond to different horizontal sizes $A$, with the number of particles $N$ adjusted accordingly ($N=\bar\rho LA$); specifically, $(A,N)=(196,5000)$, $(294,7500)$, and $(392,10000)$. The behavior of $X/L$ clearly differs between $g=0$ and $g>0$, indicating that $mgL=0$ represents a singular limit for the center of mass behavior, as expected. The numerical results for $X/L$ are consistent across the different values of $A$ studied. However, deviations from the theoretical estimate \eqref{e:xm-X-g} are observed in the weak gravity regime ($0<mgL\lesssim 0.1$), even though the theory is formulated for this regime.

To further investigate these deviations, we performed additional simulations with an increased vertical height, $L=392$ (double the previous value), using parameters $(L,A,N)=(392,196,10000)$ while keeping $\bar\rho=0.13$. Figure \ref{fig:mgL-X}(b) shows a logarithmic plot of the deviation $X-X_\infty(L)$ for $0<mgL\le 0.1$, where $X_\infty(L)$ is the theoretical value from \eqref{e:xm-X-g}, 
\begin{align}
X_\infty(L) \equiv L\lim_{\substack{L'\to\infty\\A\to \infty}}\frac{X(L',A,\bar\rho L'A,\phi)}{L'}
=L\left[\frac{1}{2}- \frac{\phi}{2|\phi|} \frac{(v^\subG_\subC(T)-\bar{v})(\bar{v}-v^\subL_\subC(T))}{\bar{v}(v^\subG_\subC(T)-v^\subL_\subC(T))}\right]
\end{align}
with $\bar{v}=1/\bar\rho$.
The plot suggests a scaling relation for this deviation in the weak gravity regime (small finite $\phi=mgL$) of the form:
\begin{align}
X(L,A,\bar\rho LA,\phi)=X_\infty(L) +L_0e^{-c\beta \phi}.	
\end{align}
Here, $L_0 \simeq 32$ is a characteristic length, and $c$ is a positive numerical constant.
This scaling behavior suggests that the deviations observed in Fig. \ref{fig:mgL-X}(a) are primarily finite-size effects related to the vertical dimension $L$. It indicates that $X/L$ converges towards the theoretical curve given by \eqref{e:xm-X-g} as the height $L$ increases. The characteristic length $L_0$ determined numerically is relatively large, though finite. The physical interpretation of this length scale $L_0$ warrants further investigation, particularly concerning its connection to weak gravity effects.

\section{Concluding remarks}

We have formulated global thermodynamics for equilibrium systems under weak gravity. Thermodynamic variables were extended from $(T,V,N)$ to $(T,V,N,mgL)$ by adopting the midpoint $\xm$ of the system as the reference for the gravitational potential. The free energy landscape revealed the presence of both global and local minima in the phase coexistence states. The principle of minimum free energy identified the global minimum and characterized the inversion of the liquid-gas configuration as a first-order transition at $g=0$. The numerical simulation results aligned well with the predictions of global thermodynamics. Before concluding this paper, we present three remarks.

First, in this paper, we focused on isothermal systems under gravity at fixed $T$. It is possible to formulate global thermodynamics in other scenarios. For instance, consider energy-conserved systems where the total energy $E = U + mg(X-\xm)$ is fixed. The maximum entropy principle determines the equilibrium state. Define entropy as a variational function in fixed $(E, V, N, mgL)$
\begin{align}
{\mathcal S}({\mathcal E}^\subl, {\mathcal V}^\subl, {\mathcal N}^\subl; E, V, N, mgL)
\equiv S({\mathcal E}^\subl, {\mathcal V}^\subl, {\mathcal N}^\subl)
+ S({\mathcal E}^\subu, V - {\mathcal V}^\subl, N - {\mathcal N}^\subl),
\end{align}
where ${\mathcal E}^\subl$ and ${\mathcal E}^\subu$ are the total energies in the lower and upper regions, respectively. These energies satisfy ${\mathcal E}^\subl = {\mathcal U}^\subl$ and ${\mathcal E}^\subu = {\mathcal U}^\subu$. Since the reference points for the gravitational potential differ between $\subl$ and $\subu$, they are related by:
\begin{align}
{\mathcal E}^\subl + {\mathcal E}^\subu = E + mgL \frac{N}{2} \left( \frac{{\mathcal V}^\subl}{V} - \frac{{\mathcal N}^\subl}{N} \right).
\end{align}
Thermodynamic entropy corresponds to the maximal value of ${\mathcal S}$ as
\begin{align}
S(E, V, N, mgL) = \max_{{\mathcal E}^\subl, {\mathcal V}^\subl, {\mathcal N}^\subl} {\mathcal S}({\mathcal E}^\subl, {\mathcal V}^\subl, {\mathcal N}^\subl; E, V, N, mgL).
\end{align}
This satisfies a fundamental relation of thermodynamics:
\begin{align}
dS = \frac{dE}{T} + \frac{\bar{p}}{T} dV - \frac{\bar{\mu}}{T} dN - \frac{N}{T} \frac{X - x_{\rm m}}{L} d(mgL),
\end{align}
where the maximum entropy principle leads to $p_\theta = \Ps(T)$ and
\begin{align}
\frac{1}{T(E, V, N, mgL)} = \pderf{S}{E}{V, N, mgL}.
\end{align}

Second, we discuss a possible non-equilibrium extension of the results presented in this paper. Our previous studies proposed global thermodynamics as a framework to extend conventional thermodynamics to heat-conducting systems \cite{Global-PRL,Global-JSP,Global-PRR,KNS}. One key prediction is the deviation of the interface temperature, proportional to the heat flux, from the equilibrium transition temperature. This paper has formulated the equilibrium version of global thermodynamics, ensuring consistency with conventional thermodynamic principles, and has demonstrated its validity. Building on this foundation, we aim to extend global thermodynamics to systems with heat conduction under gravity. 

For example, consider a system where heat baths at higher and lower temperatures are attached at $x=0$ and $x=L$, respectively. Without gravity ($g=0$), a liquid region emerges near $x=L$ in the liquid-gas coexistence state. An intriguing question arises: What happens when gravity $g>0$ is introduced? If the gravitational force is sufficiently strong, the liquid phase shifts near $x=0$. Conversely, when gravity is weak, the liquid remains near $x=L$. This behavior suggests a transition as gravitational strength changes. Numerical simulations of molecular dynamics \cite{ANS} and related studies \cite{Wagner} provide insight into this transition. Extending global thermodynamics under gravity to include heat conduction could offer quantitative predictions for such transitions \cite{Global-Heat-Gravity}.

Here, one might wonder whether the predictions of global thermodynamics are consistent with those of a continuum hydrodynamic description based on the free energy functional
\begin{align}
{\mathcal F}([\rho]) = \int_{\mathcal{D}} d^3{\mathbf{r}} \left[ f(T, \rho(\mathbf{r})) + mg x \rho(\mathbf{r}) + \frac{\kappa}{2} |\nabla \rho|^2 \right], 
\label{e:Fg-var-functional}
\end{align}
where $[\rho]$ denotes a smooth density profile $\rho(\mathbf{r})$ in a region $\mathcal{D}$, $f$ is a mesoscopic free energy density, and $\kappa$ corresponds to the surface tension at mesoscopic scales. For equilibrium systems under gravity, the continuum theory provides the same thermodynamic behavior as global thermodynamics formulated in the present paper. However, for heat conduction systems without gravity ($g = 0$), it is known that global thermodynamics predicts that the pressure at the interface deviates from the saturation pressure under heat conduction, whereas the continuum theory does not. Understanding this discrepancy remains an important and open question. 

The final remark addresses this question. We begin by reviewing the continuum description of phase coexistence states, originally proposed by van der Waals \cite{vanPress}. One significant challenge lies in achieving a microscopic understanding of the effective Hamiltonian. This difficulty arises because, for systems far from the critical point, the liquid-gas interface width is microscopic, rendering the density gradient too sharp for the van der Waals theory to apply. Interface fluctuations have been extensively studied through statistical mechanics \cite{Kirkwood_Buff,BLS,Weeks,Rowlinson_Widom,Stillinger,Napiorkowski_Dietrich,Mecke_Dietrich}. A consistent result with the van der Waals theory was derived in \cite{Weeks}, while non-analytic behavior of long-range fluctuations in the direction parallel to the steep liquid-gas interface was observed, depending on the type of molecular interactions. These fluctuations cannot be described by the van der Waals theory \cite{Napiorkowski_Dietrich,Mecke_Dietrich}. 
Experiments \cite{Daillant,Daillant2,Rice} and molecular dynamics simulations \cite{Chacon_Tarazona,Chacon_Tarazona2} further suggest limitations in the van der Waals theory. Moreover, in heat conduction states, the microscopic singularity at the liquid-gas interface becomes evident. For instance, numerical simulations have reported a microscopic temperature gap at the interface \cite{Muscatello}, suggesting a breakdown of the continuum description in this region.

Despite these uncertainties, it is anticipated that the van der Waals theory can effectively capture universal characteristics, provided that macroscopic behaviors involving interfaces are independent of microscopic details. 
For example, it is well established that the canonical distribution with the effective Hamiltonian ${\mathcal F}([\rho])$ reproduces the critical exponents near the liquid-gas critical point \cite{CP}, and furthermore, that phenomenological arguments using ${\mathcal F}([\rho])$ successfully reproduce dynamic universality classes \cite{Hohenberg_Halperin1977}.
However, outside these successful examples, the validity of this approach should be reconsidered by conducting a detailed comparison with experimental results. 
In such cases, a universal framework and corresponding mesoscopic models should also be proposed to better understand the phenomena under study. 
In this context, our recent study is worth highlighting. In Ref.~\cite{Global-modelB}, we investigated a discrete fluctuating hydrodynamics of a density field under a particle current driven by a battery. 
By applying non-equilibrium statistical mechanics \cite{Zubarev,Mclennan,Maes-rep}, we found that thermodynamic properties deviate from those predicted by the continuum description with the van der Waals theory but align with the predictions of global thermodynamics. Notably, this deviation arises only when the liquid-gas interface is assumed to be singular; it disappears when the density profile at the interface becomes smooth. 
These results suggest that global thermodynamics is the only valid framework for systems with a singularly thin interface. Constructing a consistent narrative that bridges microscopic and macroscopic scales remains a fascinating and challenging endeavor.

\section*{Acknowledgments}

The authors thank F. Kagawa and A. Hisada for useful discussions about experiments on phase coexistence; S. Yukawa and T. Nakamura for discussions on numerical simulations.
The numerical simulations were performed with LAMMPS.
This work was supported by JSPS KAKENHI Grant Numbers JP22H01144, JP23K22415, and JP25K00923.

\vspace{5mm}

\noindent{\small {\bf Data Availability}:  Data sharing  not
    applicable to this article as no datasets were
generated or analysed during the current study.}

\vspace{5mm}

\noindent{\small {\bf Conflict of Interest}:  The authors have no
  financial or proprietary interests in any material discussed
  in this article.
}

\newpage

\appendix\normalsize
\renewcommand{\theequation}{\Alph{section}.\arabic{equation}}
\setcounter{equation}{0}
\makeatletter
  \def\@seccntformat#1{%
    \@nameuse{@seccnt@prefix@#1}%
    \@nameuse{the#1}%
    \@nameuse{@seccnt@postfix@#1}%
    \@nameuse{@seccnt@afterskip@#1}}
  \def\@seccnt@prefix@section{Appendix }
  \def\@seccnt@postfix@section{:}
  \def\@seccnt@afterskip@section{\ }
  \def\@seccnt@afterskip@subsection{\ }
\makeatother

\section{Derivation of effective Hamiltonian}

We derive the probability density $\rho(\mathcal{N}^\sublR;l)$, defined in \eqref{e:rhoNl-def}, and the effective Hamiltonian in \eqref{e:eff-Hamiltonian}.

The number of particles and the center of mass for each region satisfy the following relations:
\begin{align}
&N = \hat{N}^\sublR(\Gamma;l) + \hat{N}^\subuR(\Gamma;l), \\
&N\hat X(\Gamma) = \hat{N}^\sublR(\Gamma;l) \hat X^\sublR(\Gamma;l) + \hat{N}^\subuR(\Gamma;l) \hat X^\subuR(\Gamma;l),
\end{align}
where
\begin{align}
\hat{N}^\sublR(\Gamma;l) &= \sum_{i=1}^N \int_{\botX}^{\botX + l} dx~\delta(x_i - x), \\
\hat{N}^\subuR(\Gamma;l) &= \sum_{i=1}^N \int_{\botX + l}^{\botX + L} dx~\delta(x_i - x), \\
\hat X^\sublR(\Gamma;l) &= \frac{1}{\hat{N}^\sublR(\Gamma;l)} \sum_{i=1}^N \int_{\botX}^{\botX + l} dx~x \delta(x_i - x), \\
\hat X^\subuR(\Gamma;l) &= \frac{1}{\hat{N}^\subuR(\Gamma;l)} \sum_{i=1}^N \int_{\botX + l}^{\botX + L} dx~x \delta(x_i - x).
\end{align}
Let $\xm^\sublR$ and $\xm^\subuR$ denote the midpoint of the lower and upper regions, respectively,
\begin{align}
\xm^\sublR = \botX + \frac{l}{2}, \quad \xm^\subuR = \botX + \frac{L + l}{2}.
\end{align}
Using these definitions, we find
\begin{align}
\hat{N}^\sublR(\Gamma;l) \xm^\sublR + \hat{N}^\subuR(\Gamma;l) \xm^\subuR - N \xm = \frac{1}{2}(lN - L\hat{N}^\sublR(\Gamma;l)).
\end{align}
The Hamiltonian \eqref{e:Hamiltonian} can then be rewritten as
\begin{align}
H_g(\Gamma) &= H(\Gamma) + mg\hat{N}^\sublR(\Gamma;l)(\hat X^\sublR(\Gamma;l) - \xm^\sublR) + mg\hat{N}^\subuR(\Gamma;l)(\hat X^\subuR(\Gamma;l) - \xm^\subuR) \nonumber \\
&\quad + \frac{mg}{2}(lN - L\hat{N}^\sublR(\Gamma;l)).
\label{e:Hg-def3}
\end{align}
This expression holds exactly, regardless of the choice of the length $l$ in $0 < l < L$.

The probability density given in \eqref{e:rhoNl-def} can now be rewritten as
\begin{align}
\rho(\mathcal{N}^\sublR;l) &= e^{-\beta \frac{mg}{2}(lN - L\mathcal{N}^\sublR)} \frac{1}{Z_g} \int d\Gamma~\delta(\mathcal{N}^\sublR - \hat{N}^\sublR(\Gamma;l)) \nonumber \\
&\quad \times e^{-\beta \left[ H(\Gamma) + mg\mathcal{N}^\sublR(\hat X^\sublR(\Gamma;l) - \xm^\sublR) + mg\mathcal{N}^\subuR(\hat X^\subuR(\Gamma;l) - \xm^\subuR) \right]},
\label{e:rhoNl-pre}
\end{align}
where $\mathcal{N}^\sublR$ is the number of particles in the lower region, $ \botX < x \leq \botX + l$, and $\mathcal{N}^\subuR = N - \mathcal{N}^\sublR$ is the number in the upper region, $\botX + l < x < \botX + L$.

We note that the microstate $\Gamma=(\bv{r}_1,\cdots,\bv{r}_N,\bv{p}_1,\cdots,\bv{p}_N)$ is expressed 
with all particles indexed. We then introduce a reordered expression of the microstate, $\Gamma^{\rm (re)}=(\bv{r}_1^{\rm (re)},\cdots,\bv{r}_N^{\rm (re)},\bv{p}_1^{\rm (re)},\cdots,\bv{p}_N^{\rm (re)})$, 
in which the indices for the $N$ particles are reordered from the smallest to largest values of the $x$-coordinate as $x_1^{\rm (re)}<x_2^{\rm (re)}<\cdots<x_N^{\rm (re)}$. 

Clearly, due to the symmetry of the particles, $H(\Gamma)=H(\Gamma^{\rm (re)})$. We then divide $\Gamma^{\rm (re)}$ into two as 
\begin{align}
\Gamma^{\rm (re)}=(\Gamma^\sublR,\Gamma^\subuR),
\end{align}
where 
\begin{align}
\Gamma^\sublR=(\bv{r}_1^{\rm (re)},\cdots,\bv{r}_{{\mathcal N}^\sublR}^{\rm (re)},\bv{p}_1^{\rm (re)},\cdots,\bv{p}_{{\mathcal N}^\sublR}^{\rm (re)}),\qquad \Gamma^\subuR=(\bv{r}_{{\mathcal N}^\sublR+1}^{\rm (re)},\cdots,\bv{r}_{N}^{\rm (re)},\bv{p}_{{\mathcal N}^\sublR+1}^{\rm (re)},\cdots,\bv{p}_{N}^{\rm (re)}).
\end{align}
That is, $\Gamma^\sublR$ represents the microstate for the first ${\mathcal N}^\sublR$ particles found in the lower region, 
and $\Gamma^\subuR$ corresponds to the microstate for the last ${\mathcal N}^\subuR$ particles in the upper region.

We now consider two systems, each containing ${\mathcal N}^\sublR$ and ${\mathcal N}^\subuR$ particles, separated by a virtual wall at $x=\botX+l$. 
The Hamiltonians for the two separated systems are written as
\begin{align}
&H_\sep^\sublR(\Gamma^\sublR)=\sum_{i=1}^{{\mathcal N}^\sublR}
\left[\frac{(\bv{p}_i^{\rm (re)})^2}{2m}+\sum_{j<i}\phi(|\bv{r}_i^{\rm (re)}-\bv{r}_j^{\rm (re)}|)
+\phi_{\rm w}(\bv{r}_i^{\rm (re)})
\right],\\
&H_\sep^\subuR(\Gamma^\subuR)=\sum_{i={\mathcal N}^\sublR+1}^{N}
\left[\frac{(\bv{p}_i^{\rm (re)})^2}{2m}+\sum_{{\mathcal N}^\sublR<j<i}\phi(|\bv{r}_i^{\rm (re)}-\bv{r}_j^{\rm (re)}|)
+\phi_{\rm w}(\bv{r}_i^{\rm (re)})
\right].
\end{align}
From this, we obtain
\begin{align}
H(\Gamma)=
H_\sep^\sublR(\Gamma^\sublR)+H_\sep^\subuR(\Gamma^\subuR)+o(N),
\end{align}
where the term $o(N)$ represents the energetic contribution due to interactions across the virtual wall at $x=\botX+l$.

Given that the centers of mass satisfy
\begin{align}
\hat X_\sep^\sublR(\Gamma^\sublR)=\frac{1}{{\mathcal N}^\sublR}\sum_{i=1}^{{\mathcal N}^\sublR} x_i^{\rm (re)} =\hat X^\sublR(\Gamma;l),\quad
\hat X_\sep^\subuR(\Gamma^\subuR)=\frac{1}{{\mathcal N}^\subuR}\sum_{i={\mathcal N}^\sublR+1}^{N} x_i^{\rm (re)}=\hat X^\subuR(\Gamma;l),
\end{align}
we conclude that
\begin{align}
&H(\Gamma;l)+mg{\mathcal N}^\sublR(\hat X^\sublR(\Gamma;l)-\xm^\sublR)
+mg{\mathcal N}^\subuR(\hat X^\subuR(\Gamma;l)-\xm^\subuR)\nm
&=H_\sep^\sublR(\Gamma^\sublR)+mg{\mathcal N}^\sublR(\hat X_\sep^\sublR(\Gamma^\sublR)-\xm^\sublR)
+H_\sep^\subuR(\Gamma^\subuR)+mg{\mathcal N}^\subuR(\hat X_\sep^\subuR(\Gamma^\subuR)-\xm^\subuR)
\label{e:energy-const}
\end{align}
by neglecting the $o(N)$ term.

Although the partitioning of space does not contribute energetically as shown in \eqref{e:energy-const}, 
the virtual wall at $x=\botX+l$ introduces an entropic contribution by preventing the mixing of particles between the two regions. 
In the above treatment, we reordered the particle indices and separated $\Gamma$ into $\Gamma^\sublR$ and $\Gamma^\subuR$, 
a process attributable to the index sorting mechanism. As a result, the integral over $\Gamma$ can be reformulated as the integrals over $\Gamma^\sublR$ and $\Gamma^\subuR$
\begin{align}
\int d\Gamma \delta({\mathcal N}^\sublR-\hat N^\sublR(\Gamma;l))
=\frac{N!}{{\mathcal N}^\sublR ! {\mathcal N}^\subuR !}\int d\Gamma^\sublR \int d\Gamma^\subuR.
\end{align}
The integral in \eqref{e:rhoNl-pre} then transforms as
\begin{align}
\frac{N!}{{\mathcal N}^\sublR ! {\mathcal N}^\subuR !}&\int d\Gamma^\sublR~e^{-\beta [H_\sep^\sublR(\Gamma^\sublR)+mg{\mathcal N}^\sublR (\hat X_\sep^\sublR(\Gamma^\sublR)-\xm^\sublR)]}\nm
&\times\int d\Gamma^\subuR~e^{-\beta [H_\sep^\subuR(\Gamma^\subuR)+mg{\mathcal N}^\sublR (\hat X_\sep^\subuR(\Gamma^\subuR)-\xm^\subuR)]}
\label{e:Z-divide}
\end{align}
by substituting \eqref{e:energy-const}.

Define the respective partition functions for the two systems,
\begin{align}
&Z_g^\sublR=\int d\Gamma^\sublR~e^{-\beta [H_\sep^\sublR(\Gamma^\sublR)+mg{\mathcal N}^\sublR (\hat X_\sep^\sublR(\Gamma^\sublR)-\xm^\sublR)]},\\
&Z_g^\subuR=\int d\Gamma^\subuR~e^{-\beta [H_\sep^\subuR(\Gamma^\subuR)+mg{\mathcal N}^\subuR (\hat X_\sep^\subuR(\Gamma^\subuR)-\xm^\subuR)]}.
\end{align}
These lead to the respective free energies as
\begin{align}
&F_g^\sublR=-k_B T\left(\ln Z_g^\sublR-\ln {\mathcal N}^\sublR!\right), \label{e:Fg-l}\\
&F_g^\subuR=-k_B T\left(\ln Z_g^\subuR-\ln {\mathcal N}^\subuR!\right). \label{e:Fg-u}
\end{align}
Since the two integrals on the right-hand side of \eqref{e:Z-divide} correspond to $Z_g^\sublR$ and $Z_g^\subuR$, \eqref{e:rhoNl-pre} can be identified as
\begin{align}
\rho({\mathcal N}^\sublR;l)=\frac{Z_g^\sublR Z_g^\subuR}{Z_g}\frac{N!}{{\mathcal N}^\sublR! {\mathcal N}^\subuR!}e^{-\beta\frac{mg}{2}(lN-L{\mathcal N}^\sublR)}. \label{e:rhoNl-0}
\end{align}
We now define an effective Hamiltonian ${\mathcal F}_g({\mathcal N}^\sublR; l)$ to represent the probability density $\rho({\mathcal N}^\sublR;l)$ as
\begin{align}
\rho({\mathcal N}^\sublR;l)=e^{\beta(F_g-{\mathcal F}_g({\mathcal N}^\sublR;l))},
\end{align}
where ${\mathcal F}_g$ is the effective Hamiltonian given by
\begin{align}
&{\mathcal F}_g({\mathcal N}^\sublR;~l~)=F_g^\sublR(T,l,A,{\mathcal N}^\sublR,mg)+F_g^\subuR(T,L-l,A,N-{\mathcal N}^\sublR,mg)+\frac{mg}{2}(lN-L{\mathcal N}^\sublR),
\end{align}
using \eqref{e:Fg-def}, \eqref{e:Fg-l}, and \eqref{e:Fg-u}.

\section{Uniqueness of the interface under gravity}
\label{s:uniqueness}

Let us examine the possibility of having two or more interfaces for $g\neq 0$. Taking the example of having two liquid-gas interfaces, we repeat the procedure in Secs. \ref{s:eff-Hamiltonian} and \ref{s:variational}.

We set two virtual sections at $x=\botX+l_1$ and $x=\topX-l_2$. The effective Hamiltonian is revised from \eqref{e:eff-Hamiltonian} to
\begin{align}
&{\mathcal F}_g({\mathcal N}^\sublR,{\mathcal N}^\subuR;~l_1,l_2)
=F_g^\sublR(T,l_1,A,{\mathcal N}^\sublR,mg)+
F_g^\submR(T,L-l_1-l_2,A,N-{\mathcal N}^\sublR-{\mathcal N}^\subuR,mg)\nonumber\\
&+F_g^\subuR(T,l_2,A,{\mathcal N}^\subuR,mg)
+\frac{mg}{2}\left[(L-l_2)(N-{\mathcal N}^\sublR)-(L-l_1)(N-{\mathcal N}^\subuR)\right], \label{e:eff-Hamiltonian-2}
\end{align}
where ${\mathcal N}^\sublR$ and ${\mathcal N}^\subuR$ are the numbers of particles in $x<\botX+l_1$ and $x>\topX-l_2$, respectively. The number of particles in the middle region $\botX+l_1<x<\topX-l_2$ is $N-{\mathcal N}^\sublR-{\mathcal N}^\subuR$.

The equilibrium state satisfies
\begin{align}
&(N^\sublR(l_1,l_2),N^\subuR(l_1,l_2))=\argmin_{{\mathcal N}^\sublR,{\mathcal N}^\subuR}{\mathcal F}_g({\mathcal N}^\sublR,{\mathcal N}^\subuR;l_1,l_2), \label{e:var-Nl-max2}
\end{align}
and therefore,
\begin{align}
&\pderf{{\mathcal F}_g}{{\mathcal N}^\sublR}{*}=0,\qquad \pderf{{\mathcal F}_g}{{\mathcal N}^\subuR}{*}=0.
\end{align}
The arbitrariness of $l_1$ and $l_2$ leads to
\begin{align}
&\pderf{{\mathcal F}_g}{l_1}{*}=0,\qquad \pderf{{\mathcal F}_g}{l_2}{*}=0.
\end{align}
These four equations yield the same results as in Sec. \ref{s:eff-Hamiltonian}, namely that $p(x)$ is continuous and $\mu(x)$ is linear.

We set a constraint
\begin{align}
X^\sublR=\xm^\sublR, \quad
X^\submR=\xm^\submR, \quad
X^\subuR=\xm^\subuR,
\end{align}
where $X^\sublR$, $X^\submR$, and $X^\subuR$ are the centers of mass in the respective regions, while $\xm^\sublR$, $\xm^\submR$, and $\xm^\subuR$ are the respective midpoints. The three regions are in single-phase states. 

The effective Hamiltonian yields the variational function 
\begin{align}
{\mathcal F}_g({\mathcal N}^\subl,{\mathcal N}^\subu,{\mathcal V}^\subl,{\mathcal V}^\subu)
&=
F(T,{\mathcal V}^\subl,{\mathcal N}^\subl)+
F(T,V-{\mathcal V}^\subl-{\mathcal V}^\subu,N-{\mathcal N}^\subl-{\mathcal N}^\subu)
+F(T,{\mathcal V}^\subu,{\mathcal N}^\subu)\nonumber\\
&+\frac{NmgL}{2}\left[\left(1-\frac{{\mathcal V}^\subu}{V}\right)\left(1-\frac{{\mathcal N}^\subl}{V}\right)
-
\left(1-\frac{{\mathcal V}^\subl}{V}\right)\left(1-\frac{{\mathcal N}^\subu}{V}\right)
\right], \label{e:Fg-var-lu2}
\end{align}
where $({\mathcal V}^\subl,{\mathcal N}^\subl)$ and $({\mathcal V}^\subu,{\mathcal N}^\subu)$ are the volume and number of particles in the phases located at the bottommost and topmost positions, respectively. The last term in \eqref{e:Fg-var-lu2} equals $Nmg(X-\xm)$.

The variational equations obtained from \eqref{e:Fg-var-lu2} are
\begin{align}
&\pderf{{\mathcal F}_g}{{\mathcal V}^\subl}{*}=0,\qquad \pderf{{\mathcal F}_g}{{\mathcal V}^\subu}{*}=0,\\
&\pderf{{\mathcal F}_g}{{\mathcal N}^\subl}{*}=0,\qquad \pderf{{\mathcal F}_g}{{\mathcal N}^\subu}{*}=0.
\end{align}
Substituting \eqref{e:Fg-var-lu2} into these equations, we obtain
\begin{align}
&(\bP^\subl-\bP^\subM)A=\frac{mg(N^\subl+N^\subM)}{2},\qquad
(\bP^\subM-\bP^\subu)A=\frac{mg(N^\subM+N^\subu)}{2}, \label{e:varV-2int}\\
&\bar\mu^\subl-\bar\mu^\subM=\frac{mg(V^\subl+V^\subM)}{2A},\qquad\qquad
\bar\mu^\subM-\bar\mu^\subu=\frac{mg(V^\subM+V^\subu)}{2A}. \label{e:varN-2int}
\end{align}
Here, the superscript ``$\subM$" indicates quantities in the phase located in the middle region. The two equations in \eqref{e:varV-2int} ensure the continuity of pressure at the interfaces. 
Let $p_{\theta 1}$ and $p_{\theta 2}$ be the pressures at the two interfaces.
Then, the two equations in \eqref{e:varN-2int} are transformed as
\begin{align}
\mu^\subl(T,p_{\theta 1})=\mu^\subM(T,p_{\theta 1}),\qquad
\mu^\subM(T,p_{\theta 2})=\mu^\subu(T,p_{\theta 2}),
\label{e:unstable-mu}
\end{align}
respectively. Because $\mu^\subl(T,p)$, $\mu^\subM(T,p)$, and $\mu^\subu(T,p)$ are either $\mu^\subL(T,p)$ or $\mu^\subG(T,p)$, \eqref{e:unstable-mu} indicates that the configurations $(\subL,\subG,\subL)$ and $(\subG,\subL,\subG)$ cannot be thermodynamically stable since $p_{\theta 1}\neq p_{\theta 2}$ when $g\neq 0$. 

Thus, we conclude that the possible configurations in equilibrium are limited to $(\subL,\subL,\subG)$, $(\subL,\subG,\subG)$, $(\subG,\subL,\subL)$, $(\subG,\subG,\subL)$, $(\subL,\subL,\subL)$, and $(\subG,\subG,\subG)$. This implies that the number of liquid-gas interfaces is at most one.

\newpage



\begin{thebibliography}{99}

\bibitem{Callen} 
H. B. Callen,
{\it Thermodynamics and an Introduction to Thermostatistics}, 2nd ed. (Wiley, New York, 1985)


\bibitem{Landau-Lifshitz-Statphys}
L. D. Landau and E. M. Lifshitz,
{\it Statistical Physics},
(Pergamon Press, Oxford, 1985).


\bibitem{Landau-Lifshitz-Fluid}
L. D. Landau and E. M. Lifshitz, 
{\it Fluid Mechanics}, (Pergamon Press, Oxford, 1959)


\bibitem{Prigogine-Kondepudi}
I. Prigogine and D. Kondepudi,
{\it Modern Thermodynamics : From Heat Engines to Dissipative Structures, Second Edition},(Wiley,  Chichester, 1998)


\bibitem{Global-PRL}
  N. Nakagawa and  S.-i. Sasa,
  Liquid-gas transitions in steady heat conduction,
  Phys. Rev. Lett. {\bf 119}, 260602 (2017)

\bibitem{Global-JSP}
N. Nakagawa and S.-i. Sasa, Global thermodynamics for heat
conduction states, 
J. Stat. Phys. {\bf 177}, 825-888 (2019)

\bibitem{Global-PRR}
N. Nakagawa and S.-i. Sasa, 
Unique extension of the maximum entropy principle to phase coexistence in heat conduction,
Phys. Rev. Res.  {\bf 4}, 033155 (2022)

\bibitem{KNS}
M. Kobayashi, N. Nakagawa, and S.-i. Sasa, 
Control of metastable states by heat flux
in the Hamiltonian Potts model, Phys. Rev. Lett. {\bf 130}, 247102  (2023)


\bibitem{Global-modelB}
S.-i. Sasa and N. Nakagawa,
Non-equilibrium phase coexistence in boundary-driven diffusive systems,
J. Stat. Phys., {\bf 192}, 26 (2025)

\bibitem{Zubarev} 
D. N. Zubarev,  
{\it Nonequilibrium Statistical Thermodynamics}, (Consultants Bureau, New York, 1974)


\bibitem{Mclennan}
J. A. McLennan, 
Phys. Fluids {\bf 3}, 493-502 (1960);
Introduction to Non-equilibrium Statistical Mechanics. Prentice-Hall (1988)

\bibitem{Maes-rep}
C.  Maes and  K. Neto\v{c}n\'{y},
Rigorous meaning of McLennan ensembles,
J. Math. Phys. {\bf 51}, 015219 (2010)


\bibitem{Kubo}
R. Kubo,  {\it Statistical Mechanics}, 2nd ed., (North-Holland, Amsterdam, 1988)

\bibitem{Einstein} 
  A. Einstein, A new determination of molecular dimensions
  (University of Zurich dissertation, 1905).
  A. Einstein, On the movement of small
  particles suspended in stationary liquids required by
molecular-kinetic theory of heat. Annalen der Physik 17, 549-560 (1905)


\bibitem{Onsager1931}
 L. Onsager, 
 Reciprocal relations in irreversible processes. I. Phys. Rev. {\bf 37}, 405-426 (1931);
 Reciprocal relations in irreversible processes. II. Phys. Rev. {\bf 38}, 2265-2279 (1931)
 

\bibitem{Groot-Mazur}
S. de Groot and P. Mazur, 
{\it Non-equilibrium Thermodynamics} (North-Holland, Amsterdam, 1962)


\bibitem{ANS}
A. Yoshida, N. Nakagawa, and S.-i. Sasa, 
Heat-induced liquid hovering in liquid-gas coexistence under gravity, Phys. Rev. Lett. {\bf 133},
117101  (2024).


\bibitem{Wagner}
L. E. Czelusniak, L. Cabezas-Gomez, and A. J. Wagner,
Effect of gravity on phase transition for liquid--gas simulations,
Phys. Fluids 35, 043324 (2023)


\bibitem{Global-Heat-Gravity}
N. Nakagawa and S.-i. Sasa,
Thermodynamic Variational Principle Unifying Gravity and Heat Flow, 
arXiv:2505.10380 (2025)


\bibitem{vanPress}
J. S. Rowlinson,
Translation of J. D. van der Waals' ``The thermodynamik theory of capillarity under the hypothesis of a continuous variation of density'',
J. Stat. Phys. {\bf 20} 197 (1979)


\bibitem{Kirkwood_Buff}
J. G. Kirkwood and F. P. Buff,
The Statistical Mechanical Theory of Surface Tension,
J. Chem. Phys. 17, 338 (1949)

\bibitem{BLS}
 F. P. Buff, R. A. Lovett, and F. H. Stillinger, 
 Interfacial density profile for fluids in the critical region,
  Phys. Rev. Lett. 15, 621-623 (1965)

\bibitem{Weeks}
J. D. Weeks,
Structure and thermodynamics of the liquid-vapor interface,
J. Chem. Phys. 67, 3106 (1977)

\bibitem{Rowlinson_Widom}
J. S. Rowlinson and B. Widom, 
Molecular Theory of Capillarity (Clarendon, Oxford, 1982)

\bibitem{Stillinger}
F. H. Stillinger, 
Capillary waves and the inherent density profile for the liquid-vapor interface,
J. Chem. Phys. 76, 1087 (1982)

\bibitem{Napiorkowski_Dietrich}
M. Napi\'orkowski and S. Dietrich,
Structure of the effective Hamiltonian for liquid-vapor interfaces,
Phys. Rev. E {\bf 47}, 1836 (1993)

\bibitem{Mecke_Dietrich}
K. R. Mecke and S. Dietrich,
Effective Hamiltonian for liquid-vapor interfaces,
Phys. Rev. E {\bf 59} 6766 (1999)

\bibitem{Daillant}
C. Fradin, A. Braslau, D. Luzet, D. Smilgies, M. Alba, N. Boudet, K. Mecke, and J. Daillant,
Reduction in the surface energy of liquid interfaces at short length scales,
Nature {\bf 403} 871 (2000)

\bibitem{Daillant2}
S. Mora, J. Daillant, K. Mecke, D. Luzet, A. Braslau, M. Alba, and B. Struth, 
X-Ray Synchrotron Study of Liquid-Vapor Interfaces at Short Length Scales: Effect of Long-Range Forces and Bending Energies,
Phys. Rev. Lett. {\bf 90}, 216101 (2003)

\bibitem{Rice}
D. Li, B. Yang, B. Lin, M. Meron, J. Gebhardt, T. Graber, and S. A. Rice, 
Wavelength Dependence of Liquid-Vapor Interfacial Tension of Ga,
Phys. Rev. Lett. {\bf 92}, 136102 (2004)

\bibitem{Chacon_Tarazona}
E. Chac\'on and P. Tarazona, 
Intrinsic Profiles beyond the Capillary Wave Theory: A Monte Carlo Study,
Phys. Rev. Lett. {\bf 91}, 166103 (2003)

\bibitem{Chacon_Tarazona2}
E. Chac\'on, E. M. Fern\'andez, and P. Tarazona,
Effect of dispersion forces on the capillary-wave fluctuations of liquid surfaces,
Phys. Rev. E {\bf 89}, 042406 (2014)

\bibitem{Muscatello}
J. Muscatello, E. Chac\'on, P. Tarazona, and F. Bresme,
Deconstructing Temperature Gradients across Fluid Interfaces: The Structural Origin of the Thermal Resistance of Liquid-Vapor Interfaces,
Phys. Rev. Lett. {\bf 119}, 045901  (2017)

 
\bibitem{CP}
N. Goldenfeld, 
Lectures on Phase Transitions and the Renormalization Group (Westview Press, 1992)

 \bibitem{Hohenberg_Halperin1977}
P.C. Hohenberg and B.I. Halperin,
Theory of dynamic critical phenomena,
Rev. Mod. Phys., {\bf 49}, 435--479, (1977)


\end{thebibliography}
\end{document}